\documentclass[12pt]{article}
\usepackage
[left=2.40cm, right=2.40cm, top=2.40cm, bottom=2.40cm]
{geometry}
\usepackage{setspace}

\usepackage[bbgreekl]{mathbbol}
\usepackage{amsmath,amssymb,amsfonts,amsxtra,mathrsfs,graphics,graphicx,amsthm,epsfig,ytableau,bm,longtable,float,color,tikz,mathtools,xfrac,footnote,rotating,lscape,subcaption}

\usepackage[debug,pageanchor=false]{hyperref}
\definecolor{link}{rgb}{0., 0.1, 0.5}
\hypersetup{colorlinks=true,linkcolor=link,citecolor=link,urlcolor=link,linktocpage}

\usepackage[titles]{tocloft}

\restylefloat{table}
\usepackage{dsfont}
\usepackage{multicol}
\usepackage{lipsum}
\usepackage{textgreek}
\usetikzlibrary{decorations.pathmorphing}
\usetikzlibrary{decorations.markings}
\usetikzlibrary{quotes,arrows.meta}
\usetikzlibrary{arrows, decorations.markings, calc, fadings, decorations.pathreplacing, patterns, decorations.pathmorphing, positioning}
\usepackage{tikz-cd}

\usepackage{fixmath} 
\usepackage{scalerel}
\newlength\bshft
\def\fakebold#1{\ThisStyle{\ooalign{$\SavedStyle#1$\cr%
  \kern-\bshft$\SavedStyle#1$\cr%
  \kern\bshft$\SavedStyle#1$}}}

\usetikzlibrary{positioning,shapes}
\usetikzlibrary{chains}
\usetikzlibrary{arrows,fit,decorations.pathreplacing}
\tikzstyle{every picture}+=[remember picture]
\tikzstyle{na} = [baseline=-.5ex]

\usepackage{empheq}
\usepackage{cite}
\usepackage{multirow}
\usepackage{booktabs}
\usepackage[american]{babel}

\usepackage[latin1]{inputenc}

\usepackage{array,booktabs}

\makeatletter
\newcommand{\vast}{\bBigg@{1}}
\newcommand{\Vast}{\bBigg@{5}}
\makeatother

\numberwithin{equation}{section}

\makeatletter
\@addtoreset{equation}{section}
\makeatother

\newcommand{\bea}{\begin{equation} \begin{aligned}} \newcommand{\eea}{\end{aligned} \end{equation}}

\newcommand{\ii}{\mathrm{i}}

\newcommand{\spindle}{\mathbb{\Sigma}}
\newcommand{\Morb}{\mathbb{M}}

\newcommand{\dd}{{\rm d}}
\newcommand{\im}{{\rm i}}

\providecommand{\abs}[1]{\lvert#1\rvert}

\newcommand{\cA}{\mathcal{A}}
\newcommand{\cB}{\mathcal{B}}
\newcommand{\cC}{\mathcal{C}}
\newcommand{\cD}{\mathcal{D}}

\newcommand{\cI}{\mathcal{I}}

\newcommand{\cL}{\mathcal{L}}

\newcommand{\cN}{\mathcal{N}}
\newcommand{\cO}{\mathcal{O}}
\newcommand{\cP}{\mathcal{P}}

\newcommand{\cR}{\mathcal{R}}

\newcommand{\cX}{\mathcal{X}}

\newcommand{\bC}{\mathbb{C}}

\newcommand{\bR}{\mathbb{R}}
\newcommand{\bT}{\mathbb{T}}
\newcommand{\bZ}{\mathbb{Z}}

\newcommand{\unit}{\mathbbm{1}}

\usepackage{bbold}

\usepackage[Symbol]{upgreek}
\usepackage{bm}

\usepackage{calligra}
\DeclareMathAlphabet{\mathcalligra}{T1}{calligra}{m}{n}

\makeatletter
\g@addto@macro\bfseries{\boldmath}
\makeatother

\makeatletter
\newcommand*{\rom}[1]{\expandafter\@slowromancap\romannumeral #1@}
\makeatother

\usepackage{bookmark}
\usepackage{cleveref}
\usepackage{physics}
\usepackage{tensor}
\usepackage{color}
\usepackage{enumerate}

\AtBeginEnvironment{pmatrix}{\everymath{\displaystyle}}
\AtBeginEnvironment{bmatrix}{\everymath{\displaystyle}}
\DeclareMathAlphabet{\mathpzc}{OT1}{pzc}{m}{it}

\newcommand{\iu}{\ensuremath{\mathrm{i}}}
\newcommand{\eu}{\ensuremath{\mathrm{e}}}
\newcommand{\curly}[1]{\ensuremath{\mathpzc{#1}}}

\usepackage{empheq}
\usepackage[most]{tcolorbox}

\newtcbox{\mymath}[1][]{%
    nobeforeafter, math upper, tcbox raise base,
    enhanced, colframe=blue!30!black,
    colback=blue!30, boxrule=1pt,
    #1}

\usepackage{tikz}
\usepackage{tikz-3dplot}
\tdplotsetmaincoords{60}{30} 
\usetikzlibrary{decorations.pathmorphing}

\newcommand{\xiLa}{\xi_a}
\newcommand{\XLa}{X_a}
\newcommand{\betaX}{\Theta}
\newcommand{\alphaX}{\vartheta}
\newcommand{\phiLa}{\varphi_a}
\newcommand{\rLa}{r_a}
\newcommand{\RLa}{R_a}
\newcommand{\cLa}{c_a}
\newcommand{\gR}{g_R}
\newcommand{\bLa}{b_a}
\newcommand{\D}{\cD}
\newcommand{\LB}{\cL}
\newcommand{\wa}{w^a}
\newcommand{\cxi}{c_\xi}
\newcommand{\sa}{\mathfrak s_a}

\newcommand{\Kform}{\cX}
\newcommand{\X}{\mathcal{X}}

\newcommand{\CBS}{\mathbb{C}\text{BS}}
\newcommand{\omegafuga}{\boldsymbol{\omega}_1}
\newcommand{\omegafugb}{\boldsymbol{\omega}_2}
\newcommand{\phiB}{\phi^B}

\begin{document}

%
%
	\begin{titlepage}
	~\\
		\begin{center}

		       	{\Large \bf{Equivariant localization in supergravity\\[3mm] in odd dimensions}}
			\vskip 5mm 
				\bigskip
				
					Edoardo Colombo,$^{a,b}$ Vasil Dimitrov$^{a,b}$, Dario Martelli,$^{a,b}$
					and Alberto Zaffaroni$^{c,d}$\\
					
				\bigskip
				\vskip 9mm 

{\footnotesize
{\it
 $^a$Dipartimento di Matematica,
 Universit\`a di Torino, Via Carlo Alberto 10, 10123 Torino, Italy \\
$^b$INFN, Sezione di Torino, Via Pietro Giuria 1, 10125 Torino, Italy \\
$^c$Dipartimento di Fisica, Universit\`a di Milano-Bicocca, I-20126 Milano, Italy \\
$^d$INFN, sezione di Milano-Bicocca, I-20126 Milano, Italy
}
}

       \vskip .2in
       
{\footnotesize
       {\it E-mail:}
       \href{edoardo.colombo@unito.it}{\texttt{edoardo.colombo@unito.it}},
       \href{vasilradoslavov.dimitrov@unito.it}{\texttt{vasilradoslavov.dimitrov@unito.it}},
       \href{dario.martelli@unito.it}{\texttt{dario.martelli@unito.it}},
       \href{alberto.zaffaroni@mib.infn.it}{\texttt{alberto.zaffaroni@mib.infn.it}}
}       
       \vskip 1.3in
       {\bf Abstract } 
       \vskip .1in
       \end{center}
       
\noindent        
We discuss a localization formula for certain integrals on odd-dimensional manifolds with boundaries, 
equipped with a Killing vector,  and employ this to localize  the regularised on-shell action of a large class of  supersymmetric solutions 
of five dimensional minimal gauged supergravity. Specifically, we consider asymptotically AdS$_5$ 
 solutions in the time-like class, in which the transverse K\"ahler foliation is assumed to 
be toric. We find that the background subtraction regularization method
leads to an intriguing formula for the on-shell action, in terms 
of an analytic continuation of the Martelli-Sparks-Yau  Sasakian volume. In particular, we show that the regularised on-shell action is 
 a function of the toric data of an effective compact five-dimensional manifold,  as well as of the supersymmetric Killing vector, outside the corresponding dual cone. 
As our main example we provide a derivation of the well-known entropy function of supersymmetric and rotating black holes in AdS$_5$, using only topological data. 
       \end{titlepage}

%
%

\setcounter{tocdepth}{2}

\hrule

\tableofcontents

\vspace{0.65cm}
\hrule

%
%

\section{Introduction}
\label{sect:intro}

Equivariant localization has a rich and fascinating history.  The classic result of  Duistermaat-Heckmann \cite{Duistermaat} and
the Berline-Vergne-Atiyah-Bott (BVAB) formula  \cite{berline1982classes,ATIYAH19841} in equivariant cohomology have
 numerous applications both to mathematics and physics. In the context of supersymmetric field theories,  equivariant localization is at the core of the work of Nekrasov on instanton counting \cite{nekrasov2002seibergwitten} and it is instrumental in the computation of the Martelli-Sparks-Yau (MSY) volume functional of  Sasakian manifolds \cite{Martelli:2005tp,Martelli:2006yb}, just to mention few examples. It also played a fundamental role in the evaluation of supersymmetric partition functions and indices, starting with  \cite{Pestun:2007rz}. Localization has also been applied to the calculation of the supergravity path integral \cite{Dabholkar:2010uh,Dabholkar:2014wpa}. 
 
 More recently, equivariant localization has been applied to the calculation of on-shell actions in supergravity and their applications to holography. Different approaches have been pursued in this context.  In the work of \cite{BenettiGenolini:2023kxp} the action is re-written as an integral of an equivariantly closed form that can be constructed using Killing spinor bilinears and evaluated using the BVAB formula. In \cite{Martelli:2023oqk} the universal role of the equivariant volume of the internal manifold to understand extremization problems in holography and in black hole physics is emphasized. All these investigations have applied equivariant localization to even-dimensional manifolds or orbifolds, where Killing vectors can have isolated fixed points. Less has been done for odd-dimensional spaces, where the fixed loci of the torus action are odd-dimensional\footnote{For an earlier application 
 of the method of equivariant localization for the computation of an on-shell action in field theory see \cite{Panerai:2020boq}. In particular, in this reference the authors applied the BVAB localization formula to three-dimensional manifolds.}. The approach of  \cite{Martelli:2023oqk,Colombo:2023fhu} has been applied to odd-dimensional spaces  but only in an indirect way, by considering localization on the cone over the internal manifold. On the other hand, the method of   \cite{BenettiGenolini:2023kxp} has been applied to the evaluation of the action for black holes and solitons in ungauged supergravity in the recent paper \cite{Cassani:2024kjn}, using the BVAB formula in five dimensions.  For the moment this approach has been applied only to ungauged supergravity. Moreover, localization has been performed only with respect to a subgroup of the isometry group and some explicit information about the solutions is required.
 
 In this paper, we would like to use an approach that it is only sensitive to the topology of the spaces, it allows to localize with respect to the entire isometry group and it is applicable to various holographic solutions of interest. To this purpose, we consider an alternative to the BVAB formula which have been  proposed in \cite{Goertsches:2015vga} and applies to odd-dimensional foliations. 
 The formula can be used for manifolds $M$ with a torus $\mathbb{T}$ effective action which degenerates to
an $S^1$
action on a finite number of loci $L_a$. 
 $M$ is also equipped with a vector field $\xi$ in $\mathbb{T}$ which defines a foliation. The localization formula will allow to express integrals of the form  
\begin{equation}
\int_M \eta \wedge \Phi \, ,
\end{equation}
as a sum of contributions from the degeneration loci $L_a$. Here $\eta$ is any one-form satisfying $\iota_\xi \dd\eta=0$ and $\Phi$ is a even-degree polyform which is basic with respect to the foliation,
$\iota_\xi \Phi = \cL_\xi \Phi =0$, and equivariantly closed with respect to another vector field $X$ in $\mathbb{T}$
\begin{equation}
	\dd_X\Phi = \dd\Phi -\iota_X\Phi=0 \:,\qquad\cL_X\Phi=0\:.
\end{equation} 
In typical applications, we want to compute integrals of polyforms $\eta\wedge \Phi$ whose component of highest  degree equals the dimension of $M$ so that the final result is $X$-independent.
Notice that
the final result does depend on $\xi$. As an application of their result, the authors of \cite{Goertsches:2015vga} have  given an alternative derivation of the volume of Sasakian manifolds \cite{Martelli:2005tp,Martelli:2006yb} which is a function of the Reeb vector. In our applications to supergravity, $\xi$ will be the supersymmetric Killing vector,
 constructed as a Killing spinor bilinear.

In this paper we generalize the theorem of \cite{Goertsches:2015vga} to toric manifolds with boundary and apply it to both compact and non-compact geometries. In the compact case, we review how the localization formula 
can be used to derive the Sasakian volume of \cite{Martelli:2005tp,Martelli:2006yb},  and we extend the argument   to 
the Gauntlett-Martelli-Sparks (GMS) master volume \cite{Gauntlett:2018dpc}.
Yet, our main application of the localization formula is to the calculation of the action of supersymmetric solutions  to five dimensional minimal gauged supergravity. 
We show that, using the supersymmetry conditions,  the (Euclidean) action can be written as 
\begin{equation}\label{Iaction}
	16\pi \iu G_5\cdot I[M]=-\int_M\eta\wedge(\dd\eta)^2+\int_{\partial M}\alpha(M)\:,
\end{equation}
where $\eta$ is a one-form, $\dd\eta$ is basic with respect to the supersymmetric Killing vector $\xi$ and $\alpha(M)$ is a boundary term. When $\eta$ is regular everywhere, 
we can apply the localization formula to the bulk integral in \eqref{Iaction} obtaining a sum over degeneration loci plus an extra boundary contribution.  
We will show that, quite remarkably, if we regularize the divergent result by subtracting the contribution of AdS$_5$, the total boundary piece exactly cancels. The final result for the on-shell action is just the sum of contributions of the degeneration loci $L_a$ of the supersymmetric solution minus the contribution of a single degeneration locus for AdS$_5$. We stress that these contributions can be evaluated using only topological data and  without a detailed knowledge of the solution.

The main example is the derivation of the on-shell action  of the asymptotically AdS$_5\times S^5$ supersymmetric black holes with equal electric charges,
considered in \cite{Chong:2005hr,Cabo-Bizet:2018ehj}.
The Euclidean metric is smooth although complex,
thus we denote the Euclidean 5d manifold as $\CBS$, standing for complexified black saddle in the nomenclature of \cite{Bobev:2020pjk}.
By a simple analysis
of the topology of the solution near the degeneration loci of the toric $\text{U}\qty(1)^3$ action we find 
\begin{equation}\label{IBH} \widehat{I} = \frac{\pi \ell^3}{108 G_5}\frac{\qty(\omegafuga+\omegafugb -2\pi \ii)^3}{\omegafuga\omegafugb} \, ,\end{equation}
where the hat denotes that \eqref{Iaction} is renormalized appropriately. The supersymmetric Killing vector is 
\begin{equation}\label{famousformula} \xi =\sum_{i=0}^2 \xi_i \partial_{\phi_i} = \frac{1}{\beta}( 2\pi\ii \partial_{\phi_0} - \omegafuga \partial_{\phi_1} -\omegafugb \partial_{\phi_2})\, ,\end{equation}
for a basis of $2\pi$-periodic angles. Formula \eqref{famousformula} is famous for its applications to the microstate counting of AdS$_5\times S^5$ supersymmetric black holes \cite{Cabo-Bizet:2018ehj,Choi:2018hmj,Benini:2018ywd}. It was first conjectured in \cite{Hosseini:2017mds} and then derived from an explicit computation of the action in \cite{Cabo-Bizet:2018ehj} using the Euclidean supersymmetric complex deformation of the
extremal
black hole
--- a.k.a. the $\CBS$. For the first time here we derive it from equivariant localization with purely topological arguments.

To apply the localization formula we need a globally well defined one-form
$\eta$.
This poses, at the moment, some restrictions on the class of solutions we can consider. 
We will make a prediction for the on-shell action of a  general class of toric asymptotically AdS$_5$ backgrounds characterised by the absence of magnetic fluxes. As we will see, these 
can be equivalently described as satisfying a certain ``Calabi-Yau'' condition, $\sum_{i=0}^2 \nu_i V^a_i=1$,  where $V^a$ are the vectors of the fan defined in section \ref{Five-dimensional toric geometries}, and $\nu_i$ are some rational numbers.
The subtraction of the degeneration locus of AdS$_5$ is effectively equivalent to a sort of {\it compactification} of the geometry and we have an intriguing prediction for the on-shell action
\begin{equation}
  \widehat{I} =\frac{\big(\frac{\ell}3\,\nu_i\,\xi_i\big)^3}{2\pi\ii\,G_5}\:\text{Vol}_Y(\xi)\:,
\end{equation}
where $\text{Vol}_Y(\xi)$ is the Sasakian volume of the compact five-dimensional manifold with fan $V^a$. In the case of AdS$_5\times S^5$ supersymmetric black holes the manifold $Y$ is just $S^5$ whose Sasakian volume is 
\begin{equation} \text{Vol}_{S^5}(\xi)=\frac{\pi^3}{\xi_0\xi_1\xi_2}\, ,\end{equation}
and $\nu_i=1$ for $i=0,1,2$.

Our formalism still needs to be extended to cases where the supergravity gauge field has a non-trivial topology and we plan to do it in the near future. This would allow us to include solutions like the topological soliton discussed in \cite{Chong:2005hr,Cassani:2015upa}, as well as general solutions (still to be found) corresponding to multi black holes and black lenses. We considered only toric solutions of minimal gauged supergravity but we are certain that the odd-dimensional localization formula that we have presented can be useful to study non-toric examples and general supergravities with vector and hypermultiplets. All this is left for future work.
 
 The paper is organized as follows. In section \ref{sect:theorem} we first re-derive and generalize the theorem of \cite{Goertsches:2015vga} to toric manifolds with boundary. We then discuss the main ingredients of the localization formula for a class of general compact and non-compact toric geometries in five dimensions. To make contact with the previous literature  we review how the localization formula 
can be used to derive the Sasakian volume of \cite{Martelli:2005tp,Martelli:2006yb}, following  \cite{Goertsches:2015vga}, and we extend the argument to derive  a useful generalization of the Sasakian volume,
the GMS master volume \cite{Couzens:2018wnk,Gauntlett:2018dpc}. In section \ref{sec:sugra5d}, we rewrite the action of minimal five-dimensional gauged supergravity as the integral of an equivariant form for a general supersymmetric solution with a time-like Killing vector. We also discuss the toric geometry structure of such solutions. 
In section \ref{sect:on shell localization} we discuss the localization of the on-shell action. We first discuss the regularity conditions we have to impose on the one-form $\eta$. We then derive the on-shell action  for the asymptotically AdS$_5\times S^5$ supersymmetric black holes with equal electric charges by using only topological arguments. We also provide a general prediction for the on-shell action of a class of toric backgrounds, finding a remarkable connection with the Sasakian volume of compact manifolds. We finish the section by proving that, using the background subtraction method, all boundary contributions to the action cancel out.  In section \ref{sect:Discussion} we provide conclusions and a discussion of the results obtained. The paper contains four appendices. In appendix \ref{app:toric} we review the elements of toric geometry that have been used in the text. In appendix \ref{app:known} {we summarise some aspects of the complexified supersymmetric black hole solution of \cite{Chong:2005hr} in different sets of coordinates. In appendix \ref{comparisonwithlucietti} we discuss the toric geometry and some global aspects of such solution from a four-dimensional perspective. Finally, in appendix  
\ref{app:FG expansions} we give details about the asymptotic expansion of the solution,
needed to demonstrate the vanishing of the overall boundary contribution.

\newpage

\section{Equivariant localization in odd dimensions}

\label{sect:theorem}

In this section we present an equivariant localization formula for odd dimensional manifolds, which generalizes
the theorem of \cite{Goertsches:2015vga} to manifolds with boundary.
This formula assumes toric symmetry, meaning that we consider manifolds equipped with an action of the torus $\bT=U(1)^{\frac{D+1}2}$,
where $D$ is the dimension of the manifold.
An element of the algebra of $\bT$ generates orbits in $M$, and the collection of all such orbits is a foliation of $M$.
This foliation plays an important role in the localization formula.
We will discuss the geometric set up and the precise statement of the localization theorem in section \ref{Geometric set up and statement of the theorem},
while the derivation of the formula will be in section \ref{Derivation of the localization formula}.
As a warm-up, in section \ref{sec:S3} we discuss a very simple three-dimensional example involving $S^3$ and its hemispheres.
At last in \ref{Five-dimensional toric geometries} we discuss five-dimensional geometries, both compact (re-deriving the expression for the
Sasakian volume \cite{Martelli:2005tp,Martelli:2006yb} and the GMS master volume \cite{Couzens:2018wnk,Gauntlett:2018dpc})
and non-compact.

\subsection{Geometric set up and statement of the theorem}
\label{Geometric set up and statement of the theorem}

Let $M$ be a $D$-dimensional orientable manifold with boundary, where $D$ is an odd number.
We take $M$, with the inclusion of the boundary $\partial M$, to be compact,
and we assume that there is a torus $\bT= \text{U}(1)^{\frac{D+1}2}$ effective action on $M$ 
that does not have any fixed points on $\partial M$.
We will denote with $L_a$ the loci where the $\bT$ action degenerates to a circle action;
the $L_a$ are thus submanifolds diffeomorphic to $S^1$ where $\bT$ acts with a $\text{U}(1)^{\frac{D-1}2}$ isotropy subgroup.

It is convenient to define a Riemannian metric $\gR$ on $M$ such that the $\bT$ action is contained in the isometry group of $(M,\gR)$.
Since the integral that we will consider does not involve $\gR$, the choice of this metric is arbitrary;
in particular when we will apply the localization formula presented in this section to supergravity solutions,
$\gR$ does not need to be related to the physical metric of the solution.%
\footnote{Although for certain steps of the computation it will indeed be convenient to relate them.}
Sufficiently close to $L_a$ the metric is approximately flat and can be written as
\begin{equation}
\label{flat_metric}
	\gR\big|_{\text{near }L_a}\approx\big(\RLa\,\dd\phiLa^0\big)^2+\sum_{i=1}^{\frac{D-1}2}
		\Big[\big(\dd\rLa^i\big)^2+(\rLa^i)^2\big(\dd\phiLa^i+\cLa^i\dd\phiLa^0\big)^2\Big]\:,
\end{equation}
where $\RLa$ and $\cLa^i$ are real constants, the $\rLa^i$ are radial coordinates, while the $\phiLa^i$ are $2\pi$-periodic angular coordinates,
such that the torus $\bT$ acts by rotating them.
We can always choose these coordinates so that the ordering $(\phiLa^0,\rLa^1,\phiLa^1,\rLa^2,\phiLa^2,\,\ldots)$
matches the convention for positive orientation on $M$.

Let $\xi$ be a vector field on $M$ generated by an element of the algebra of $\bT$ (thus a Killing vector for $\gR$), 
and let us consider the foliation whose leaves are the orbits of $\xi$.
A differential form $\alpha$ is basic (with the respect to the foliation induced by $\xi$) if its contraction with $\xi$ and Lie derivative along $\xi$ vanish:
\begin{equation}
	\iota_\xi\alpha=0=\cL_\xi\alpha\:,\qquad\text{iff }\alpha\text{ is a basic form.}
\end{equation}
We can expand $\xi$ in the $\partial_{\phiLa^i}$ basis,
\begin{equation}
\label{xiLa}
	\xi={\sum}_{i=0}^{\frac{D-1}2}\xiLa^i\,\partial_{\phiLa^i}\:,
\end{equation}
where $\xiLa^i$ are constants. 
We will assume that $\xiLa^0\ne0$ for all $L_a$.
\footnote{In \cite{Goertsches:2015vga} a stronger condition is demanded, that is that $\xi$ has finitely many closed orbits,
		in which case the closed orbits are exactly the $L_a$. For only finitely many orbits to exist, the ratios of the components of $\xi$ in any base
		of $2\pi$-periodic coordinates must be irrational numbers. The authors of \cite{Goertsches:2015vga} then comment that the result obtained
		with their formula can be extended to more generic choices of $\xi$ by continuity.}

We will use equivariant localization to compute the integral $\int_M\eta\wedge\Phi$, where $\eta$ is a any one-form on $M$ such that
\begin{equation}
\label{eta_property}
	\iota_\xi\dd\eta=0\:,
\end{equation}
while $\Phi$ is a basic even-degree polyform on $M$, that is $\Phi$ is the formal sum of basic $2k$-forms $\Phi_{2k}$:
\begin{equation}
	\Phi=\Phi_{D-1}+\Phi_{D-3}+\ldots+\Phi_0\:.
\end{equation}
By definition the integral of a polyform corresponds to the integral of its $D$-form component;
in particular the integral of $\eta\wedge\Phi$ is just the integral of $\eta\wedge\Phi_{D-1}$.
The forms $\Phi_{D-1},\ldots,\Phi_{0}$ are introduced in order to have a polyform that is closed in the equivariant basic de Rham cohomology,
which we will now introduce.

Let us consider an arbitrary vector field $X$ on $M$ generated by an element of the algebra of $\bT$.
The only conditions we require on $X$ are that $X$ and $\xi$ are not collinear in $M\smallsetminus\bigcup_aL_a$, and that $\XLa^0\ne0$ for all $L_a$,
where $X=\sum_{i=0}^{(D-1)/2}\XLa^i\,\partial_{\phiLa^i}$.%
\footnote{Only a measure-zero set of vectors generated by elements of the algebra of $\bT$ does not satisfy these conditions.}
We introduce the equivariant differential $\dd_X$, which acts on polyforms and whose square is (minus) the Lie derivative along $X$, by Cartan's formula:
\begin{equation}
	\dd_X\equiv\dd-\iota_X\:,\qquad\dd_X^2=-\cL_X\:.
\end{equation}
The equivariant differential generates a cohomology on polyforms that are invariant along the orbits of $X$.
We require the basic polyform $\Phi$ to be equivariantly closed:
\begin{equation}
\label{equivariantly_closed}
	\dd_X\Phi=0\:,\qquad\cL_X\Phi=0\:.
\end{equation}
In all the applications that we will consider, we will first fix $\Phi_{D-1}$ so that $\eta\wedge\Phi_{D-1}$
matches the
integrand that we want to integrate, then choose an arbitrary $X$ and construct the other components of $\Phi$ so that \eqref{equivariantly_closed}
is verified. The localization formula recasts $\int_M\eta\wedge\Phi_{D-1}$ as a sum over various localized contributions, each one depending on $X$;
when all the contributions are taken into account, the $X$-dependence drops out, as expected.

We can finally state the equivariant localization formula for the foliation generated by $\xi$, which recasts the integral of $\eta\wedge\Phi$
as a sum over contributions localized at the $L_a$ plus integrals on the boundary $\partial M$:
\begin{empheq}[box={\mymath[colback=white, colframe=black]}]{equation}
\label{localization_formula}
	\int_M\eta\wedge\Phi\,=\,(-2\pi)^{\frac{D-1}2}\sum_a\frac{\Phi_0\big|_{L_a}\cdot\int_{L_a}\eta}
		{\,{\displaystyle\prod}_{\,i=1}^{\,\frac{D-1}2}\!\left(\frac{\XLa^0}{\xiLa^0}\:\xiLa^i-\XLa^i\right)}\:
		+\:\sum_{j=0}^{\frac{D-3}2}\int_{\partial M}\eta\wedge\betaX\wedge\big(\dd\betaX\big)^j\wedge\Phi_{D-2j-3}\:.
\end{empheq}
At the numerator of the contribution localized at $L_a$ we have the function $\Phi_0$ evaluated at $L_a$: since $\cL_X\Phi=0$,
$\Phi_0$ is constant along the orbits of $X$, 
which implies that $\Phi_0\big|_{L_a}$ is a constant, because $L_a$ is an orbit of $X$.
The form $\betaX$ appearing in the boundary terms is a one-form on $M\smallsetminus\bigcup_aL_a$ defined by
\begin{equation}
\label{betaX_def}
	\betaX=\frac{\alphaX}{\alphaX(X)}\:,
\end{equation}
 where $\alphaX$ is an arbitrary $\bT$-invariant basic one-form on $M$ satisfying
\begin{equation}
	\alphaX_p(X)\geq0\quad\forall\,p\in M\:,\qquad\alphaX_p(X)=0\quad\text{iff}\quad p\in\textstyle\bigcup_aL_a
\end{equation}
A simple way to construct such a form $\alphaX$ is to define the vector $Y=\gR(\xi,\xi)\,X-\gR(\xi,X)\,\xi$,
and then setting $\alphaX$ as the dual form to $Y$, $\alphaX_\mu=(\gR)_{\mu\nu}Y^\nu$.
By Cauchy-Schwartz inequality $\alphaX(X)=\gR(\xi,\xi)\gR(X,X)-\gR(\xi,X)^2\geq0$, and the equal sign holds only where $\xi$ and $X$ are collinear,
which must happen at the $L_a$ since the $\bT$ action degenerates to a single circle and does not happen on $M\smallsetminus\bigcup_aL_a$
by assumption. It is then easy to verify that $\iota_\xi\alphaX=0$ and that the Lie derivative along any vector generated by an element of the algebra of $\bT$
vanishes.

An important observation is that the localization formula \eqref{localization_formula} is also valid for a \emph{complex valued} polyform
$\eta\wedge\Phi\in\big[\bigoplus_k\Omega^k(M)\big]_\bC$, where the complex valued one-form $\eta\in\Omega^1(M)_\bC$ satisfies
$\iota_\xi\dd\eta$ with respect to a complex vector $\xi$, which is generated by an element of the complexified Lie algebra Lie$(\bT)_\bC$.
The one-form $\alphaX$ (and by extension $\betaX$) can also be taken to be complex valued, and the condition $\alphaX(X)\geq0$
can be relaxed to Re$\,\alphaX(X)\geq0$, while still imposing $\alphaX_p(X)=0$ for $p\in\textstyle\bigcup_aL_a$.

\subsection{Derivation of the localization formula}
\label{Derivation of the localization formula}

In this section we show how to derive the equivariant localization formula \eqref{localization_formula}.
We will not attempt to generalize the cohomological approach of \cite{Goertsches:2015vga},
but rather we will follow a different route, which closely resembles Witten's method for deriving the BVAB formula \cite{Witten:1982im}.
The key idea is that it is possible to insert the term $\exp(t\,\dd_X\alphaX)$ in the integrand without affecting the integral (up to boundary contributions,
which we carefully keep track of in a similar manner as \cite{Couzens:2024vbn}),
where $t$ is an arbitrary real parameter. The localization formula is then recovered in the $t\to+\infty$ limit.

Using that $\dd_X\Phi=0$ and $\dd_X^2\alphaX=-\cL_X\alphaX=0$ we can write
\begin{equation}
	\int_M\eta\wedge\Phi\wedge(\dd_X\alphaX)^n=-\int_M\dd_X\Big(\eta\wedge\Phi\wedge\alphaX\wedge(\dd_X\alphaX)^{n-1}\Big)+
		\int_M(\dd_X\eta)\wedge\Phi\wedge\alphaX\wedge(\dd_X\alphaX)^{n-1}\:.
\end{equation}
The integrand of the second term on the right hand side vanishes when contracted with $\xi$, since $\Phi$ and $\alphaX$ are basic
and $\eta$ satisfies \eqref{eta_property}; hence the polyform integrand cannot have a $D$-form component%
\footnote{The contraction of a non-zero $D$-form with a non-zero vector is never zero.
		This is true also for complex valued $D$-form and vector.}
and its integral is zero by definition.
As for the other integral, we can replace the overall $\dd_X$ with $\dd$ (the $\iota_X$ piece must integrate to zero by a similar argument) and apply Stokes,
finding
\begin{equation}
	\int_M\eta\wedge\Phi\wedge(\dd_X\alphaX)^n=-\int_{\partial M}\eta\wedge\Phi\wedge\alphaX\wedge(\dd_X\alphaX)^{n-1}\:.
\end{equation}
The polyform $\dd_X\alphaX$ is invertible on $M\smallsetminus\bigcup_aL_a$ and its inverse is
\begin{equation}
\label{polyform_inverse}
	(\dd_X\alphaX)^{-1}=-{\sum}_{\,j=0}^{\,\frac{D-1}2}\:\frac{(\dd\alphaX)^j}{\alphaX(X)^{j+1}}\:.
\end{equation}
Since $\partial M\cap L_a=\varnothing$, we can write $(\dd_X\alphaX)^{n-1}$ as $(\dd_X\alphaX)^{-1}\wedge(\dd_X\alphaX)^{n}$.
Acting on both sides with $\sum_{n=1}^\infty\frac{t^n}{n!}$ for an arbitrary $t$ we get
\begin{equation}
	\int_M\eta\wedge\Phi\wedge\Big(\eu^{t\,\dd_X\alphaX}-1\Big)=
		-\int_{\partial M}\eta\wedge\Phi\wedge\alphaX\wedge(\dd_X\alphaX)^{-1}\wedge\Big(\eu^{t\,\dd_X\alphaX}-1\Big)\:.
\end{equation}
The key observation is that the above expression is true for any $t$, therefore we take the $t\to+\infty$ limit to simplify it.
In this limit the exponential term is suppressed where $\alphaX(X)>0$, since it can be rewritten as
\begin{equation}
\label{exponentiated_polyform}
	\eu^{t\,\dd_X\alphaX}=\eu^{t\,\dd\alphaX}\,\eu^{-t\,\alphaX(X)}=\eu^{-t\,\alphaX(X)}\,{\sum}_{\,n=0}^{\,\frac{D-1}2}\:\frac{t^n}{n!}\,(\dd\alphaX)^n\:.
\end{equation}
Considering that $\alphaX(X)\geq0$ everywhere (or more generally Re$\,\alphaX(X)\geq0$) and $\alphaX(X)=0$ only at the $L_a$, we find
\begin{equation}
	\int_M\eta\wedge\Phi=
		\sum_{a}\lim_{t\to+\infty}\bigg[\int_{\text{near }L_a}\!\!\!\!\!\!\!\!\!\!\!\!\eta\wedge\Phi\wedge \eu^{t\,\dd_X\alphaX}\bigg]
		-\int_{\partial M}\eta\wedge\alphaX\wedge(\dd_X\alphaX)^{-1}\wedge\Phi\:.
\end{equation}
Using \eqref{polyform_inverse} and \eqref{betaX_def}, the boundary term in the above expression matches the one of 
the localization formula \eqref{localization_formula}.
It remains to compute the $t\to+\infty$ limit of the contribution localized at $L_a$.

\subsubsection{Expansion near the localization loci}

Using \eqref{exponentiated_polyform} we can make explicit which are the integrals that we need to evaluate in the limit of large $t$:
\begin{equation}
\label{localized_contribution}
	\lim_{t\to+\infty}\int_{\text{near }L_a}\!\!\!\!\!\!\!\!\!\!\!\!\eta\wedge\Phi\wedge \eu^{t\,\dd_X\alphaX}=
		\lim_{t\to+\infty}{\sum}_{\,n=0}^{\,\frac{D-1}2}\:\,\frac{t^n}{n!}
		\int_{\text{near }L_a}\!\!\!\!\!\!\!\!\!\!\!\! \eu^{-t\,\alphaX(X)}\:\eta\wedge\Phi_{D-2n-1}\wedge(\dd\alphaX)^n\:.
\end{equation}
First, let us work out the expansion of the form $\alphaX$ near $L_a$.
Since $\iota_\xi\alphaX=0$ and $\xi$ has components $\xiLa^i$ in the $\partial_{\phiLa^i}$ basis \eqref{xiLa},
the angular components of $\alphaX$ must involve the combinations $\xiLa^0\,\dd\phiLa^i-\xiLa^i\,\dd\phiLa^0$, with $i=1,\ldots,\frac{D-1}2$.
Furthermore, none of the components of $\alphaX$ can depend on the angles, considering that $\alphaX$ is $\bT$-invariant.
We can then write $\alphaX$ as follows:
\begin{equation}
	\alphaX={\sum}_{\,i=1}^{\,\frac{D-1}2}\bigg[\,\bLa^i\,(\rLa^i)^2\big(\xiLa^0\,\dd\phiLa^i-\xiLa^i\,\dd\phiLa^0\big)+
		\widehat\bLa^i\,\dd\big((\rLa^i)^2\big)\Big]\big(1+\cO(\rLa^i)\big)\:,
\end{equation}
where $\bLa^i$ and $\widehat\bLa^i$ are constant.%
\footnote{As an exercise it is possible to work out the value of $\bLa^i$ and $\widehat\bLa^i$ if one chooses to set $\alphaX_\mu=(\gR)_{\mu\nu}Y^\nu$,
		where $Y=\gR(\xi,\xi)\,X-\gR(\xi,X)\,\xi$. Using \eqref{flat_metric}, they are $\bLa^i=\xiLa^0\,\XLa^i-\xiLa^i\,\XLa^0$ and $\widehat\bLa^i=0$.
		Ultimately, the value of these constants does not affect the result for the $L_a$ contribution.}
Note that we are implicitly using that $\alphaX$ must be a regular form at $L_a$,
meaning that the differentials of the angles require at least a radius squared factor in front, and so forth.
From the above expression we can immediately find the expansion of $\alphaX(X)$ and $\dd\alphaX$ near $L_a$:
\begin{equation}
\begin{aligned}
	\alphaX(X)\,\big|_{\text{near }L_a}\,&\approx{\sum}_{\,i=1}^{\,\frac{D-1}2}\,\bLa^i\,(\rLa^i)^2\big(\xiLa^0\,\XLa^i-\xiLa^i\,\XLa^0\big)\:,\\
	\dd\alphaX\,\big|_{\text{near }L_a}\,&\approx{\sum}_{\,i=1}^{\,\frac{D-1}2}\,\bLa^i\,\dd\big((\rLa^i)^2\big)\wedge
		\big(\xiLa^0\,\dd\phiLa^i-\xiLa^i\,\dd\phiLa^0\big)\:.
\end{aligned}
\end{equation}
The requirement $\alphaX(X)\geq0$ fixes the sign of $\bLa^i$ to be the same as the sign of $\xiLa^0\,\XLa^i-\xiLa^i\,\XLa^0$
(more generally, for complex $\alphaX$ we have Re$\,[\bLa^i(\xiLa^0\,\XLa^i-\xiLa^i\,\XLa^0)]\geq0$).

Only the region around $L_a$ with $\rLa^i\lesssim t^{-1/2}$ gives a meaningful contribution to the integral
of $\exp\big(-t\,\alphaX(X)\big)\cdot(D\text{-form})$.
For sufficiently large $t$ we can then ignore  the dependence on the radial and angular (except $\phiLa^0$) coordinates of the $D$-form,
and we can extend the range of $\rLa^i$ to be $(0,+\infty)$. Thus 
\begin{equation}
\label{asymptotic_integral}
	\int_{\text{near }L_a}\!\!\!\!\!\!\!\!\!\!\!\! \eu^{-t\,\alphaX(X)}(D\text{-form})\approx
		\int_0^{2\pi}\!\!\!\!\!\dd\phiLa^0\,(\text{function of }\phiLa^0)\cdot{\prod}_{\,i=1}^{\,\frac{D-1}2}\,
		2\pi\int_0^\infty\!\!\!\!\dd\big((\rLa^i)^2\big)\: \eu^{-t\,\bLa^i\,(\rLa^i)^2\big(\xiLa^0\,\XLa^i-\xiLa^i\,\XLa^0\big)}\:,
\end{equation}
and the integral of the exponential term is easily computed:
\begin{equation}
	\int_0^\infty\!\!\!\!\dd\big((\rLa^i)^2\big)\: \eu^{-t\,\bLa^i\,(\rLa^i)^2\big(\xiLa^0\,\XLa^i-\xiLa^i\,\XLa^0\big)}=
		\Big(t\,(\bLa^i)\,\big(\xiLa^0\,\XLa^i-\xiLa^i\,\XLa^0\big)\Big)^{-1}\:.
\end{equation}
An immediate consequence of this is that the $n=\frac{D-1}2$ term in the right hand side of \eqref{localized_contribution} gives a non-vanishing
contribution for $t\to+\infty$, since each integral brings a factor of $t^{-\frac{D-1}2}$. We only need to compute
\begin{equation}
\label{localized_contribution2}
	\lim_{t\to+\infty}\int_{\text{near }L_a}\!\!\!\!\!\!\!\!\!\!\!\!\eta\wedge\Phi\wedge \eu^{t\,\dd_X\alphaX}=
		\lim_{t\to+\infty}t^{\frac{D-1}2}
		\int_{\text{near }L_a}\!\!\!\!\!\!\!\!\!\!\!\! \eu^{-t\,\alphaX(X)}\:\Phi_0\cdot\big(\tfrac{D-1}2\,!\,\big)^{-1}\,\eta\wedge(\dd\alphaX)^{\frac{D-1}2}\:.
\end{equation}
If we name the components of $\eta$ as follows,
\begin{equation}
	\eta=\eta_a^0\,\dd\phiLa^0+{\sum}_{\,i=1}^{\,\frac{D-1}2}\,\eta_a^i\:\big(\xiLa^0\,\dd\phiLa^i-\xiLa^i\,\dd\phiLa^0\big)
		+(\dd\rLa^i\text{ terms})\:,
\end{equation}
where each $\eta_a^i$ is a function, we can write the expansion of the integrand of \eqref{localized_contribution2} near $L_a$ as
\begin{equation}
	\big(\tfrac{D-1}2\,!\,\big)^{-1}\,\eta\wedge(\dd\alphaX)^{\frac{D-1}2}\,\big|_{\text{near }L_a}\approx\,
		\eta_a^0\,\dd\phiLa^0\:{\bigwedge}_{\,i=1}^{\frac{D-1}2}\,\xiLa^0\:\bLa^i\:\dd\big((\rLa^i)^2\big)\wedge\dd\phiLa^i\:.
\end{equation}
If we proceed the same way as in \eqref{asymptotic_integral} and put all the pieces together, we arrive at
\begin{equation}
\label{localized_contribution3}
	\lim_{t\to+\infty}\int_{\text{near }L_a}\!\!\!\!\!\!\!\!\!\!\!\!\eta\wedge\Phi\wedge \eu^{t\,\dd_X\alphaX}=
		\frac{\Phi_0\big|_{L_a}\cdot\int_0^{2\pi}\!\eta_0^a\,\big|_{L_a}\dd\phiLa^0}
		{\,{\displaystyle\prod}_{\,i=1}^{\,\frac{D-1}2}\!\left(\frac{\XLa^0}{\xiLa^0}\:\xiLa^i-\XLa^i\right)}\:.
\end{equation}
The constants $\bLa^i$ have canceled, as expected. The integral of $\eta_0^a\,|_{L_a}$ over $\phiLa^0$ is the same as $\int_{L_a}\!\eta\:$:
the $\dd\rLa^i$ pullback to zero on $L_a$, and for $\eta$ to be regular at $L_a$ we must have $\eta_a^i=\cO\big((\rLa^i)^2\big)$,
and thus $\eta_i^a\,|_{L_a}=0$. Then \eqref{localized_contribution3} is exactly the contribution at $L_a$ of the localization formula 
\eqref{localization_formula}, which concludes its derivation.

 \subsection{A simple example: the three-sphere, whole and cut into a half}\label{sec:S3}

As a simple  application of the theorem we  compute the (Sasakian) volume of  $S^3$ and of the manifolds with boundaries obtained by cutting $S^3$ along the equator and keeping only the North or South hemisphere.

The  three-sphere is the set of points $(z_1,z_2)\subset \mathbb{C}^2$  with $|z_1|^2+|z_2|^2=1$. We can parameterize 
\begin{align}z_1= \sin \frac{\theta}{2} e^{i\phi_1}\, ,\qquad\qquad z_2= \cos \frac{\theta}{2} e^{i\phi_2}\, , \end{align}
where $0\le \theta\le \pi$ and $\phi_1$ and $\phi_2$ are $2\pi$-periodic angles.  In these coordinates it is obvious that $S^3$ is a $\mathbb{T}=$\text{U}(1)$^2$ fibration over the segment $\theta\in[0,\pi]$.
There are two loci where the $\mathbb{T}$-action degenerates to a circle action, corresponding to the endpoints of the segment.
At $\theta=0$ ($z_1=0$) the angle $\phi_1$ degenerates and the points $(0,e^{i\phi_2})$ parameterize a circle $L_1$. At $\theta=\pi$ ($z_2=0$) the angle $\phi_2$ degenerates and the 
circle $L_2$ is given by $(e^{i\phi_1},0)$.
 The canonical Riemannian metric  of the round three-sphere is obtained by restricting the flat metric of $\mathbb{C}^2$ 
 and reads
\begin{align}\label{eq:roundS3}
\dd s^2=\frac14 \dd \theta^2 + \sin^2 \frac{\theta}{2}\, \dd\phi_1^2 + \cos^2 \frac{\theta}{2}\, \dd\phi_2^2 \, .
\end{align}
The North and South hemispheres are defined by restricting the manifold to the segments $\theta\in[0,\frac{\pi}{2}]$ and $\theta\in[\frac{\pi}{2},\pi]$, respectively. 

We consider the vector field $\xi=\xi^1 \partial_{\phi_1}+\xi^2 \partial_{\phi_2}$ and the contact one-form
\begin{align}\label{eq:contact}
\eta = \frac{ \sin^2 \frac{\theta}{2}\, \dd\phi_1 + \cos^2 \frac{\theta}{2}\, \dd\phi_2}{\sin^2 \frac{\theta}{2} \, \xi^1 + \cos^2 \frac{\theta}{2} \, \xi^2 } \, ,
\end{align}
satisfying  $\eta(\xi)=1$. The two-form $\dd\eta$ is basic, $\iota_\xi \dd\eta=0$ and we are interested in computing
the Sasakian volume 
\begin{align} \label{eq:vol3} \text{Vol}_M(\xi) = \frac12 \int_{M} \eta \wedge \dd \eta \end{align}
where $M$ is $S^3$ or one of hemispheres.  For $\xi^1 = \xi^2 =1$, $\frac 12 \eta \wedge \dd \eta$ is just the volume form of the canonical metric \eqref{eq:roundS3} and $\text{Vol}_M(\xi) $ is the Riemannian volume of the round sphere. For generic $\xi$, $\text{Vol}_M(\xi) $ can be interpreted as the volume of a family of Sasakian metrics $g_\xi$ on $S^3$ \cite{Martelli:2005tp,Martelli:2006yb}.\footnote{ A  metric $g$ is Sasaki if the cone over it, $\dd r^2 +r^2 g$, is K\"ahler. The   K\"ahler form on the cone is given by $\omega =\frac 12 \dd(r^2\eta)$. See section \ref{The Sasakian volume} for more details.} 

The integral \eqref{eq:vol3} can be computed by elementary methods but we want to use this simple example to show how localization in odd dimensions works. The two-form 
\begin{align} \dd_X \eta =\dd \eta -\eta(X) \end{align}
is clearly equivariantly closed, since $\mathcal{L}_X \eta=0$, and 
\begin{align} \label{eq:vol3II} \text{Vol}_M(\xi) = \frac12 \int_{M} \eta \wedge \dd \eta  = \frac12 \int_{M} \eta \wedge \dd_X \eta \, ,\end{align}
since the additional terms introduced by the equivariantization are forms of degree less than three. 

We will now apply the localization theorem \eqref{localization_formula} with $\Phi=\dd_X\eta$.
At the localization locus $L_1$ we have $\varphi_{L_1}^0=\phi_2$ and $\varphi_{L_1}^1=\phi_1$, and
\begin{align} \Phi_0 \Big |_{L_1} = -\frac{X^2}{\xi^2}\, ,\qquad \int_{L_1} \eta=\frac{2\pi}{\xi^2} \, , \qquad  \frac{X_{L_1}^0}{\xi_{L_1}^0}\:\xi_{L_1}^1-X_{L_1}^1 = \frac{X^2}{\xi^2} \xi^1- X^1 \, ,\end{align}
while at $L_2$ we have $\varphi_{L_2}^0=\phi_1$ and $\varphi_{L_2}^1=\phi_2$, and
\begin{align} \Phi_0 \Big |_{L_2} = -\frac{X^1}{\xi^1}\, ,\qquad \int_{L_2} \eta=\frac{2\pi}{\xi^1} \, \qquad  \frac{X_{L_2}^0}{\xi_{L_2}^0}\:\xi_{L_2}^1-X_{L_2}^1 = \frac{X^1}{\xi^1} \xi^2- X^2 \, .\end{align}

For $M=S^3$ we just add the contributions of the  loci $L_1$ and $L_2$ obtaining
\begin{align} \label{eq:vol3III} \text{Vol}_{S^3}(\xi)   = \frac12 \int_{M} \eta \wedge \dd_X \eta = 2\pi^2 \left (\, \frac{1}{\xi^1 X^2 -\xi^2 X^1} \frac{X^2}{\xi^2} -\frac{1}{\xi^1 X^2 -\xi^2 X^1} \frac{X^1}{\xi^1} \right )=\frac{2\pi^2}{\xi^1 \xi^2}\, ,\end{align}
which is the well-known expression for the Sasakian volume of $S^3$ \cite{Martelli:2005tp,Martelli:2006yb}. Notice that the $X$-dependence cancels 
when the two localized contributions are summed.

When $M$ is the North hemisphere $\theta\in[0,\frac{\pi}{2}]$ we include the contribution 
of $L_1$ and the boundary term in \eqref{localization_formula}. 
We use  $\alphaX_\mu=(\gR)_{\mu\nu}Y^\nu$  where $Y=\gR(\xi,\xi)\,X-\gR(\xi,X)\,\xi$ and $g_R$ is the canonical metric \eqref{eq:roundS3}. The result must be independent of $X$ and $g_R$. A short computation indeed shows that
\begin{align} \label{eq:NS} \text{Vol}_{S_N^3}(\xi)   = 2\pi^2 \left (\, \frac{1}{\xi^1 X^2 -\xi^2 X^1} \frac{X^2}{\xi^2} -\frac{1}{\xi^1 X^2 -\xi^2 X^1} \frac{X^1+X^2}{\xi^1+\xi^2} \right )=\frac{2\pi^2}{\xi^2(\xi^1+ \xi^2)}\, .\end{align}
Similarly, for the South hemisphere $\theta\in[\frac{\pi}{2},\pi]$, including the contribution of $L_2$ and the boundary term (which has now opposite orientation), we obtain
\begin{align} \label{eq:SS} \text{Vol}_{S_S^3}(\xi)   = 2\pi^2 \left (\, -\frac{1}{\xi^1 X^2 -\xi^2 X^1} \frac{X^1}{\xi^1} +\frac{1}{\xi^1 X^2 -\xi^2 X^1} \frac{X^1+X^2}{\xi^1+\xi^2} \right )=\frac{2\pi^2}{\xi^1(\xi^1+ \xi^2)}\, .\end{align}

Notice that, as expected,
\begin{align}  \text{Vol}_{S_N^3}(\xi) +\text{Vol}_{S_S^3}(\xi) = \text{Vol}_{S^3}(\xi) \, .\end{align}

\subsection{Five-dimensional toric geometries}
\label{Five-dimensional toric geometries}

In this section we focus on $D=5$ dimensions, which is the case of interest for this paper.
The consideration that we will make are applicable not only to the manifolds with boundary that can arise from solutions to five-dimensional supergravity,
but also to closed manifolds like compact toric Sasakian geometries.
We provide two examples, relevant to holography: 
in subsection \ref{The Sasakian volume} we review how the localization formula \eqref{localization_formula}
can be used to derive the Sasakian volume of \cite{Martelli:2005tp,Martelli:2006yb},
while in subsection \ref{The GMS master volume} we discuss a generalization of the Sasakian volume,
namely the GMS master volume \cite{Gauntlett:2018dpc}.
The Sasakian volume is properly defined for compact manifolds without boundary, but as we will discuss in section \ref{sect:on shell localization}
there is a surprising connection with the on-shell action of the supergravity solutions that we consider.

We can describe the five-dimensional geometry in terms of degeneration loci of the $\mathbb{T}$ action. This will define a set of three-dimensional vectors $V^a_i$ with integer entries,
which are the coordinates in a $2\pi$-periodic basis of angles of  vector fields degenerating on codimension two submanifolds. Using the familiar terminology of toric manifolds, we will refer to the set of vectors $V^a$ as the {\it fan} of the five-dimensional geometry.

We will make the following (quite general) assumptions on the geometry. We denote with $\D_a$ the three-dimensional submanifolds where the $\bT$-action has a $\text{U}(1)$ isotropy subgroup.
Given a set $2\pi$-periodic angular coordinates $(\phi_0,\phi_1,\phi_2)$ on $M$ such that $\bT$ acts by rotating each $\phi_i$,
the action of the isotropy group at $\D_a$ can always be written as
\begin{equation}
\label{U(1)_isotropy}
	\phi_i\:\longrightarrow\:\phi_i+V_i^a\,\varphi\:,\qquad\varphi\in\bR/2\pi\bZ\:,
\end{equation}
for $V^a_i\in\bZ$ such that gcd$(V^a_0,V^a_1,V^a_2)=1$.%
\footnote{If the $V^a_i$ had a common factor the action would not be effective. These integers are defined up to an overall sign, which we will fix later.}
The localization circles $L_a$ are the submanifolds with $\text{U}(1)^2$ isotropy subgroup, and can be found as the intersections of the closures of
two of the loci with $\text{U}(1)$ isotropy, namely
\begin{equation}
	L_a=\overline{\D_{a-1}}\cap\overline{\D_a}\:,\quad a=1,\ldots,d\:.
\end{equation}
If $M$ is a compact manifold without boundary, we must have $\D_0\equiv\D_d$.
When $\partial M\ne\varnothing$ we have $\D_0\ne\D_d$, and both of them intersect the boundary.
In terms of the adapted coordinates near $L_a$, for which a Riemannian metric $\gR$ can be approximated as
\begin{equation}
\label{flat_metric_5d}
	\gR\big|_{\text{near }L_a}\approx\big(\RLa\,\dd\phiLa^0\big)^2+\sum_{i=1,2}
		\Big[\big(\dd\rLa^i\big)^2+(\rLa^i)^2\big(\dd\phiLa^i+\cLa^i\dd\phiLa^0\big)^2\Big]\:,
\end{equation}
we have the identifications $\D_{a-1}\cap\{$neighborhood of $L_a\}\equiv\{\rLa^1=0\}$ and $\D_a\cap\{$neighborhood of $L_a\}\equiv\{\rLa^2=0\}$.

To apply the localization formula we need to know, in addition to the restrictions of $\Phi_0$,  the weights at the denominator of the localization formula, $\XLa^0\,\xiLa^i\,(\xiLa^0)^{-1}-\XLa^i$, and the integral of $\eta$ on $L_a$. All these quantities can be expressed solely in terms of the $V^{a-1}_i$,
$V^a_i$ and the components of $\xi$, $X$ in the $\partial_{\phi_i}$ basis.
Furthermore, the weights are related to the restrictions of the equivariant Chern classes of the four-dimensional base of the foliation. We discuss all these points in the following.

 \paragraph*{I) The weights in terms of the $V^a$.}

The vector\footnote{We make use of Einstein notation for indices taking values in $\{0,1,2\}$.} $V^a\equiv V^a_i\,\partial_{\phi_i}$
is the generator of the action of the $\text{U}(1)$ subgroup of $\bT$ that keeps fixed the points of $\D_a$. Thus $V^a$ vanishes on $\D_a$.
By imposing that the norms of $V^{a-1}$ and $V^a$ vanish on $\D_{a-1}$ and $\D_a$ respectively using the near $L_a$ metric \eqref{flat_metric_5d}, we find
\begin{equation}
\begin{aligned}
	\gR(V^{a},V^{a})\,\big|_{\D_{a}}=0\quad&\implies\quad
		\iota_{V^{a}}\,\dd\phiLa^0=0=\iota_{V^{a}}\big(\dd\phiLa^1+\cLa^1\dd\phiLa^0\big)\:,\\
	\gR(V^{a-1},V^{a-1})\,\big|_{\D_{a-1}}=0\quad&\implies\quad
		\iota_{V^{a-1}}\,\dd\phiLa^0=0=\iota_{V^{a-1}}\big(\dd\phiLa^2+\cLa^2\dd\phiLa^0\big)\:,
\end{aligned}
\end{equation}
which are equivalent to the following conditions:
\begin{equation}
\label{phiLa_conditions}
	V^a_i\,\partial_{\phi_i}\phiLa^0=V^a_i\,\partial_{\phi_i}\phiLa^1=V^{a-1}_i\,\partial_{\phi_i}\phiLa^0=V^{a-1}_i\,\partial_{\phi_i}\phiLa^2=0\:.
\end{equation}
The above are solved by%
\footnote{Up to redefining the sign of $\phiLa^0$ and one of the $\phiLa^i$ at the same time.}
\begin{equation}
\label{phiLa}
	\phiLa^0=\sa\cdot(\phi,V^{a-1},V^a)\:,\quad\phiLa^1=(\phi,V^a,\wa)\:,\quad\phiLa^2=(\phi,\wa,V^{a-1})\:,
\end{equation}
where we are using the shorthand notation
\begin{equation}
\label{determinant_shorthand}
	(\phi,V^{a-1},V^a)\equiv\det
	\begin{pmatrix}
		\phi_0\:\: & V^{a-1}_0\: & V^a_0\\
		\phi_1\:\: & V^{a-1}_1\: & V^a_1\\
		\phi_2\:\: & V^{a-1}_2\: & V^a_2
	\end{pmatrix}\:,
\end{equation}
and we have defined an arbitrary $\wa\in\bZ^3$ such that $(\wa,V^{a-1},V^a)=1$ and the sign $\sa\in\{-1,+1\}$ which must be selected
so that the ordering $(\phiLa^0,\rLa^1,\phiLa^1,\rLa^2,\phiLa^2)$ corresponds to the positive orientation on $M$.%
\footnote{In particular the signs $\sa$ must be chosen so that the forms $\sa\dd\rLa^1\wedge\dd\rLa^2$, with $a=1,\ldots,d$, differ by
		multiplication with a positive function. In appendix \ref{app:toric} we show that, under general conditions,  one can fix $\sa=+1$ for all $a$.}
For $\sa=+1$ the change of coordinates \eqref{phiLa} is a $\text{SL}(3,\bZ$) transformation, while for $\sa=-1$ it is a $\text{SL}(3,\bZ$) followed by a reflection.
Different choices of $\wa\in\bZ^3$ all lead to coordinates for which the metric has the form \eqref{flat_metric_5d} near $L_a$, with different
$\cLa^1,\cLa^2\in\bR$.

From \eqref{phiLa} it is immediate to find the components of $\xi$ in the $\partial_{\phiLa^i}$ basis:
\begin{equation}
\label{xiLa_determinant}
	\xiLa^0=\sa\cdot(\xi,V^{a-1},V^a)\:,\quad\xiLa^1=(\xi,V^a,\wa)\:,\quad\xiLa^2=(\xi,\wa,V^{a-1})\:,
\end{equation}
where the components of $\xi$ in the $\partial_{\phi_i}$ basis are the ones entering the determinant.
The components of $X$ in the $\partial_{\phiLa^i}$ basis can also be found in the same manner,
which leads to the following expression for the weights appearing at the denominator of the localization formula \eqref{localization_formula}:
\begin{equation}\label{weights}
	\frac{\XLa^0}{\xiLa^0}\,\xiLa^1-\XLa^1=-\frac{(\xi,X,V^a)}{(\xi,V^{a-1},V^a)}\:,\qquad
		\frac{\XLa^0}{\xiLa^0}\,\xiLa^2-\XLa^2=\frac{(\xi,X,V^{a-1})}{(\xi,V^{a-1},V^a)}\:
\end{equation}
where we have used $(\wa,V^{a-1},V^a)=1$ and the  identity\footnote{The identity  follows from the fact that for any $w^0,\ldots,w^n\in\bR^n$ we have
\begin{align}
\label{linearly_dependent}
	\sum_{i=0}^n(-1)^i (w^0,\ldots,\widehat w^i,\ldots,w^n)\cdot w^i=0\:.
\end{align}
by taking $n=3$ and the determinant of the above expression with $w^3$ and $w^4$.}
\begin{align}
\label{determinant_product_formula}
	(w^1,w^2,w^3) (w^0,w^3,w^4) + (w^2,w^0,w^3) (w^1,w^3,w^4) + (w^0,w^1,w^3) (w^2,w^3,w^4)=0\:,
\end{align}
valid for generic vectors $w^i\in\bR^3$.

\paragraph*{II) The weights as restrictions of basic equivariant Chern classes.}

For simplicity we will restrict ourselves to the case where $\xi$ can be written as
\begin{equation}
\label{orbifold_assumption}
	\xi=\cxi\,n_i\,\partial_{\phi_i}\:,\qquad\cxi\in\bR\:,\quad n_i\in\bZ\:,\quad\text{gcd}(n_0,n_1,n_2)=1\:.
\end{equation}
Since the set of vectors $\xi$ that can be written in the above manner is a dense subset of all the possible choices of $\xi$,
we will be able to extrapolate our conclusions to a generic foliation by continuity.
The advantage of assumption \eqref{orbifold_assumption} is that every orbit of $\xi$ closes, and thus the trajectory of each integral curve of $\xi$
is diffeomorphic to $S^1$. The foliation whose leaves are the orbits of $\xi$ can now be seen as a $S^1$ fibration over a four-dimensional orbifold.
We will assume that this orbifold is 
symplectic and that its quotient by $\text{U}(1)^2$ is \emph{convex}, so that it is possible to apply the notions of toric geometry
reviewed in appendix \ref{app:toric}.

In order to define $2\pi$-periodic angular coordinates $\phi^B_1,\phi^B_2$ on the four-dimensional base, we need to choose
$B\in$ $\text{SL}(3,\bZ$) such that
\begin{equation}
	B_{ij}\,n_j=\delta_{0\,i}\:,
\end{equation}
so that we can set $\phi^B_1=B_{1\,i}\,\phi_i$ and $\phi^B_2=B_{2\,i}\,\phi_i$.
Each $\D_a\subset M$ is a $S^1$ fibration over one of the toric divisors $D_a$ of the base, and the isotropy subgroup of the $\text{U}(1)^2$ toric action at $D_a$
can be found from \eqref{U(1)_isotropy} as
\begin{equation}
\label{small_v_def}
	\phi^B_i\:\longrightarrow\:\phi^B_i+v^a_i\,\varphi\:,\qquad\varphi\in\bR/2\pi\bZ\:,\qquad v_i^a\equiv B_{ij}\,V^a_j\:,\qquad i=1,2\:.
\end{equation}
the $v^a_i$ are then the vectors of the toric fan of the four-dimensional base, and we can assume without loss of generality that the $v^a_i$
as two-component vectors of integers are ordered in anti-clockwise order.%
\footnote{This is possible considering that the $V^a$ where defined up to a sign; this assumption fixes their sign completely.
		The integer gcd($v^a_1,v^a_2$) is the label associated to the divisor $D_a$.}

Let $p_a$ be the point of intersection of the divisors $D_{a-1}$ and $D_a$; $p_a$ is then a fixed point of the $\text{U}(1)^2$ toric action, and
the fiber over it is the circle $L_a$. The order of the orbifold singularity at $p_a$ can be computed as
\begin{equation}
\label{u_defining_property}
	d_a=\big|\det(v^{a-1},v^a)\:\big|=\det(v^{a-1},v^a)=\cxi^{-1}\cdot(\xi,V^{a-1},V^a)\:,
\end{equation}
since $B_{ij} \xi_j=c_\xi \delta_{i0}$  for all $i$ and $B_{ij} V_j^a =v_i^a$ for  $i=1,2$.
We can now introduce the vectors
\begin{equation}
	(u^1_a)_i=\cxi^{-1}\:(B^{-1})_{ji}\cdot(\xi,e_j,V^a)\:,\quad(u^2_a)_i=\cxi^{-1}\:(B^{-1})_{ji}\cdot(\xi,V^{a-1},e_j)\:\qquad i=1,2\:,
\end{equation}
where $(e_j)_i=\delta_{ij}$, which satisfy%
\footnote{Using the identity $(B^{-1})_{j0}\cdot(\xi,e_j,V^a)=\cxi^{-1}\cdot(\xi,\xi,V^a)=0$, it is possible to promote the sum over $i=1,2$
		to a sum over $i=0,1,2$, and then cancel $B$ with its inverse.} 
\begin{equation}
 u^1_a\cdot v^{a-1}=d_a\:,\quad u^1_a\cdot v^a=0\:,\quad
		u^2_a\cdot v^{a-1}=0\:,\quad u^1_a\cdot v^a=d_a\:.
\end{equation}
These vectors determine the restrictions of the equivariant Chern classes at $p_a$:
the Chern class of the line bundle $\LB_a$ associated to the divisor $D_a$%
\footnote{$\LB_a$ by definition restricts to the normal bundle $\cN D_a$ over $D_a$ and and it is a trivial line bundle over the orbifold with $D_a$ removed.}
is expressible in terms of ``moment maps" $\mu^a_i$,
\begin{equation}
\label{c_1_def}
	c_1(\LB_a)=\sum_{i=1,2}\dd(\mu_i^a\,\dd\phi^B_i)\:,
\end{equation}
and the moment maps at $p_a$ are given in terms of the $u^{1,2}_a$ as (see appendix \ref{app:toric} for more details)
\begin{equation}\label{mm}
	\mu_i^{a-1}\,\big|_{p_a}=-\frac1{2\pi}\,\frac{(u^1_a)_i}{d_a}\:,\qquad
		\mu_i^a\,\big|_{p_a}=-\frac1{2\pi}\,\frac{(u^2_a)_i}{d_a}\:,\qquad\mu^b_i\,\big|_{p_a}=0\quad\text{if }b\ne a-1,a\:.
\end{equation}
If we extend $c_1(\LB_a)$ to a basic form on $M$, we can consider the equivariantly closed polyform $c_1^\bT(\LB_a)$ obtained
by substituting $\dd\to\dd_X$ in \eqref{c_1_def},%
\footnote{The $\mu^a_i$ are $\bT$-invariant and thus the vanishing of $\cL_X$ is automatically verified.}
\begin{equation}
	c_1^\bT(\LB_a)=\sum_{i=1,2}\dd_X(\mu_i^a\,\dd\phi^B_i)=\sum_{i=1,2}\Big(\dd(\mu_i^a\,\dd\phi^B_i)-\mu_i^a\,B_{ij}\,X^j\Big)\:,
\end{equation}
whose restrictions on $L_a$ are
\begin{equation}
\begin{aligned}
\label{weights_VS_Chern}
	c_1^\bT(\LB_{a-1})\,\big|_{L_a}&=\sum_{i=1,2}\frac1{2\pi}\,\frac{(u^1_a)_iB_{ij}X^j}{d_a}=\frac1{2\pi}\frac{(\xi,X,V^a)}{(\xi,V^{a-1},V^a)}=
		-\frac1{2\pi}\bigg(\frac{\XLa^0}{\xiLa^0}\,\xiLa^1-\XLa^1\bigg)\:,\\
	c_1^\bT(\LB_a)\,\big|_{L_a}&=\sum_{i=1,2}\frac1{2\pi}\,\frac{(u^2_a)_iB_{ij}X^j}{d_a}=-\frac1{2\pi}\frac{(\xi,X,V^{a-1})}{(\xi,V^{a-1},V^a)}=
		-\frac1{2\pi}\bigg(\frac{\XLa^0}{\xiLa^0}\,\xiLa^2-\XLa^2\bigg)\:,
\end{aligned}
\end{equation}
and $c_1^\bT(\LB_b)\,|_{L_a}=0$ for $b\ne a-1,a$.
The above relations show the proportionality between the restrictions of the equivariant Chern classes and  the weights of the localization formula.
There is also an alternative derivation of this relation which we discuss in appendix \ref{app:toric}.

\paragraph*{III) Computing $\int_{L_a}\eta$ and the $L_a$-restriction of $\dd_X\eta$.}

The localization theorem requires $\eta$ to be a one-form on $M$ that satisfies $\iota_\xi\dd\eta=0$.
No other condition is needed for the localization formula to be applied.
However, in all the cases and examples that we will consider in this paper, we will
focus on integrands that are invariant under the symmetry group $\bT$, which in particular implies that 
the form $\eta$ also satisfies $\cL_\xi\eta=0=\cL_X\eta$\:.
By Cartan's formula, $\iota_\xi\dd\eta=0$ and $\cL_\xi\eta=0$ together imply that $\iota_\xi\,\eta$ is constant over $M$:
\begin{equation}
	\iota_\xi\,\eta=c_\eta\:,\quad c_\eta\in\bR\:.
\end{equation}
Considering that $\xi\,|_{L_a}=\xiLa^0\,\partial_{\phiLa^0}$, the above constraint fixes completely the only component of $\eta$
with a non-vanishing pullback on $L_a$, and we find
\begin{equation}\label{inteta}
	\int_{L_a}\eta\,=\int_0^{2\pi}\frac{c_\eta}{\xiLa^0}\,\dd\phiLa^0\,=\,2\pi\,\frac{c_\eta}{\xiLa^0}\,=\,\frac{2\pi\,c_\eta}{(\xi,V^{a-1},V^a)}\:,
\end{equation}
where we have used \eqref{xiLa_determinant} and the fact that $\sa=+1$ for the geometries that we are considering.

We can construct the following simple example of a basic equivariant polyform:
\begin{equation}
	\dd_X\eta=\dd\eta-\eta(X)\:.
\end{equation}
The condition $\iota_\xi\dd\eta=0$ and the additional constraints $\cL_\xi\eta=0=\cL_X\eta$ ensure that the above is both basic and equivariantly closed.
The above polyform is missing the degree-four component, but we can construct a polyform $\Phi$ with a non-trivial $\Phi_4$ as
\begin{equation}
\label{Phi_ddeta}
	\Phi=(\dd_X\eta)^2\quad\implies\quad\Phi_4=(\dd\eta)^2\:,\quad\Phi_2=-2\,\eta(X)\,\dd\eta\:,\quad\Phi_0=\eta(X)^2\:.
\end{equation}
This polyform is also both basic and equivariantly closed, and we can use it to compute the integral of $\eta\wedge(\dd\eta)^2$ with the
localization formula \eqref{localization_formula}.

We can find the restriction of $\Phi_0$ given by \eqref{Phi_ddeta} at the localization locus $L_a$
by considering that the vector $X-(\XLa^0/\xiLa^0)\,\xi$ vanishes at $L_a$, and thus its contraction with $\eta$ must vanish as well:
\begin{equation}
	\bigg(\eta(X)-\frac{\XLa^0}{\xiLa^0}\,c_\eta\bigg)\,\Big|_{L_a}=0\:.
\end{equation}
This leads to
\begin{align}\label{deta}
\dd_X \eta \big|_{L_a}= - \eta(X) \big|_{L_a} = - c_\eta\,\frac{\XLa^0}{\xiLa^0}\,= - c_\eta \,\frac{(X,V^{a-1},V^a)}{(\xi,V^{a-1},V^a)}\:.
\end{align}
and
\begin{equation}
\label{Phi0_expression}
	\Phi_0\,\big|_{L_a}=\bigg(c_\eta\,\frac{\XLa^0}{\xiLa^0}\,\bigg)^2=\big(c_\eta\big)^2\,\frac{(X,V^{a-1},V^a)^2}{(\xi,V^{a-1},V^a)^2}\:.
\end{equation}

\subsubsection{The Sasakian volume}
\label{The Sasakian volume}

As an example, we consider the case of five-dimensional compact toric Sasaki-Einstein  manifolds $M_5$ and we review the derivation of the Sasakian volume  given in \cite{Goertsches:2015vga}.  We use the notations of \cite{Martelli:2005tp,Martelli:2006yb} to which we also refer for more details.

We consider a family of Sasakian metrics $g$ on $M_5$, defined by the condition that  the metric of the cone $C(M_5)$
\begin{align}\label{met:cone} \dd s^2 = \dd r^2 +r^2 g \, ,\end{align} is K\"ahler. They come equipped with a Reeb vector $\xi$\footnote{On the cone, the Reeb vector shows up as the partner under complex structure of the dilatation vector $r\partial_r$.}  and a (globaly defined) contact one-form $\eta$, with $\iota_\xi \eta=1$ and $\iota_\xi \dd\eta=0$, such that the   K\"ahler form on the cone is given by 
\begin{align} \Kform=\frac 12 \dd(r^2\eta)  \, . \end{align} 
We are interested in computing the Riemannian volume of $M_5$ with metric $g$, which is a function of $\xi$ and can be expressed as \cite{Martelli:2005tp,Martelli:2006yb}
\begin{align} \label{eq:vol5} \text{Vol}_{M_5}(\xi) = \frac18 \int_{M_5} \eta \wedge (\dd \eta)^2 \, .\end{align}
By introducing the equivariantly closed form $\dd_X \eta =\dd \eta -\eta(X)$ we can equivalently compute
\begin{align} \label{eq:vol5II} \text{Vol}_{M_5}(\xi) =\frac18 \int_{M_5} \eta \wedge (\dd_X \eta)^2 \, ,\end{align}
since only the top degree terms contribute to the integral. 

In order to use the localization theorem we need to understand the degeneration loci of the torus action on $M_5$.   For a Sasaki manifold, the fan that we introduced in section \ref{Five-dimensional toric geometries} is nothing else than the fan  of the six-dimensional toric cone  $C(M_5)$,  which is perhaps familiar to physicists. Let us briefly recall the toric description of $C(M_5)$.
 On the cone,  we can introduce  symplectic coordinates  $y_i$ and $\phi_i$ such that
\begin{align} \Kform = \sum_{i=0}^2 \dd y_i \wedge \dd \phi_i   \, , \end{align}
 where $\phi_i$ are $2\pi$-periodic angles. The image of the moment maps $y_i$ is a three-dimensional non compact convex conical polytope in $\mathbb{R}^3$ and $C(M_5)$ can be seen as a   $\mathbb{T}=\,$\text{U}(1)$^{3}$ fibration over it.  The facets of the polytope are given by the linear equations $\mathcal{F}_a=\{ V^a_i y_i=0\}$ and correspond to the loci where the vector field $V^a=V^a_i \partial_{\phi_i}$ degenerates. The set of vectors $V^a$ with $a=1,\ldots d$ define the fan of the toric variety $C(M_5)$. The Calabi-Yau condition requires that the $V^a$ lie on a plane. 
 The manifold $M_5$ is obtained by restricting the cone to $r=1$, which can be described as the plane $\xi_i y_i=\frac12$ \cite{Martelli:2005tp,Martelli:2006yb}.  $M_5$ itself is then a $\mathbb{T}$ fibration over a two-dimensional compact convex polytope obtained by intersecting the non-compact conical polytope with the plane. The counter-images of the sides of the two-dimensional polytope are the loci ${\cal D}_a$ in $M_5$ with $\text{U}(1)$ isotropy group where $V^a=V^a_i \partial_{\phi_i}$ degenerates while the counter-images of the vertices are the localization circles $L_a$.

As an example, we can consider the case of a five-sphere, $M_5=S^5$ which is the set of points $(z_0,z_1,z_2)\subset \mathbb{C}^3$  with $|z_0|^2+|z_1|^2+|z_2|^3=1$. There exist  three localization circles $L_a$ corresponding to the points where $z_b=0$ for $b\ne a$. The cone over the five-sphere is just  $C(S^5)=\mathbb{C}^3$. The simplectic coordinates are defined by $y_i=\frac{|z_i|^2}{2}$ and $\phi_i=\text{arg}(z_i)$ for $i=0,1,2$.
The non compact polytope  of $\mathbb{C}^3$ is just the first octant $\{ y_0>0,y_1>0,y_2>0\}$ of $\mathbb{R}^3$. On the $a$-th facet $\mathcal{F}_a=\{y_a=0\}$  the vector $\partial_{\phi_a}$ degenerates. The vectors $V^a$ are thus given by
\begin{align}\label{fanS5}  V^1=(1,0,0)\, ,\qquad V^2=(0,1,0)\, ,\qquad V^3=(0,0,1)\, ,\end{align}
and correctly lie on a plane. A family of Sasakian metrics on $S^5$ with Reeb vector $\xi=\sum_{i=0}^2 \xi_i \partial_{\phi_i}$ is obtained by considering a K\"ahler metric on $\mathbb{C}^3$ and by taking  the radial coordinate on $C(S^5)$ to be $r^2 = 2 \sum_{i=0}^2 \xi_i y_i =\sum_{i=0}^2 \xi_i |z_i|^2$ \cite{Martelli:2005tp,Martelli:2006yb}. 
Regularity requires $\xi_i>0$. The compact polytope of $S^5$ is then the triangle obtained by intersecting the octant $\{ y_0>0,y_1>0,y_2>0\}$ with the plane $\xi_i y_i=\frac 12$. The three localization circles $L_a$ correspond to the three vertices of this polytope as shown in figure \ref{fig:S5}.\footnote{A similar construction applies as well as $S^3$ and it is compatible with the discussion in \ref{sec:S3}. The  cone is $\mathbb{C}^2$ and the corresponding polytope the first quadrant of $\mathbb{R}^2$. The polytope of $S^3$ is obtained by intesecting it with the plane $\xi_i y_i=\frac12$ and it is a segment. The endpoints of the segment are the two localization loci $L^1$ and $L^2$.}

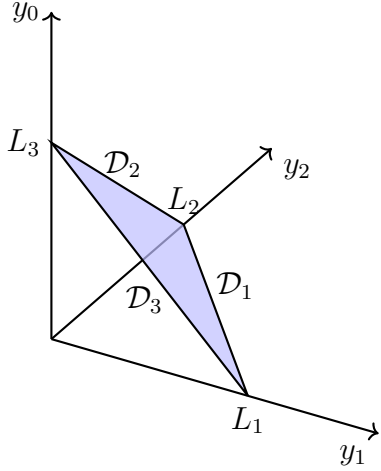
\begin{figure}[h]
\begin{tikzpicture}[tdplot_main_coords,scale=1]
  \def\axlength{5}
  \def\trianglepos{3}
  \def\yaxiscoeff{3.5/3}
  
  \draw[->,thick] (0,0,0) -- (\axlength,0,0) node[anchor=north east] {$y_1$};
  
  \draw[->,thick] (0,0,0) -- (0,{\yaxiscoeff*\axlength},0) node[anchor=north west] {$y_2$};
  
  \draw[->,thick] (0,0,0) -- (0,0,\axlength) node[anchor=east] {$y_0$};

  \coordinate (A) at (\trianglepos,0,0);
  \coordinate (B) at (0,{\yaxiscoeff*\trianglepos},0);
  \coordinate (C) at (0,0,\trianglepos);

  \fill[fill=blue!25, opacity=0.7] (A) -- (B) -- (C) -- cycle;
  \draw[thick]
    (A) -- node[midway, above right, xshift=-4pt] {$\mathcal{D}_1$} (B)
    -- node[midway, above, xshift=2pt, yshift=-2pt] {$\mathcal{D}_2$} (C)
    -- node[midway, below, xshift=-2pt, yshift=-2pt] {$\mathcal{D}_3$} (A);

  \node at (A) [below] {$L_1$};
  \node at (B) [above] {$L_2$};
  \node at (C) [left] {$L_3$};
\end{tikzpicture}
\centering
\caption{The two-dimensional polytope of $S^5$ obtained by cutting the first octant with the plane $\xi_i y_i=\frac 12$ and the three localization loci $L_a$. Notice that to have a compact polytope, the Reeb vector, which is the normal to the plane,  must lie inside the octant.}
\label{fig:S5}
\end{figure}
  
 We now have all the ingredients to apply the localization formula \eqref{localization_formula} to a generic Sasaki manifold. The localization circle are associated with the pair of vectors $(V^{a-1},V^a)$ for $a=1,\ldots d$ where we use the convention that $V^0\equiv V^d$ and $V^{d+1}\equiv V^1$. For a Sasakian structure $c_\eta= \iota_\xi \eta=1$ by definition. It is important to notice that $c_\eta$ will be different from one in other examples in this paper. Using \eqref{inteta}, \eqref{deta}  and \eqref{weights}, we find \cite{Goertsches:2015vga}
\begin{align}\label{sasvolX}
\text{Vol}_{M_5}(\xi)  \,=\,\frac{\pi^2}{2}\sum_{a=1}^d \frac{\eta(X)^2\big|_{L_a}\cdot\int_{L_a}\eta}
		{\,{\displaystyle\prod}_{\,i=1}^{2}\!\left(\frac{\XLa^0}{\xiLa^0}\:\xiLa^i-\XLa^i\right)}\: = \,\pi^3\sum_{a=1}^d \frac{(X,V^{a-1},V^a)^2}{(\xi,V^{a-1},V^a) \, (\xi,X,V^a)\, (\xi,V^{a-1},X)} \, . 
\end{align}

Although not immediately obvious, this expression is $X$-independent and reproduces the Sasakian volume given in \cite{Martelli:2005tp,Martelli:2006yb}. For example, for the five-sphere, using \eqref{fanS5}, we find
\begin{align}\label{volS5}
\text{Vol}_{S^5}(\xi) =\frac{\pi^3}{\xi_0 \xi_1 \xi_2} \, .
\end{align} 
The $X$-independence of \eqref{sasvolX} can be  proved as follows.
 In order to satisfy the  Calabi-Yau condition the $V^a$ must lie on a plane, $\nu_i V_i^a=1$ for all $a=1,\ldots d$, where $\nu_i$ are integer numbers. Using the determinant identity
 \eqref{linearly_dependent}
\begin{align}
	\det(X,V^{a-1},V^a)\cdot\xi-\det(\xi,V^{a-1},V^a)\cdot X+\det(\xi,X,V^a)\cdot V^{a-1}-\det(\xi,X,V^{a-1})\cdot V^a=0\:,
\end{align}
and  contracting with the CY direction $\nu_i$ we obtain
\begin{align}
\label{determinant_c1_vs_eta}
	\frac{\det(\xi,X,V^a)}{\det(\xi,V^{a-1},V^a)}-\frac{\det(\xi,X,V^{a-1})}{\det(\xi,V^{a-1},V^a)}=
		-(\xi_i\,\nu_i)\,\frac{\det(X,V^{a-1},V^a)}{\det(\xi,V^{a-1},V^a)}+X_i\,\nu_i\:,
\end{align}
which, using \eqref{weights_VS_Chern} and \eqref{deta} with $c_\eta=1$, is equivalent to the relation
\begin{align} \label{detac1}
	2\pi\sum_{b=0}^dc_1^\bT(\cL_b)\,\big|_{L_a}= (\xi_i \nu_i) \: (\dd_X \eta)\,\big|_{L_a} +X_i\,\nu_i\:.
\end{align}
Since the equivariant forms $2\pi\sum_{b=0}^dc_1^\bT(\cL_b)$ and $(\xi_i \nu_i) \: (\dd_X \eta)$ differ by a constant at the localization loci, we can use replace one with the other when computing the integral of polyforms of top degree. Using the determinant identity \eqref{determinant_product_formula}, we easily compute 
\begin{align} &\text{Vol}_{{\cal D}_a}(\xi) = \frac 12 \int _{{\cal D}_a}  \eta \wedge  \dd \eta= \frac 12 \int _{M_5} \eta \wedge  \dd \eta\wedge c_1(\cL_a) = \frac 12 \int _{M_5} \eta \wedge  \dd_X \eta\wedge c_1^\bT(\cL_a)= \\
&-2\pi^2\left (\frac{(X,V^{a-1},V^a)}{(\xi,V^{a-1},V^a)(\xi,X,V^a)}+\frac{(X,V^{a},V^{a+1})}{(\xi,V^{a},V^{a+1})(\xi,V^a,X)} \right )= \frac{2\pi^2(V^{a-1},V^a,V^{a+1})}{(\xi,V^{a-1},V^a)(\xi,V^{a},V^{a+1})}
 \end{align}
which is indeed the Sasakian volume of the three-cycle ${\cal D}_a$ associated with $V^a$  \cite{Martelli:2005tp}. Notice that $X$ nicely cancels out. Using \eqref{detac1} and the fact that $\int_{M_5} \eta\wedge \dd_X\eta=0$, being the integral of a polyform of maximum degree three,\footnote{This can be also easily checked using the localization formula \eqref{localization_formula}.} we now find
\begin{align}
 \label{eq:vol5II} \text{Vol}_{M_5}(\xi) =\frac18 \int_{M_5} \eta \wedge (\dd_X \eta)^2 = \frac{\pi}{4 (\xi_i \nu_i)}\sum_{a=1}^d \int_{M_5} \eta \wedge \dd_X \eta \wedge c_1^\bT(\cL_a) \\  = \frac{\pi}{2 (\xi_i \nu_i)} \sum_{a=1}^d \text{Vol}_{{\cal D}_a}(\xi)=  \frac{\pi^3}{\xi_i \nu_i} \sum_{a=1}^d \frac{(V^{a-1},V^a,V^{a+1})}{(\xi,V^{a-1},V^a)(\xi,V^{a},V^{a+1})}  \, ,\end{align}
reproducing the Sasakian volume of $M_5$ \cite{Martelli:2005tp,Martelli:2006yb}.

 \subsubsection{The GMS master volume}
\label{The GMS master volume}

We can analogously use odd-dimensional localization to compute a generalization of the Sasakian volume,
the GMS master volume which has important applications to holography \cite{Couzens:2018wnk,Gauntlett:2018dpc}.
Given the Reeb vector $\xi$ and the associated contact form $\eta$ for the Sasakian metric of $M_5$, the master volume is defined by
\begin{align} \label{eq:master5} {\cal V}_{M_5}(\xi,\lambda_a) = \frac12 \int_{M_5} \eta \wedge \Kform^2 \, ,\end{align}
where $\Kform$ is a transverse K\"ahler form. The $\lambda_a$ are the K\"ahler parameters of the geometry defined by expanding the cohomology class of $\Kform$ in the Chern classes $c_1(\cL_a)$ (see appendix \ref{app:toric} for more details)
\begin{align}  [ \Kform ]= -2\pi \sum_{a=1}^d \lambda_a c_1(\cL_a)\, .\end{align}

Using  the localization formula \eqref{localization_formula}, we can easily compute 
 \begin{align} \int_{L_{a}}  \eta   =\int _{M_5} \eta \wedge  c_1(\cL_{a-1})\wedge c_1(\cL_a) =  \int _{M_5} \eta \wedge  c_1^\bT(\cL_{a-1})\wedge c_1^\bT(\cL_a) = \frac{2\pi}{(\xi,V^{a-1},V^a)} \, ,\end{align} 
 which is consistent with \eqref{inteta}, taking into account that $c_\eta=1$ for a Sasaki manifold. Using the determinant identities, we also have
 \begin{align} &\int _{M_5} \eta \wedge  c_1(\cL_{a})\wedge c_1(\cL_a) =  \int _{M_5} \eta \wedge  c_1^\bT(\cL_{a})\wedge c_1^\bT(\cL_a)  =\\
&\frac{2\pi(\xi,V^{a-1},X)}{(\xi,V^{a-1},V^a)(\xi,X,V^a)}+\frac{2\pi(\xi,X,V^{a+1})}{(\xi,V^{a},V^{a+1})(\xi,V^a,X)} =- \frac{2\pi (\xi, V^{a-1},V^{a+1})}{(\xi,V^{a-1},V^a)(\xi,V^{a},V^{a+1}))} \, . \end{align}
We then see that
\begin{align}\label{master}  {\cal V}_{M_5}(\xi,\lambda_a) = \frac{(2\pi)^3}{2}  \sum_{a=1}^d \frac{\lambda_a \left ( \lambda_{a-1} (V^a,V^{a+1},\xi) -\lambda_{a} (V^{a-1},V^{a+1},\xi)+\lambda_{a+1} (V^{a-1},V^{a},\xi)\right)}{(V^{a-1},V^{a},\xi) (V^a,V^{a+1},\xi)} \, ,\end{align}
reproducing the expression for the master volume computed with a different method in \cite{Gauntlett:2018dpc}.

\section{Minimal gauged supergravity in five dimensions}
\label{sec:sugra5d}
In this section we will rewrite the on-shell action of supersymmetric solutions of minimal gauged supergravity in five dimensions in a form which is amenable to be evaluated using localization. Bosonic supersymmetric solution of this theory were classified in \cite{Gauntlett:2003fk}. We will focus on the class of supersymmetric solutions with a time-like Killing vector, leaving the null case for future work. We will also discuss the toric geometry associated with the K\"ahler base of such solutions.

\subsection{Supersymmetric solutions in the time-like class}
\label{sec:timelike_susy}

5d minimal gauged supergravity has the following bosonic action, written as an integral over a 5d Lorentzian manifold $M$
\begin{align}\label{eq:action}
  S[M]={}& \frac{1}{16\pi G_5} \int_{M}^{}\cL\:,\qquad
	\cL=\star_g \qty(R_g + \frac{12}{\ell^2}) - \frac{\curly{x}^2}{2} F \wedge \star_g F - \frac{\curly{x}^3}{3 \sqrt{3}} A \wedge F \wedge F \,, 
\end{align}
where $\ell$ is the $\text{AdS}_5$ scale, $R_g$ and $\star_g$ are the 5d Ricci scalar and Hodge star with respect to the 5d metric $g$, $A$ is an Abelian gauge field with field strength $F = \dd[]{A}$, and $\curly{x}$ is a constant that we use to keep track differing normalizations of the gauge field across the literature. The equations of motion of this action read
\begin{align}\label{eq:eom}
  \begin{aligned}
    (R_g)_{\mu\nu} + \frac{4}{\ell^2} g_{\mu\nu} - \frac{\curly{x}^2}{2} F_{\mu\rho} F\indices{_\nu^\rho} + \frac{\curly{x}^2}{12} g_{\mu\nu} F_{\rho\sigma}F^{\rho\sigma} ={}& 0 \,, \\
    \dd[]{ \star_g F} + \frac{\curly{x}}{\sqrt{3}} F \wedge F ={}& 0\,. 
  \end{aligned}
\end{align}
Bosonic supersymmetric solution of this theory were classified in \cite{Gauntlett:2003fk}. Here we follow the conventions of \cite{Cassani:2015upa}. A solution is supersymmetric if there exists a non-zero Dirac spinor $\epsilon$ satisfying
\begin{align}\label{eq:kse}
  \qty[(\nabla_g)_\mu - \frac{\iu \, \curly{x}}{8 \sqrt{3}} \qty(\Gamma_{\mu\nu\rho} - 4 g_{\mu\nu} \Gamma_\rho) F^{\nu\rho} - \frac{1}{2 \ell} \qty(\Gamma_\mu + \iu \sqrt{3} \, \curly{x} A_\mu)] \epsilon ={}& 0 \,,
\end{align}
where the covariant derivative acting on spinors reads
\begin{align}
  (\nabla_g)_\mu ={}& \partial_\mu + \frac{1}{4} \omega_{\mu AB} \Gamma^{AB} \,,
\end{align}
with $A,B,\dotso = 0,\dots, 4$ denoting the 5d flat indices. Further, in \cite{Gauntlett:2003fk} it was shown that all such solutions admit a Killing vector $\xi$ that is either timelike or null. Focusing on the timelike case we choose adapted coordinates such that $\xi = \partial_y$ and the 5d metric takes the form
\begin{align}\label{eq:5d_metric}
  \dd[]{s}^2_g ={}& - f^2 (\dd[]{y} + \omega)^2 + f^{-1} \dd[]{s}^2_\gamma \,, 
\end{align}
where $\gamma_{mn}$ is the metric on the base with $m,n,\dotso$ running over the base coordinates, $f$ is a positive function on the base and $\omega$ is a basic one-form. We define the normalized dual form to the Killing vector $\xi$
\begin{align}
  \eta_0 = - f^{-2} \xi^\flat ={}& - f^{-2} g_{\mu\nu} \xi^\mu \dd[]{x^\nu} = \dd[]{y} + \omega \,,
\end{align}
whose differential is basic
\begin{align}\label{ddomega}
  \dd[]{\eta_0} ={}& \dd[]{\omega} \,, \qquad \iota_\xi \dd[]{\eta_0} = 0 \,. 
\end{align}
Supersymmetry dictates that the base is K\"ahler, meaning that it admits a real, non-degenerate, closed two form $\X^1$, such that $(\X^1)\indices{_m^n}$ is an integrable complex structure. The 4d base also admits a complex two form $\X^2 + \iu \X^3$ satisfying
\begin{align}\label{eq:P_condition}
  \qty(\nabla_\gamma + \iu P)_m \qty(\X^2 + \iu \X^3)_{np} ={}& 0 \,,
\end{align}
where $P$ is the potential for the Ricci form $\mathcal{R}$
\begin{align}
  \mathcal{R} ={}& \dd[]{P} \,, \quad \mathcal{R}_{mn} = \frac{1}{2} (R_\gamma)_{mnpq} (\X^1)^{pq} \,,
\end{align}
and the real two-forms $\X^I$, $I = 1,2,3$ satisfy  
\begin{align}\label{eq:XI_conditions}
  (\X^I)\indices{_m^p} (\X^J)\indices{_p^n} ={}& - \delta^{IJ} \delta\indices{_m^n} + \varepsilon^{IJK} (\X^K)\indices{_m^n} \,, \quad \star_\gamma \X^I = - \X^I \,.
\end{align}
The volume element on the base is given by
\begin{align}\label{eq:orientation_base}
  \star_\gamma 1 ={}& - \frac{1}{2} \X^1 \wedge \X^1 = - \frac{1}{2} \X^2 \wedge \X^2 = - \frac{1}{2} \X^3 \wedge \X^3 \,. 
\end{align}
The geometry of the K\"ahler base determines the entire 5d solution through
\begin{align}\label{eq:susy_relations}
  f = - \frac{24}{\ell^2 R_\gamma} \,, \quad F = - \frac{\sqrt{3}}{\curly{x}} \dd[]{\qty[ f \eta_0 + \frac{\ell}{3} P]} \,, 
  \end{align}
  and 
  \begin{align}
  \label{domegaeq}
   f \dd[]{\omega} = G^+ + G^- \,, \qquad  G^+ = - \frac{\ell}{2}\mathcal{R} - \frac{3}{\ell} f^{-1} \X^1 \,,
\end{align}
where $\star_\gamma G^\pm = \pm G^\pm$ are self-dual and anti-self-dual two forms, respectively, with $G^-$  determined up to an anti-holomorphic function; see \cite{Gauntlett:2003fk, Cassani:2015upa} for details.
Finally the integrability condition for (\ref{domegaeq}) reads \cite{Cassani:2015upa}
\begin{align}
\label{nastyPDE}
  \nabla_\gamma^2 \left[ \frac{1}{2} \nabla_\gamma^2 R_\gamma + \frac{2}{3} (R_\gamma)_{mn}(R_\gamma)^{mn} - \frac{1}{3} R_\gamma^2  \right] + \nabla^m_\gamma \qty((R_\gamma)_{mn} \partial^n R_\gamma) ={}& 0 \,. 
\end{align}
This is a very complicated PDE for the metric $\gamma$, which is the main stumbling block for finding explicit solutions.

\subsection{Symplectic toric description}
\label{sect:symptoric}

Having established  that supersymmetric solutions of minimal gauged supergravity may be characterised in terms of K\"ahler geometry, it is natural to wonder how much milage 
one can get by making the additional assumption that the K\"ahler metric admits two commuting isometries, so that one can invoke the machinery of symplectic toric geometry, that we reviewed in section 
\ref{sect:theorem} (see also appendix \ref{app:toric}). This idea played an important role in the context of Sasaki-Einstein (SE) manifolds 
\cite{Martelli:2005tp} and more recently in the framework of GK geometry \cite{Gauntlett:2018dpc,Gauntlett:2019roi,Hosseini:2019ddy}, which are indeed instances of 
odd-dimensional manifolds charaterised by a transverse K\"ahler foliation. In fact, the analogy of the present set  up with these goes further as we also have a canonical Killing vector (that we might occasionally refer to as ``Reeb'' vector) that plays a central role in characterising the full odd-dimensional geometry. Precisely as in the SE and GK contexts, the transverse foliation may not be
 globally defined, even if the odd-dimensional geometry is smooth and in order to encompass generic solutions  one needs to work at least in the orbifold category.

The toric assumption implies that there exist canonical (Darboux) coordinates $y_i,\tau^i$, in which the  
four-dimensional metric $\dd s^2_\gamma$ in (\ref{eq:5d_metric}) takes the form
\begin{equation}
\label{toricmetric}
\dd s^2 =G_{ij} (y) \dd y_i \dd y_j + G^{ij} (y) \dd\tau^i \dd\tau^j \, , 
\end{equation}
where $G_{ij} = \partial_{y_i}\partial_{y_j} G(y)$ and $G^{ij}$ is its inverse matrix. In principle, the metric is fully determined by the 
function $G(y)$, that is referred to as symplectic potential. The angular coordinates $\tau^i$ are $2\pi$-periodic and the K\"ahler form  reads
\begin{equation}
\label{X1toric}
\Kform^1 = \dd y_i \wedge \dd \tau^i\, , 
\end{equation}
showing that  $y_i$ are the moment maps associated to the torus action.
The five-dimensional solution is then reconstructed as follows.  The function $f$  is determined in terms of the symplectic potential using the known expression for the Ricci scalar
\begin{equation}
R_\gamma = -\partial_{y_i}\partial_{y_j}  G^{ij}\, . 
\end{equation}
The Ricci potential $P$ reads
\begin{equation}
P  = - \frac{1}{2}\partial_{y_i}  G^{ij} \dd \tau^j\, , 
\end{equation}
and the one-form $\omega$  is  $\omega = \omega_i \dd \tau^i$, however this cannot be written explicitely in terms of the metric $G_{ij}$.

In principle the symplectic potential $G$ characterises a solution uniquely, but unfortunately it obeys a 
unwieldy PDE following from (\ref{nastyPDE}), for which at present there are no available results. 
This  hinders progress in finding explicit local solutions and undermines classification programs analogous to those available in the ungauged theory \cite{Breunholder:2018roc}. 
Nevertheless, some useful information can be extracted from analysing the behaviour of the symplectic potential near to the loci (partially) fixed by the torus action.
For example, in  \cite{Lucietti:2022fqj} it is shown  that an extremal horizon corresponds to a singular point in the moment polytope.\footnote{This extends to gauged supergravity previous results obtained in other contexts \cite{Hollands:2007aj,Hollands:2007qf,Figueras:2009ci}.}
One can make some progress by removing the assumption that the K\"ahler base is a smooth manifold  \cite{Lucietti:2022fqj},  allowing at least for orbifold singularities. Moreover, one can also 
incorporate ``non-extremal horizons'', working in the more general context of Euclidean complex solutions. 
In any case, ultimately one will be confronted with (\ref{nastyPDE}). In this respect, our results may be regarded as a significant step towards the formulation of 
necessary and sufficient conditions for the existence of solutions to this equation, subject to suitable boundary conditions. 
Notably, we will show that the action can be written as a function of the Reeb vector and some topological (toric) data, precisely as in SE and GK geometry.

In the black hole literature   \cite{Hollands:2007aj,Hollands:2007qf,Figueras:2009ci,Breunholder:2018roc} traditionally the topological data of solutions with toric symmetry 
has been describred in terms of a ``rod structure'', which are one-dimensional diagrams, with various labels attached to the ``rods''.  This  structure is essentially equivalent to that of planar non-compact polytopes specified by the normals to its edges (the fan)  \cite{Lucietti:2022fqj}. However, in order to specify in full the five-dimensional geometry 
 one needs a \emph{three-dimensional} fan, of which the two-dimensional one is the projection on a direction transverse to the Reeb vector, as discussed in section \ref{Five-dimensional toric geometries}.

 While the two-dimensional polytope can be determined from the image of the moment maps $y_i$ (see appendix \ref{comparisonwithlucietti} for an explicit example) 
 there is no notion of three-dimensional moment map. However, the three-dimensional fan may be extracted from the geometry as follows.
  Let us denote by 
\begin{align}
\label{4dvectorfields}
S^a = v^a_1\partial_{ \tau_1}+v^a_2\partial_{\tau_2}\,
\end{align}
a Killing vector field that  degenerates on an edge (divisor) of the two-dimensional polytope. This means that $||S^a||_\gamma|_{D_a}=0$. 
Assuming the K\"ahler base is an orbifold,  we have $v^a\in \mathbb{Z}^2$. 
Consider the vector field 
\begin{align}
\label{defKa}
V^a = t^a  \partial_y +  S^a \, ,
\end{align}
where $t^a$ is constant. The norm of $V^a$ in the five-dimensional metric $g$ reads
\begin{align}
  ||V^a||_g^2 &  =  -  f^2  (t^a - v^a_i \omega_i)^2+  v^a_i v^a_j   f^{-1}G_{ij}\, .
\end{align}
For Lorentzian extremal black holes, one can show that $V^a=\xi=\partial_y$ generates the extremal Killing horizon,
and thus $f$ vanishes at the horizon, since $g(\xi,\xi)=-f^2$  \cite{Lucietti:2022fqj}.
In Euclidean signature, we will consider a larger family of supersymmetric solutions  where $\xi$ is different from the Killing vector degenerating at the horizon and $f$ is never vanishing.
This is the case, for example, of the supersymmetric non-extremal black hole solutions considered in \cite{Cabo-Bizet:2018ehj}.
In any case, one can  argue that the function  $v^a_i \omega_i$ is constant on the divisor $D_a$, so that  defining 
 $t^a\equiv v^a_i \omega_i$, we have that   $||V^a||_g^2=0$ on ${\cal D}_a$ despite $f$ is generically complex.
 Thus  $V^a=(t^a,v^a)$ is the three-dimensional lift of the two-dimensional fan $v^a$.
  To see that $v^a_i \omega_i$ is constant on $D_a$,
recall from \eqref{ddomega} that $\dd\omega=\dd\eta_0$.
Since $S^a$ generates a symmetry of the solution, we then have 
   \begin{align}
\dd (v^a_i \omega_i ) = \dd (\iota_{S^a}\omega ) = - \iota_{S_a} \dd \omega =   - \iota_{S_a} \dd \eta_0\:,
\end{align}
and considering that $\dd\eta_0$ is a regular form on the base under the assumption  that $f _{D_a}\neq 0$,
\footnote{If $f|_{D_a}=0$ then $K^a=\partial_y$ generates an extremal horizon  \cite{Lucietti:2022fqj}.
We shall assume that $f$ is never vanishing for the Euclidean solutions of interest.
Considering that $\eta_0$ is (minus) a Killing spinor bilinear divided by the square of $f$, $\eta_0$ is regular under the assumption of never-vanishing $f$.}
we conclude that  $\dd (v^a_i \omega_i )|_{D_a}=0$.

 The relationship between the four-dimensional and five dimensional points of view is further illustrated in appendices \ref{app:known} and \ref{comparisonwithlucietti}.

\subsection{Rewriting of the on-shell action}
\label{sect:rewriting}

In this section we impose the equations of motion and some supersymmetry relations to rewrite the action \eqref{eq:action} as a sum of a bulk and a boundary contribution, where the bulk contribution ends up being a Chern-Simons type of term. First, we make the choice of opposite relative orientation between the total space and the base
\begin{align}\label{eq:opposite_orientation}
  \star_g 1 ={}& - f^{-1} \eta_0 \wedge \star_\gamma 1 \,.
\end{align}
This implies that for a basic one-form $\theta_{(1)}$ and a basic two-form $\theta_{(2)}$ we have the relations
\begin{align}\label{eq:relation_5d_4d_hodge}
  \star_g \qty(\eta_0 \wedge \theta_{(1)}) ={}& f^{-2} \star_\gamma \theta_{(1)} \,, \qquad \star_{g} \, \theta_{(2)} = - f \eta_0 \wedge \star_\gamma \, \theta_{(2)} \,. 
\end{align}
Next, we trace the Einstein equation in \eqref{eq:eom} and substitute the result for $\star_g R_g$ into the Lagrangian density \eqref{eq:action} to obtain
\begin{align}\label{eq:osa_after_remove_ricci}
  	\cL=- \frac{8}{\ell^2} \star_g 1 - \frac{\curly{x}^2}{3} F \wedge \star_g F - \frac{\curly{x}^3}{3 \sqrt{3}} A \wedge F \wedge F\:.
\end{align}
When supersymmetry is present the 5d volume element can be written as
\begin{align}\label{eq:grav_term}
  - \frac{8}{\ell^2} \star_g 1 ={}& \frac{4}{3\ell} \eta_0 \wedge \qty(G^+ + \frac{\ell}{2} \mathcal{R}) \wedge \X^1 = \frac{2}{3} \eta_0 \wedge \mathcal{R} \wedge \X^1 \,,
\end{align}
where to obtain the first equality we have used \eqref{eq:orientation_base} and \eqref{domegaeq}. To obtain the second equality we have used that $G^+ \wedge \X^1 = 0$, due to $G^+$ being self-dual and $\X^1$ being anti-self-dual. Using \eqref{eq:susy_relations}, \eqref{eq:relation_5d_4d_hodge}, $\star_\gamma G^{\pm} = \pm G^{\pm}$, and \eqref{domegaeq} we obtain
\begin{align}\label{eq:stargF}
  \star_g F ={}& - \frac{\sqrt{3}}{\curly{x}} \qty[ - f^{-2} \star_\gamma \dd[]{f} + f \eta_0 \wedge \qty(f \dd[]{\omega} + \frac{2}{3} \ell \mathcal{R} + \frac{2}{\ell} f^{-1} \X^1)] \,,
\end{align}
meaning that the Maxwell term can be written as
\begin{align}\label{eq:maxwell_term}
  - \frac{\curly{x}^2}{3} F \wedge \star_g F ={}& \dd[]{\qty(f \eta_0)} \wedge f^{-2} \star_\gamma \dd[]{f} - f \eta_0 \wedge \qty(f \dd[]{\omega} + \frac{\ell}{3} \mathcal{R}) \wedge \qty(f \dd[]{\omega} + \frac{2}{3} \ell \mathcal{R}) \notag \\
     & - \frac{2}{\ell} \eta_0 \wedge \qty(f \dd[]{\omega} + \frac{\ell}{3} \mathcal{R}) \wedge \X^1 \,. 
\end{align}
We also impose the Maxwell equation through
\begin{align}\label{eq:maxwell_relation}
  \dd[]{ \star_g F} + \frac{\curly{x}}{\sqrt{3}} F \wedge F ={}& 0 & & \Longleftrightarrow & & \dd[]{\qty(f^{-2} \star_\gamma \dd[]{f})} = \frac{2}{\ell} \dd[]{\omega} \wedge \X^1 - \qty(\frac{\ell}{3} \mathcal{R})^2 \,.
\end{align}
Plugging \eqref{eq:grav_term}, \eqref{eq:maxwell_term} and \eqref{eq:maxwell_relation} into the action \eqref{eq:osa_after_remove_ricci}, the Lagrangian density 
becomes
\begin{align}\label{eq:Lagrangian_intermediate}
  \mathcal{L} ={}&  - f \eta_0 \wedge \qty(\qty(f \dd[]{\omega} + \frac{\ell}{3} \mathcal{R})^2 + \qty(f \dd[]{\omega} + \frac{\ell}{3} \mathcal{R}) \wedge \frac{\ell}{3} \mathcal{R} + \qty(\frac{\ell}{3} \mathcal{R})^2) - \frac{\curly{x}^3}{3 \sqrt{3}} A \wedge F \wedge F \notag \\
                 &+ \dd[]{} \qty[f \eta_0 \wedge f^{-2} \star_\gamma \dd[]{f}] \,. 
\end{align}
Now we define a new $\eta$ as
\begin{align}\label{eq:new_eta}
  \eta ={}& f\eta_0 + \frac{\curly{x}}{\sqrt{3}} A \,, 
\end{align}
whose differential, similarly to $\eta_0$, is basic 
\begin{align}
  \dd[]{\eta} ={}& - \frac{\ell}{3} \mathcal{R} \,, \qquad \iota_\xi \dd[]{\eta} = 0 \,. 
\end{align}
From the definition of $F$ in \eqref{eq:susy_relations} we have $f \dd[]{\omega} + \frac{\ell}{3} \mathcal{R} = - \qty(\dd[]{f} \wedge \eta_0 + \frac{\curly{x}}{\sqrt{3}} F)$ and the Lagrangian \eqref{eq:Lagrangian_intermediate} can be massaged to 
\begin{align}
  \mathcal{L} ={}& - \eta \wedge (\dd[]{\eta})^2 + \dd[]{} \qty[f \eta_0 \wedge f^{-2} \star_\gamma \dd[]{f} + \eta \wedge \frac{\curly{x}}{\sqrt{3}}A \wedge \qty(\dd[]{\eta} + \frac{\curly{x}}{\sqrt{3}}F)] \,. 
\end{align}
Finally, using \eqref{eq:new_eta}, we can express this result entirely in terms of $\eta_0, \eta$ and the function $f$  
\begin{empheq}[box={\mymath[colback=white, colframe=black]}]{equation}\label{eq:Lagrangian_final}
  \mathcal{L} = - \eta \wedge \qty(\dd[]{\eta})^2 + \dd[]{} \qty[f \eta_0 \wedge f^{-2} \star_\gamma \dd[]{f} + 2 f \eta_0 \wedge \eta \wedge \dd[]{\eta} + \eta \wedge f \eta_0 \wedge \dd[]{} (f\eta_0)] 
\end{empheq}

In our main applications we will analytically continue the solution to Euclidean signature and we will consider (generically complex) smooth geometries.
The Euclidean action would be
\begin{equation}
	I[M]=-\frac{\ii}{16\pi G_5}\int_M\cL
\end{equation}
where $\cL$ has been analytically continued to a complex five-form, and it is understood that $S[M]$ in \eqref{eq:action} takes as an argument a Lorentzian manifold, while $I[M]$ takes as an argument a Euclidean manifold.
We will also assume that we can work in a gauge where $A$, which is generically given by
\begin{align}
  A ={}& - \frac{\sqrt{3}}{\curly{x}} \qty[f \eta_0 + \frac{\ell}{3}P + \dd[]{\Lambda}] \,, 
\end{align}
is everywhere regular on $M$. 
In section  \ref{sect:reg gauge} we will show how to fix such a gauge, when it is possible.
The form $\eta$ is a linear combination of $f\eta_0$, which is everywhere regular for the Euclidean solutions of interest,
and the gauge field $A$.
Ensuring that $A$ is everywhere regular also makes $\eta$ everywhere regular.
Then the total derivative term in \eqref{eq:Lagrangian_final} is everywhere regular and the action can be written as
\begin{equation}
\label{OSA_rewritten}
	16\pi\ii\,G_5\cdot I[M]=-\int_M\eta\wedge(\dd\eta)^2+\int_{\partial M}\alpha(M)\:,
\end{equation}
where we have defined the four-form $\alpha(M)$ as
\begin{equation}
\label{alpha_expression}
	\alpha(M)=f \eta_0 \wedge f^{-2} \star_\gamma \dd[]{f} + 2 f \eta_0 \wedge \eta \wedge \dd[]{\eta} + \eta \wedge f \eta_0 \wedge \dd[]{} (f\eta_0)\:.
\end{equation}
The first piece in \eqref{OSA_rewritten} is the integral over $M$ of the Chern-Simons form $\eta\wedge(\dd\eta)^2$, which can easily be recast in
the form required for the application of the localization formula \eqref{localization_formula}, as we are going to discuss in the next section.

\section{Localization of the on-shell action}
\label{sect:on shell localization}

In this section we show how to evaluate the on-shell action of five-dimensional minimal gauged supergravity
using the equivariant localization theorem presented in section \ref{sect:theorem}.

In section \ref{sect:rewriting} we have shown that the on-shell action can be recast as an integral of the Chern-Simons form $\eta\wedge(\dd\eta)^2$
on the manifold $M$, plus an integral over the boundary $\partial M$.
This rewriting relies on a choice of gauge that makes $A$ regular everywhere on $M$; in section \ref{sect:reg gauge}
we will discuss the existence of such a gauge and its implications.
The localization formula \eqref{localization_formula} can then be applied to the Chern-Simons term,
recasting the integral over $M$ into a sum over a finite number of localized contributions in the bulk plus a boundary integral:
\begin{equation}
\label{CS_integral}
	-\int_M\eta\wedge(\dd\eta)^2=\sum_a\cI_a[M;X]+\int_{\partial M}\beta(M;X)\:,
\end{equation}
where $X$ is an arbitrary Killing vector. Each term separately depends on $X$, but the $X$-dependence drops out when they are all summed.
We will give the expression for bulk contributions $\cI_a[M;X]$ and the four-form $\beta(M;X)$ in section \ref{sect:bulk localization};
for the moment let us focus on a bird's-eye view of the strategy that we will implement throughout this section.

Putting together \eqref{OSA_rewritten} and \eqref{CS_integral} we obtain the following expression for the on-shell action $I[M]$:
\begin{equation}
\label{I[M]}
	16\pi\ii\,G_5\cdot I[M]=\sum_a\cI_a[M;X]+\int_{\partial M}\Big(\alpha(M)+\beta(M;X)\Big)\:,
\end{equation}
where the four-form $\alpha(M)$ is given by \eqref{alpha_expression}.
The boundary contributions
in the above expression are divergent%
\footnote{The $\alpha(M)$ and $\beta(M;X)$ terms are separately divergent, and the divergences do not cancel when summed together.}
in the limit where the distance of the boundary from the center goes to infinity, and thus requires regularization.
In holography one usually deals with such divergences using either holographic renormalization or background subtraction. Here we implement the latter procedure and define the following renormalized quantity
\begin{align}
  \widehat{I} ={}& I[M] - I[N] \,,
\end{align}
where $N$ is a suitably chosen subtraction manifold.

In this paper we will consider the case of conformally flat asymptotic boundaries $\partial M$, $\partial N$, leaving the more general case for future work.
With this assumption, in \cref{sect:boundary} we will show that
as long as the subtraction manifold $N$ has the same leading asymptotics as $M$, the total boundary contribution to the on-shell action vanishes:
\begin{equation}
\label{bdy_cancellation}
	16\pi\ii\,G_5\cdot \big(I_{\text{bdy}}[\partial M]-I_{\text{bdy}}[\partial N]\big)=
		\int_{\partial M}\Big(\alpha(M)+\beta(M;X)\Big)-\int_{\partial N}\Big(\alpha(N)+\beta(N;X)\Big)=0\:.
\end{equation}
Then the renormalized on-shell action will be given by the difference of the localized contributions in the bulk coming from $M$ and $N$ respectively:
\begin{equation}
\label{hat_I_fmla}
	16\pi\ii\,G_5\cdot \widehat I=\sum_a\cI_a[M;X]-\sum_{a'}\cI_{a'}[N;X]\:.
\end{equation}
In section \ref{sect:black hole localization} we will first consider the case of the supersymmetric black hole, taking the subtraction manifold $N$
to be pure AdS$_5$, and we will show that our procedure correctly reproduces the on-shell action of the known solution \cite{Cabo-Bizet:2018ehj}
without relying on knowledge of the explicit metric.
In section \ref{sect:bulk localization} we will discuss more general geometries.

\subsection{Regularity condition for the gauge}
\label{sect:reg gauge}

In this section we discuss the conditions for the regularity of the gauge field $A$ and when it is possible to have $A$ regular everywhere on $M$.
Regularity of $A$ is equivalent to regularity of $\eta$
\begin{align}\label{eq:eta_gauge_shift}
  \eta ={}& f\eta_0 + \frac{\curly{x}}{\sqrt{3}} A= -\frac{\ell}3P-\dd\Lambda\,, 
\end{align}
where $\dd\Lambda$ is a gauge shift to be determined.

As in section \ref{Five-dimensional toric geometries}, we denote with $\D_a$ the loci where the $\bT= \text{U}(1)^3$ isometry has a $\text{U}(1)$ isotropy subgroup,
and we denote with $V^a$ the vector on $M$ that generates this subgroup, and thus vanishes at $\D_a$.
A necessary condition for the regularity of $\eta$ is that its contraction with $V^a$ vanishes at $\D_a$,
\begin{equation}
	\iota_{V^a}\,\eta\,\big|_{\D_a}=0\:,
\end{equation}
since $V^a$ vanishes at $\cD_a$, and thus the above contraction would not be zero only if $\eta$ was divergent.
As we will argue in a moment, the Ricci potential $P$ satisfies%
\footnote{$P$ can always be redefined by adding a basic closed one-form. However there is a particularly convenient ``gauge" for $P$, which is the one
		that satisfies \eqref{P_irregularity}.}
\begin{equation}
\label{P_irregularity}
	\iota_{V^a}\,P\,\big|_{\D_a}=-1\:,
\end{equation}
and thus $P$ has a non-vanishing component along the angle that degenerates at $\cD_a$, making $P$ irregular.
We also anticipate that this is the only source of irregularity of $P$ at the $\D_a$.
Therefore we can take $\dd\Lambda=\frac\ell3\nu_i\,\dd\phi_i$ with $\nu_i$ constants, without loss of generality,
and conclude that $\eta$ is regular at $\D_a$ if and only if 
\begin{equation}
\label{gauge_regularity}
	0=\iota_{V^a}\,\eta\,\big|_{\D_a}=\frac\ell3\,(1-\nu_iV^a_i)\:.
\end{equation}
Imposing that $\eta$ is regular everywhere is then equivalent to demanding the existence of constants $\Lambda_i$
that verify the above condition for all $\D_a$. Notice that requiring \eqref{gauge_regularity} for all $a$ implies that the vectors $V^a$ lie on a plane. In the compact case, the analogous condition with $\nu_i$ integers identifies Sasaki-Einstein manifolds where the vectors $V^a$ form the fan of the corresponding six-dimensional Calabi-Yau cone. We will therefore sometime refer to imposing \eqref{gauge_regularity} for all $a$  as the ``Calabi-Yau'' condition. Physically, this  implies that the integral of $F$ over any two-cycle in the geometry must be zero so that there are no magnetic fluxes. Moreover, using  the relation (\ref{eq:susy_relations}) and the fact that $f$ and $\eta_0$ are globally regular, it follows that the pullback of ${\cal R}$ to the five-dimensional geometry, $\pi^* {\cal R}$, is trivial in cohomology, $[\pi^* {\cal R}]=0$, which would be the standard definition of a Calabi-Yau condition on a K\"ahler manifold.

Let us now discuss the irregularity of the Ricci potential $P$ and relation \eqref{P_irregularity}.
From the cohomological identity, valid on the base,
\begin{equation}
	[\cR]=2\pi\sum_a\big[c_1(\cL_a)\big]\:,
\end{equation}
together with expression \eqref{c_1_def} for $P$, we can write 
\begin{equation}
	P=2\pi\sum_a\sum_{i=1,2}\mu^a_i\dd\phiB_i+(\text{regular one-form})\:.
\end{equation}
From the above expression we can already conclude that the only possible source of irregularity of $P$ at $\cD_a$ would be a non-vanishing
component along the angle that degenerates.
The contraction $\iota_{V^a}P$ at $\D_a$ is expressible in terms of the $v^a$ introduced in \eqref{small_v_def} as
\begin{equation}
	\iota_{V^a}\,P\,\big|_{\D_a}=2\pi\sum_{i=1,2}\mu_i^av_i^a\,\big|_{\D_a}\:.
\end{equation}
We can argue that $\iota_{V^a}P$ is constant along $\D_a$ as follows: since the moment maps $\mu^a_i$ do not depend on the angles, we have
\begin{equation}
	0=\iota_{V^a}\sum_{i=1,2}\dd(\mu^a_i\dd\phiB_i)\,\big|_{\D_a}=-\dd\sum_{i=1,2}\mu^a_iv_i^a\,\big|_{\D_a}\:,
\end{equation}
which implies that $\iota_{V^a}P\,|_{\D_a}$ is constant. Since the closure of $\cD_a$ includes the loci $L_a$, $L_{a+1}$, we have
\begin{equation}
	\iota_{V^a}\,P\,\big|_{\D_a}=\iota_{V^a}\,P\,\big|_{L_a}=2\pi\sum_{i=1,2}\mu_i^av_i^a\,\big|_{L_a}=
		-2\pi\sum_{i=1,2}\frac{(u^2_a)_iv^a_i}{2\pi\,d_a}=-1\:,
\end{equation}
where we have used \eqref{mm} and \eqref{u_defining_property}. It is also easy to check that if we had used 
$\iota_{V^a}P|_{\D_a}=\iota_{V^a}P|_{L_{a+1}}$ instead we would have reached the same conclusion.

\subsection{The black hole on-shell action from localization}
\label{sect:black hole localization}

Before considering a general geometry, let us focus first on one of the most relevant examples where we can apply our localization procedure,
the supersymmetric black hole, found in \cite{Chong:2005hr} as a (real) Lorentzian solution that is also extremal,
and analytically continued to a family of (complex) Euclidean non-extremal supersymmetric solutions in \cite{Cabo-Bizet:2018ehj}. Some general aspects
of these solutions are reviewed in appendix \ref{app:known}.
In this section we will reproduce the on-shell action of these Euclidean solutions, without using any explicit knowledge of the metric,
but rather starting from simple topological considerations.

We work with three $2\pi$-periodic angular coordinates $\phi_0$, $\phi_1$, $\phi_2$, defined as follows:
$\partial_{\phi_0}$, $\partial_{\phi_1}$, $\partial_{\phi_2}$
are the generators of the $\text{U}(1)^3$ isometry, and in relation to the asymptotic boundary $\partial M$, which has topology $S^1\times S^3$,
the angle $\phi_0$ is the coordinate along $S^1$, while $\phi_1,\phi_2$ are angular coordinates in $S^3$.
More precisely, $\phi_1,\phi_2$ are such that the vector fields $\partial_{\phi_2}$ and $\partial_{\phi_1}$ become zero at antipodal circles in the
asymptotic $S^3$, which we refer to as the North and South poles respectively.

In the bulk, $\partial_{\phi_2}$ and $\partial_{\phi_1}$
vanish on three-dimensional non-compact submanifolds, which we call $\D_0$ and $\D_2$ respectively,
that connect the N/S poles of the asymptotic $S^3$ to the N/S poles of the ``horizon".
At the ``horizon" of these Euclidean black holes, the time circle $S^1$ shrinks to zero, as prescribed by the Gibbons-Hawking procedure
for analytic continuation of black hole solutions \cite{Gibbons:1976ue}. We call $\D_1$ this ``horizon" region, and since the time circle
is parametrized by $\phi_0$, we have that the vector field $\partial_{\phi_0}$ vanishes at $\D_1$.
In summary, we have the following three-dimensional submanifolds where the action of the $\text{U}(1)^3$ isometry has a $\text{U}(1)$ isotropy subgroup: 
\begin{equation}
\label{BH_divisors}
	\D_0\::\quad\partial_{\phi_2}\,|_{\D_0}\,=\,0\:,\qquad\quad\D_1\::\quad\partial_{\phi_0}\,|_{\D_1}\,=\,0\:,\qquad\quad
		\D_2\::\quad\partial_{\phi_1}\,|_{\D_2}\,=\,0\:.
\end{equation}
We have ordered them in such a way that $\overline{\D_a}\cap\overline{\D_b}$ is non-empty only if $a$, $b$ are consecutive.
Denoting with $V^a\equiv V^a_i\partial_{\phi_i}$ the Killing vectors that vanish at each $\D_a$ as per \eqref{BH_divisors},
their components in the $\partial_{\phi_i}$ basis are%
\footnote{With some abuse of notation, we are denoting with $V^a$ both the 5d vector field and its three-component restriction to the 
		$\partial_{\phi_i}$ basis.}
\begin{equation}
\label{BH_vectors}
	V^0=(0,0,1)\:,\qquad V^1=(1,0,0)\:,\qquad V^2=(0,1,0)\:.
\end{equation}
The above three vectors encode all the information required for the localization computation, together with the following parametrization for
the supersymmetric Killing vector $\xi$
\begin{equation}
\label{xi_param}
	\xi=\frac1\beta\,\big(2\pi\ii\,\partial_{\phi_0}-\omegafuga\,\partial_{\phi_1}-\omegafugb\,\partial_{\phi_2}\big)\:.
\end{equation}
in order to match the conventions of  \cite{Cabo-Bizet:2018ehj}.
Notice that for the Euclidean non-extremal supersymmetric solutions, the supersymmetric Killing vector $\xi$ is different from the Killing vector $V^1$ that degenerates at the "horizon" $\D_1$. This ensures that the norm $g(\xi,\xi)=f^2$ is non-vanishing at the localization loci and $\eta$ is well-defined.  

The localization loci are the circles where the $\text{U}(1)^3$ has a $\text{U}(1)^2$ isotropy subgroup. For the Euclidean black holes there are two of them,
$L_1=\overline{\D_0}\cap\overline{\D_1}$ and $L_2=\overline{\D_1}\cap\overline{\D_2}$, corresponding respectively with the North and South
poles of the $S^3$ horizon $\D_1$. At each $L_a$ we choose adapted $2\pi$-periodic coordinates $\varphi_{L_a}^0$, $\varphi_{L_a}^1$, $\varphi_{L_a}^2$,
such that $\varphi_{L_a}^0$ parametrizes the circle $L_a$, while the pairs $(\varphi_{L_a}^0,\varphi_{L_a}^2)$ and $(\varphi_{L_a}^0,\varphi_{L_a}^1)$
are angular coordinates on $\D_{a-1}$ and $\D_a$ respectively.
This is easily done as follows:
\begin{equation}
\begin{aligned}
\label{L1_L2_coord}
	\,&L_1\::\quad\varphi_{L_1}^0=\phi_1\:,\quad\varphi_{L_1}^1=\phi_2\:,\quad\varphi_{L_1}^2=\phi_0\:,
		\quad(\text{North pole of }S^3\text{ horizon})\\[1mm]
	\,&L_2\::\quad\varphi_{L_2}^0=\phi_2\:,\quad\varphi_{L_2}^1=\phi_0\:,\quad\varphi_{L_2}^2=\phi_1\:.\,\quad(\text{South pole of }S^3\text{ horizon})
\end{aligned}
\end{equation}
Since our strategy to compute the on-shell action involves background subtraction, let us discuss the subtraction geometry, which is pure AdS$_5$.
In AdS$_5$ we have the analogue of $\D_0$ and $\D_2$, with $\partial_{\phi_2}$ and $\partial_{\phi_1}$ vanishing of each respectively,
but other then these AdS does not have any other locus with $\text{U}(1)$ isotropy subgroup of the torus isometry action.
AdS has only one localization locus, the circle $L_0=\overline{\D_0}^{\text{AdS}}\cap\overline{\D_2}^{\text{AdS}}$
corresponding to the center of AdS.
The $2\pi$-periodic coordinates adapted to $L_0$ are the following
\begin{equation}
\label{L0_coord}
	L_0\::\quad\varphi_{L_0}^0=-\,\phi_0\:,\quad\varphi_{L_0}^1=\phi_1\:,\quad\varphi_{L_0}^2=\phi_2\:,\quad(\text{AdS center})
\end{equation}
where the minus sign in $\varphi_{L_0}^0$ is introduced in order for the ordering $(\varphi_{L_a}^0,r_{L_a}^1,\varphi_{L_a}^1,r_{L_a}^2,\varphi_{L_a}^2)$
to always be compatible with the overall convention for the orientation, with $r_{L_a}^1$ and $r_{L_a}^2$ being radial coordinates becoming zero
at $\D_{a-1}$ and $\D_a$ respectively.
By convention the index $a$ loops, so when $a=0$ we have $a-1\equiv2$.
We observe that the coordinates \eqref{L1_L2_coord}, \eqref{L0_coord} can also be determined from the general formula \eqref{phiLa} by setting
$\mathfrak s_1=\mathfrak s_2=1=-\,\mathfrak s_0$ and $(w^a)_i=\delta_{a,i}$.

Let us now discuss the gauge choice that makes the one-form $\eta$ regular everywhere.
As explained in section \ref{sect:reg gauge}, we need to find three constants $\nu_i$ such that $V^a_i\,\nu_i=1$ for $a=0,1,2$,
with $V^a_i$ given by \eqref{BH_vectors}.
This is easily solved by setting $\nu_0=\nu_1=\nu_2=1$ so that $\eta$ is given by
\begin{equation}
	\eta=-\frac{\ell}3\,(P+\nu_i\,\dd\phi_i)=-\frac{\ell}3\,\big(P+\dd\phi_0+\dd\phi_1+\dd\phi_2\big)\:.
\end{equation}
In AdS$_5$ the same exact gauge choice is picked. The contraction $\iota_\xi\eta$, both in the case of the black hole and AdS$_5$, is given by
\begin{equation}
	c_\eta = \iota_\xi\eta=-\frac{\ell}{3\beta}\,\big(2\pi\ii-\omegafuga-\omegafugb\big)\:.
\end{equation}

We apply the localization formula \eqref{localization_formula} to the following integral:
\begin{equation}
\label{BH_integral}
	-\int\eta\wedge(\dd\eta)^2=-\int\eta\wedge\Phi\:,\qquad\Phi=(\dd_X\eta)^2\:,
\end{equation}
where the integration manifold is the Euclidean black hole in one case and Euclidean AdS in the other.
The polyform $\Phi$ has the same shape as \eqref{Phi_ddeta}, so we can borrow equations \eqref{inteta} and \eqref{Phi0_expression}:
\begin{equation}
	\int_{L_a}\eta\,
		=\,2\pi\,\frac{\iota_\xi\eta}{\xiLa^0}\:,\qquad
		\Phi_0\,\big|_{L_a}=\bigg(\iota_\xi\eta\,\frac{\XLa^0}{\xiLa^0}\,\bigg)^2\:.
\end{equation}
Then the on-shell action of the Euclidean black holes regularized with background subtraction is then given by formula \eqref{hat_I_fmla}, which reads
\begin{equation}
	\widehat I=\frac1{16\pi\ii\,G_5}\Big(\cI_1[X]+\cI_2[X]-\cI_0[X]\Big)\:,
\end{equation}
where $\cI_a[X]$ denotes the contribution to the integral \eqref{BH_integral} coming from the localization locus $L_a$,
with respect to an arbitrary choice of Killing vector $X$.
The above formula uses that all the boundary contributions to the on-shell action cancel, which we will prove in section \ref{sect:boundary}.
The localization formula \eqref{localization_formula} gives the following expression for $\cI_a[X]$,
\begin{equation}
	\cI_a[X]=-(2\pi)^2\frac{\Phi_0\big|_{L_a}\cdot\int_{L_a}\eta}
		{\,{\displaystyle\prod}_{\,i=1,2}\!\left(\frac{\XLa^0}{\xiLa^0}\:\xiLa^i-\XLa^i\right)}=
		\Big(\frac{2\pi\ell}{3\beta}\Big)^3\frac{\big(2\pi\ii-\omegafuga-\omegafugb\big)^3\,(\XLa^0)^2}
			{\xiLa^0\big(\XLa^0\,\xiLa^1-\XLa^1\,\xiLa^0\big)\big(\XLa^0\,\xiLa^2-\XLa^2\,\xiLa^0\big)}\:,
\end{equation}
where $\xiLa^i$ and $\XLa^i$ are the components of $\xi$ and $X$ in the adapted coordinates \eqref{L1_L2_coord}, \eqref{L0_coord},
explicitly: $\xi=\xiLa^i\partial_{\phi_{L_a}^i}$ and $X=\XLa^i\partial_{\phi_{L_a}^i}$.
The final expression for the renormalized on-shell action of the Euclidean black hole that we obtain is $X$-independent:
\begin{equation}\label{IBH}
	\widehat I=\frac{\pi\,\ell^3}{108\,G_5}\,\frac{\big(\omegafuga+\omegafugb-2\pi\ii\big)^3}{\omegafuga\omegafugb}
\end{equation}
This expression perfectly matches the result of the  explicit computation of the action given in \cite{Cabo-Bizet:2018ehj}.

Let us notice that the fan for the complexified non-extremal Euclidean black hole \eqref{BH_vectors} is the same as the fan \eqref{fanS5} of the compact manifold $S^5$.
This analogy extends further since the on-shell action \eqref{IBH} is proportional to the Sasakian volume of $S^5$ 
\begin{equation}
	\widehat I=\frac{\ell^3 \big(\xi_1+\xi_2+\xi_3)^3}{54\pi\ii\,G_5}\:\text{Vol}_{S^5}(\xi)\:,
\end{equation} 
as one can see by comparing with \eqref{volS5}. The formal reason for this analogy will be explained in the next section. It is also intriguing to observe that the toric structure of the both the Euclidean black hole and the subtraction manifold can be obtained by intersecting the polytope of $\mathbb{C}^3=C(S^5)$  with a plane. When the normal to the plane lies inside the first octant we obtain the compact two-dimensional polytope of $S^5$ as in \cref{fig:S5}. When $\xi$ is continued outside the octant we obtain non-compact two-dimensional polytopes corresponding to our geometries as shown in \cref{fig:CBSmAdS}.

\begin{figure}[h]
\centering
\begin{tikzpicture}[tdplot_main_coords,scale=1]
  \def\axlength{5.5}
  \def\trianglepos{2.5}
  \def\yaxiscoeff{3.5/3}
  \def\lengthA{3}
  \def\lengthB{2}
  \def\unit{1}
  
  \draw[->,thick] (0,0,0) -- (\axlength,0,0) node[anchor=north east] {$y_0$};
  
  \draw[->,thick] (0,0,0) -- (0,{\yaxiscoeff*\axlength},0) node[anchor=north west] {$y_2$};
  
  \draw[->,thick] (0,0,0) -- (0,0,\axlength) node[anchor=east] {$y_1$};
  
  \coordinate (B) at (0,{\yaxiscoeff*\trianglepos},0);
  \coordinate (C) at ({\trianglepos},0,0);
  \coordinate (A) at (0,{\yaxiscoeff*\axlength},2.5);
  \coordinate (D) at ({\axlength},0,2.5);

  \draw[thick] (A) -- (B) node[midway, left] {$\mathcal{D}_0$};
  \draw[decorate, decoration = snake] (B) -- (C) node[midway, left] {$\mathcal{D}_1$};
  \draw[thick] (C) -- (D) node[midway, right, yshift=-3] {$\mathcal{D}_2$};
  \fill[blue!25, opacity=0.7] (A) -- (B) -- (C) -- (D) -- cycle;

  \coordinate (B0) at (0,0,{\trianglepos+1.5});
  \coordinate (A0) at ($(B0) + (A) - (B)$);
  \coordinate (D0) at ($(B0) + (D) - (C)$);

  \draw[thick] (A0) -- (B0) node[midway, above, yshift=3, xshift=-7] {$\mathcal{D}_0^{\text{AdS}}$} -- (D0) node[midway, right, xshift=-2, yshift=-4] {$\mathcal{D}_2^{\text{AdS}}$};
  \fill[blue!25, opacity=0.7] (A0) -- (B0) -- (D0) -- cycle;

  \node at (B) [left, xshift=2, yshift=4] {$L_1$};
  \node at (C) [below] {$L_2$};
  \node at (B0) [left] {$L_0$};
\end{tikzpicture}
\caption{The two-dimensional polytopes of the Euclidean black hole and of AdS$_5$ are obtained by cutting the first octant, the polytope of $C(S^5)$, with the plane $\xi_i y_i=\frac 12$ where $\xi$ lies outside the octant itself. Different positions of the plane give different topologies. In the figure we are assuming that the components of $\xi$ are real. To obtain the complex geometry of the non-extremal Euclidean black holes an analytic continuation is needed.}
\label{fig:CBSmAdS}
\end{figure}
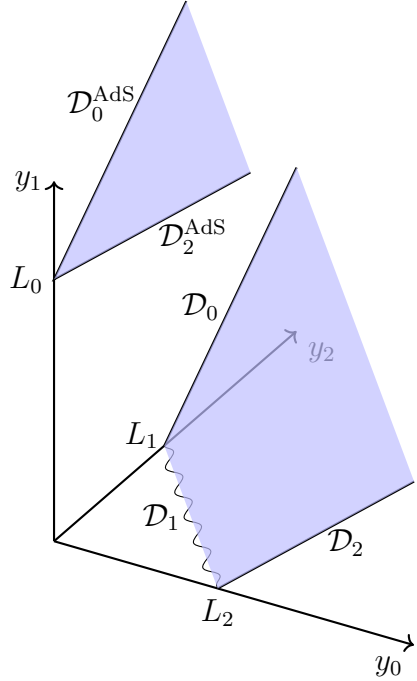

\subsection{A general class of  toric geometries}
\label{sect:bulk localization}

In this section we will consider a general Euclidean solution to five-dimensional gauged supergravity asymptotic to conformally flat $S^1\times S^3$,
with $\text{U}(1)^3$ isometry and a single everywhere regular gauge patch. We compute the on-shell action with our localization procedure,
by taking as a subtraction manifold pure AdS$_5$.

As in the previous section where we considered the example of Euclidean black holes, we work with $2\pi$-periodic coordinates $\phi_0$, $\phi_1$, $\phi_2$
such that $\phi_0$ parametrizes the $S^1$ at the asymptotic boundary, while $\phi_1$, $\phi_2$ parametrize the circles at the North and South poles
of the asymptotic $S^3$ respectively.

Let $\D_0,\D_1,\ldots,\D_d$ be the three-dimensional submanifolds where the $\text{U}(1)^3$  isometry has a $\text{U}(1)$ isotropy subgroup,
ordered so that  $\overline{\D_a}\cap\overline{\D_b}$ is non-empty only if $a$, $b$ are consecutive.
Then $L_a=\overline{\D_{a-1}}\cap\overline{\D_a}$ are the circles with $\text{U}(1)^2$ isotropy, and play the roles of localization loci.
All the $\overline{\D_a}$ are compact except for $\overline{\D_0}$, $\overline{\D_d}$, which intersect the boundary at $S^1\times\{\text{N/S pole of }S^3\}$.
Defining the $V^a$ to be the Killing vector fields that vanish at each $\D_a$, we must have $V^0=\partial_{\phi_2}$, $V^d=\partial_{\phi_1}$.
Writing the $V^a=V^a_i\partial_{\phi_i}$, we have (with some abuse of notation)
\begin{equation}
\label{fan_general}
	V^0=(0,0,1)\:,\quad\ldots\quad,\:V^d=(0,1,0)\:.
\end{equation}
The above fan encodes all the relevant information for localizing, together with a parametrization of $\xi$, which we take to be the same as \eqref{xi_param}:
\begin{equation}
	\xi=\frac1\beta\,\big(2\pi\ii\,\partial_{\phi_0}-\omegafuga\,\partial_{\phi_1}-\omegafugb\,\partial_{\phi_2}\big)\:.
\end{equation}
For the subtraction manifold AdS$_5$ the same $\xi$ as above is taken. The fan of AdS only contains $V^0$ and $V^d$, and we denote the single
localization locus in AdS as $L_0$, corresponding to the AdS center.

The condition for the existence of a single everywhere defined gauge patch is equivalent to demanding that there exist constants $\nu_i$
such that $\nu_i\,V^a_i=1$ for all $a$.
Then $\eta=-\frac{\ell}3(P+\nu_i\,\dd\phi_i)$ is regular everywhere and
\begin{equation}
	c_\eta=\iota_\xi\eta=-\frac\ell3\,\nu_i\,\xi_i\:,\qquad\text{with }\nu_i\text{ satisfying}\quad\nu_i\,V^a_i=1\:.
\end{equation}
We can then compute the integral
\begin{equation}
	-\int\eta\wedge(\dd\eta)^2=-\int\eta\wedge\Phi\:,\qquad\Phi=(\dd_X\eta)^2\:,
\end{equation}
where the integration manifold is the solution of interest in one case, and Euclidean AdS$_5$ in the other (for AdS the same gauge for $\eta$ is chosen).
Since $\Phi$ is the same as \eqref{Phi_ddeta}, so we can use \eqref{inteta} and \eqref{Phi0_expression},
\begin{equation}
	\int_{L_a}\eta\,=-\,\sa\,\frac{2\pi\,(\frac\ell3\,\nu_i\,\xi_i)}{(\xi,V^{a-1},V^a)}\:,\qquad
		\Phi_0\,\big|_{L_a}=\big(\tfrac\ell3\,\nu_i\,\xi_i\big)^2\,\frac{(X,V^{a-1},V^a)^2}{(\xi,V^{a-1},V^a)^2}\:,
\end{equation}
where $(\xi,V^{a-1},V^a)$ is a shorthand notation for the determinant of the components of $\xi$, $V^{a-1}$, $V^a$ in the $\partial_{\phi_i}$ basis,
whereas the signs $\sa$ are all $\sa=+1$ for $a\ne0$, and $\mathfrak s_0=-1$, in order to account properly for the orientation at the AdS$_5$ center,
with the same logic as in section \ref{sect:black hole localization}.
When $a=0$ we are using the convention that $a-1\equiv d$.

The on-shell action renormalized with background subtraction is given by \eqref{hat_I_fmla},
\begin{equation}
\label{hat_I_general}
	\widehat I=\frac1{16\pi\ii\,G_5}\bigg(\sum_{a=1}^d\cI_a[X]-\cI_0[X]\bigg)\:,
\end{equation}
where $X$ is an arbitrary Killing vector and the localized contributions at each $a=0,\ldots,d$ are given by
\begin{align}
\label{Ia_general}
	\cI_a[X]=-(2\pi)^2\frac{\Phi_0\big|_{L_a}\cdot\int_{L_a}\eta}
		{\,{\displaystyle\prod}_{\,i=1,2}\!\left(\frac{\XLa^0}{\xiLa^0}\:\xiLa^i-\XLa^i\right)}=
		 \frac{\sa\,\big(2\pi\frac\ell3\,\nu_i\,\xi_i\big)^3\,(X,V^{a-1},V^a)^2}{(\xi,V^{a-1},V^a) \, (\xi,X,V^a)\, (\xi,V^{a-1},X)}\:.
\end{align}
Notice that we have used the expression \eqref{weights} for the weights of the localization formula \eqref{localization_formula}.
The formula \eqref{hat_I_general} for $\widehat I$ is valid because all the boundary contributions cancel each other, as we will show in section
\ref{sect:boundary}.
The orientation sign $\sa$ is $-1$ only for $a=0$, and thus cancels the subtraction sign in \eqref{hat_I_general}.
The expression that we obtain for $\widehat I$ is then proportional to \eqref{sasvolX},%
\footnote{With the index $a$ running from 0 to $d$ instead of starting form 1.}
and we can use similar manipulations to show that it is independent of the choice of $X$, giving us
\begin{empheq}[box={\mymath[colback=white, colframe=black]}]{equation}
\label{CY_general_result}
	\widehat I=\frac{\big(\frac{\ell}3\,\nu_i\,\xi_i\big)^3}{2\pi\ii\,G_5}\:\text{Vol}_Y(\xi)\:,\qquad
		\text{Vol}_Y(\xi)=\frac{\pi^3}{\nu_i\,\xi_i}
		\sum_{a=0}^d \frac{(V^{a-1},V^a,V^{a+1})}{(\xi,V^{a-1},V^a)(\xi,V^{a},V^{a+1})}\:.
\end{empheq}
Remarkably, the result is expressed in terms of the analytic continuation%
\footnote{For a Sasakian manifold $Y$ the Reeb $\xi$ should be real valued and its components should belong to the cone dual to its polytope,
		but here we are considering complex valued $\xi$.}
 of the Sasakian volume \eqref{eq:vol5II} $\text{Vol}_Y(\xi)$ of a compact geometry $Y$
obtained form a six-dimensional K\"ahler cone whose fan is the same as \eqref{fan_general}.

Let us be thorough and discuss the steps needed to go from \eqref{Ia_general} to \eqref{CY_general_result},
since there are a few differences from section \ref{The Sasakian volume},
and some of the observations that we make are also preparatory for section \ref{sect:boundary}.
Using the determinant identity \eqref{determinant_c1_vs_eta} and the relations%
\footnote{The first identity can be obtained by trivially rearranging the terms in the second sum. In order to prove the second identity,
		one can define $\cA_{a,b}=(\xi,V^{a},V^b)/[(\xi,X,V^b)(\xi,V^a,X)]$ and notice that identity \eqref{determinant_product_formula}
		implies that $\cA_{a,b}+\cA_{b,c}=\cA_{a,c}$. By iteration, $\cA_{d,0}+\cA_{0,1}+\cA_{1,2}+\ldots+\cA_{d-1,d}=\cA_{d,d}=0$,
		which proves the identity.}
\begin{equation}
\label{equivalence_of_equivariantizations}
	\sum_{a=0}^d\frac1{(\xi,X,V^a)}+\sum_{a=0}^d\frac1{(\xi,V^{a-1},X)}\,=\,0\,=\,\sum_{a=0}^d\frac{(\xi,V^{a-1},V^a)}{(\xi,X,V^a)(\xi,V^{a-1},V)}\:,
\end{equation}
we can write the renormalized on-shell action \eqref{hat_I_general} as
\begin{align}
	\,&\widehat I=\frac{(2\pi\frac\ell3)^3}{16\pi\ii\,G}\,(\nu_i\,\xi_i)\sum_{a=0}^d\big(\cB_{a-1,\,a}-\cB_{a,\,a-1}-2\,\cC_a\big)\:,\\
	\,&\cB_{a,b}\equiv\frac{(\xi,X,V^a)}{(\xi,V^a,V^b)(\xi,X,V^b)}\:,\qquad\cC_a\equiv\frac1{(\xi,V^{a-1},V^a)}\:.
\end{align}
Since the index $a$ is cyclic mod $d+1$, we can rearrange the sum as
\begin{equation}
	\sum_{a=0}^d\big(\cB_{a-1,\,a}-\cB_{a,\,a-1}-2\,\cC_a\big)=\sum_{a=0}^d\big(\cB_{a-1,\,a}-\cB_{a+1,\,a}-\cC_a-\cC_{a+1}\big)\:,
\end{equation}
\begin{equation}
	\cB_{a-1,\,a}-\cB_{a+1,\,a}-\cC_a-\cC_{a+1}=\frac{(\xi,V^{a-1},V^{a+1})}{(\xi,V^{a-1},V^a)(\xi,V^a,V^{a+1})}-\cC_a-\cC_{a+1}=
		\frac{-\:(\nu_i\,\xi_i)\:(V^{a-1},V^a,V^{a+1})}{(\xi,V^{a-1},V^a)(\xi,V^a,V^{a+1})},
\end{equation}
where we have used \eqref{determinant_product_formula} in the first step and \eqref{determinant_c1_vs_eta}
with $X$ replaced by $V^{a+1}$ in the second step. This concludes the algebraic derivation of \eqref{CY_general_result}.

Let us observe that the determinant identity \eqref{determinant_c1_vs_eta} is equivalent to
\begin{align}
\label{different_equivariantization}
	(\dd_X \eta)\,\big|_{L_a}=-\frac\ell3\,2\pi\sum_{b=0}^dc_1^\bT(\cL_b)\,\big|_{L_a}+\frac\ell3\,\nu_i\,X_i\:,
\end{align}
where the restrictions of the Chern classes are given by \eqref{weights_VS_Chern}.
Notice that this differs from \eqref{detac1} just by numerical factors, due to the different value of
$\iota_\xi\eta$ compared to the setup of section \ref{The Sasakian volume}.
Considering that $\dd\eta=-\frac\ell3\cR$ and $[\cR]=2\pi\sum_a[c_1(\cL_a)]$, then relation \eqref{different_equivariantization}
is expected, since two different ways to extend $\dd\eta$ into an equivariant polyform by giving it a zero-form component can only differ by a constant.
Since the polyforms
\begin{equation}
	\Phi=(\dd_X\eta)^2\:,\qquad\Phi'=\frac{\ell^2}{9}(\dd_XP)^2\, , 
\end{equation}
are both equivariantly closed and have the same top degree component, either of them can be used to compute the integral of the Chern-Simons form
$\eta\wedge(\dd\eta)^2$. Generically, changing the polyform has an impact on the value of the localized contributions $\cI_a$ and the boundary integrals
taken separately.
But when we substitute the analogue of \eqref{different_equivariantization} into \eqref{Ia_general}, the constant shift $\frac\ell3\,\nu_i\,X_i$
vanishes completely due to \eqref{equivalence_of_equivariantizations}, meaning that in our subtraction scheme there is no difference between using
either $\Phi$ or $\Phi'$ when it comes to the overall bulk contributions to $\widehat I$.
In particular, when in section \cref{sect:boundary} we show that all the boundary contributions cancel each other, we are free to use either $\Phi$ or $\Phi'$ as we please,
and the latter is more convenient. When using $\Phi'$ the boundary contributions can be written as
\begin{equation}
\label{beta_expression}
	-\int_M\eta\wedge\Phi'=(\text{bulk contributions})+\frac{\ell^2}9\int_M\eta\wedge\betaX\wedge\big(2P(X)\cR-P(X)^2\dd\betaX\big)\:,
\end{equation}
with $\betaX$ defined by \eqref{betaX_def}, up to a suitable choice of $\alphaX$. This is the expression that we will use in section \ref{sect:boundary}.

\subsection{Vanishing of the boundary term}
\label{sect:boundary}

In order to complete the localization computation, we will now show that the overall boundary contributions in our scheme are vanishing,
as already anticipated in \eqref{bdy_cancellation}.
For simplicity we will perform this computation in Lorentzian signature (it works out just the same in Euclidean)
and set the $\text{AdS}_5$ scale to $\ell = 1$ throughout this section.

The boundary integrals take the form
\begin{equation}
	S_{\text{bdy}}[\partial M]=\frac1{16 \pi G_5}\int_{\partial M}\Big(\alpha(M)+\beta(M;X)\Big)\:,
\end{equation}
where the integrands can be read from \eqref{alpha_expression} and \eqref{beta_expression},
\begin{align}
  \begin{aligned}
\label{bdy_integrands}
  \alpha(M) ={}& \eval{\qty[f \eta_0 \wedge f^{-2} \star_\gamma \dd[]{f} + 2 f \eta_0 \wedge \eta \wedge \dd[]{\eta} + \eta \wedge f \eta_0 \wedge \dd[]{} (f\eta_0)]}_{M} \,,  \\
  \beta(M;X) ={}& \frac19\eval{\qty[\,\eta\wedge\betaX\wedge\big(2P(X)\cR-P(X)^2\dd\betaX\big)]}_{M} \,,
  \end{aligned}
\end{align}
with the following $\betaX$:
\begin{align}
  \betaX=\frac{Y^\flat}{\iota_XY^\flat}=\frac{Y^\mu g_{\mu\nu}\dd x^\nu}{g(X,Y)} \,, \qquad Y = g(\xi,\xi) X - g(\xi, X) \xi \,.
\end{align}
Here $g$ is the physical metric: we will see that this choice of $\betaX$ ensures overall boundary cancellation already at the level of the overall integrand.

First, let us introduce some local geometrical properties in the vicinity of the boundary. We assume a generic asymptotically $\text{AdS}$ supersymmetric solution to 5d minimal supergravity, which possesses a $\text{U}\qty(1)^3$ isometry. In Fefferman-Graham coordinates $(t,\rho,\psi,z, \bar{z})$ the boundary is located at $\rho = 0$ and the three $\text{U}\qty(1)$'s are generated by $(\partial_t, \partial_\psi, \partial_\varphi)$, with $z = \abs{z} \eu^{-\iu \varphi}$.
The field configuration assumes the following expansion in the vicinity of the boundary
\begin{align}
  \begin{aligned}
    \dd[]{s}_g^2 ={}& \frac{\dd[]{\rho}^2}{\rho^2} + \frac{1}{\rho^2} \qty(h^{(0)}_{ij} + h^{(2)}_{ij} \rho^2 + h^{(4)}_{ij} \rho^4 + \dots ) \dd[]{x^i} \dd[]{x^j} \,, \\
    A ={}& \qty(A^{(0)}_i + A^{(2)}_i \rho^2 + \dots) \dd[]{x^i} \, , 
  \end{aligned}
\end{align}
where $x^i = (t,\psi,z, \bar{z})$ label the coordinates on a hypersurface of constant $\rho$. In \cref{app:FG expansions} we show how to impose the supersymmetry conditions on the solution order by order in the Fefferman-Graham expansion. As a result of this analysis the conformally flat boundary is specified by
\begin{align}\label{eq:conformal_boundary}
  \begin{aligned}
    h_{ij}^{(0)} \dd[]{x^i} \dd[]{x^j} ={}& - \dd[]{t}^2 + (\dd[]{\psi} + a)^2 + 4 \eu^{w} \dd[]{z} \dd[]{\bar{z}} \,, \\
    A^{(0)}_i \dd[]{x^i} ={}& - \frac{\sqrt{3}}{\curly{x}} \qty[ - \frac{1}{6} u \qty(\dd[]{\psi} + a) - \frac{1}{6} \iu \qty(\dd[]{\bar{z}} \partial_{\bar{z}} - \dd[]{z} \partial_z) w + \qty(\frac{u}{12} - f_0) \dd[]{t} + \dd[]{\lambda}] \,, 
  \end{aligned}
\end{align}
where conformal flatness demands
\begin{align}\label{eq:u_and_w}
  u ={}& \text{const} \,, \quad \eu^{w/2} = \frac{2}{u(1 + z \bar{z})} \,,
\end{align}
and the one-form $a = a_z \dd[]{z} + \overline{a_{z}} \dd[]{\bar{z}}$ satisfies
\begin{align}\label{eq:gauge_choice_a}
  \dd[]{a} ={}& \iu \, u \, \eu^w \dd[]{z} \wedge \dd[]{\bar{z}} \,. 
\end{align}
Overall the boundary data is specified by
\begin{align}
  u \,, \quad a_z(z, \bar{z}) \,, \quad\dd\Lambda\equiv f_0\,\dd y+\dd\lambda(t + \psi, z, \bar{z}) \, ,  
\end{align}
where $\dd\Lambda$ is the gauge shift that we fixed in section \ref{sect:reg gauge}, which we have now rewritten in terms of 
a constant $f_0$ and a basic function $\lambda$. The arguments of $\lambda$ appear as such, since the procedure outlined in \cref{app:FG expansions} involves the coordinate transformation \eqref{eq:toFG1}. Independently of supersymmetry, the first sub-leading order in the Fefferman-Graham expansion of the metric is fully determined by the leading one \cite{deHaro:2000vlm}
\begin{align}\label{eq:second_order_induced_metric}
  h_{ij}^{(2)} \dd[]{x^i} \dd[]{x^j} ={}& \frac{1}{12} \qty(R_{h^{(0)}} h_{ij}^{(0)} - 6 \qty(R_{h^{(0)}})_{ij}) \dd[]{x^i} \dd[]{x^j} \notag \\
  ={}& \frac{8 \Box w + u^2}{96} \dd[]{t}^2 - \frac{8 \Box w + 7 u^2}{96} \qty(\dd[]{\psi} + a)^2 + \frac{16 \Box w + 5 u^2}{24} e^{w} \dd[]{z} \dd[]{\bar{z}} \,,
\end{align}
where we have introduced the Laplacian $\Box = e^{-w} \partial_z \partial_{\bar{z}}$ of the 2d boundary metric $\dd[]{s}^2_{\text{2d}} = 4 \eu^{w} \dd[]{z} \dd[]{\bar{z}}$. The sub-sub-leading order is not determined by the boundary data, but is constrained by supersymmetry and involves the four non-boundary functions $k_I(z, \bar{z})$, $I = 1,2,3,4$, introduced in \cref{app:FG expansions}. In the supplementary \texttt{Mathematica} notebook we store the full
expressions for $\qty(h^{(0)}_{ij}, h^{(2)}_{ij}, h^{(4)}_{ij}, A^{(0)}_i, A^{(2)}_i)$ in the variable \texttt{FGexpansion}. Note that since the conformal boundary is flat the field strength vanishes at leading order
\begin{align}
  F ={}& \dd[]{A} = \order{\rho} \,,
\end{align}
but starts deviating from zero already at $\order{\rho}$ through the appearance of the non-boundary functions $k_I(z, \bar{z})$.

We are now ready to construct the integrands $\alpha(M)$ and $\beta(M;X)$ given by \eqref{bdy_integrands}
 in the coordinate system $(y,r, \hat{\psi}, z, \bar{z})$ discussed in \cref{app:FG expansions}, in which
\begin{align}
  \begin{aligned}
  f ={}& \frac{12 r^2 U^2 W^2}{\qty[\partial_z \partial_{\bar{z}} \log{W} + \partial_r \qty(rW \partial_r \qty(r^3 W)) + W \partial_r \qty(r^3 W)]} \,, \\
  P ={}& - \frac{1}{U^2 W} \partial_r \qty(r^3 W) (\dd[]{\hat{\psi}} + \phi) - \iu \qty(\dd[]{\bar{z}} \partial_{\bar{z}} - \dd[]{z} \partial_z) \log{W} \,,  \\
  \omega ={}& c (\dd[]{\hat{\psi}} + \phi) + C_z \dd[]{z} + \overline{C_z} \dd[]{\bar{z}} \,, \\
  \phi ={}& \phi_z \dd[]{z} + \overline{\phi_z} \dd[]{\bar{z}} \,. 
  \end{aligned}
\end{align}
$\star_\gamma \dd[]{f}$ is calculated with respect to the metric on the base \eqref{eq:base_metric} and  
\begin{align}
  \eta_0 ={}& \dd[]{y} + \omega \,, \qquad \eta = - \frac{1}{3}P + f_0 \dd[]{y} - \dd[]{\lambda} \,. 
\end{align}
In this coordinate system the supersymmetric Killing vector is simply $\xi = \partial_y$ and the vector $X$ is taken as an arbitrary linear combination of the three $\text{U}\qty(1)$'s
\begin{align}
  X ={}& X^0 \, \partial_y + X^1 \, \partial_{\hat{\psi}} + X^2 \, \partial_{\varphi} = X^0 \, \partial_y + X^1 \, \partial_{\hat{\psi}} + X^2 \, \iu \qty(\bar{z} \partial_{\bar{z}} - z \partial_z) \,,
\end{align}
with $X^{0,1,2}$ constants. The resulting form $\alpha(M)$ depends on the functions $(U,W,\phi_z,c,C_z)$ and $\beta(M;X)$ additionally depends on the arbitrary vector $X$. Then we implement the asymptotic expansion \eqref{eq:large_r_expansion} followed by the coordinate transformations \eqref{eq:toFG1} and \eqref{eq:toFG2} to write down these forms in the Fefferman-Graham coordinates $(t,\rho,\psi, z, \bar{z})$ discussed above. Schematically, at the hypersurface $\rho = \epsilon$ we obtain
\begin{align}
  \begin{aligned}
    \alpha(M) ={}& \alpha_{-4}(u, f_0) \, \epsilon^{-4} + \alpha_{-2}[u, f_0, \lambda; k_I] \, \epsilon^{-2} + \alpha_{0}[u, f_0, \lambda; k_I] \,, \\
    \beta(M;X) ={}& \beta_{-4}(u, f_0) \, \epsilon^{-4} + \beta_{-2}[u, a_z, f_0, \lambda; k_I; X] \, \epsilon^{-2} + \beta_{0}[u, a_z, f_0, \lambda; k_I; X] \,, \\
  \end{aligned}
\end{align}
where the square brackets indicate that we have dependence both on the functions $a_z(z, \bar{z})$, $\lambda(t + \psi, z, \bar{z})$, $k_i(z, \bar{z})$ and on their derivatives. The final expressions for $\alpha(M)$ and $\beta(M;X)$ are stored in the supplementary \texttt{Mathematica} notebook under the variable \texttt{inducedFourForms}.
Spectacularly, when we consider
\begin{align}
  \alpha(M) + \beta(M;X) ={}& \qty(\alpha + \beta)_{-4}(u) \, \epsilon^{-4} \notag \\
                            &+ \qty(\alpha + \beta)_{-2}[u, a_z, f_0, \lambda; X] \, \epsilon^{-2} \notag \\
  & + \qty(\alpha + \beta)_{0}[u, a_z, f_0, \lambda; X]
\end{align}
the dependence on the non-boundary functions $k_I(z, \bar{z})$ drops out! In particular, this means that as long as the subtraction manifold $N$ has the same conformal boundary as the original manifold in question $M$ and we use the same vector $X$ when constructing the forms $\beta(M;X)$ and $\beta(N;X)$, we have
\begin{align}
  \alpha(M) + \beta(M;X) - \alpha(N) - \beta(N;X) ={}& 0 \,,
\end{align}
thus $S_{\text{bdy}}[\partial M]-S_{\text{bdy}}[\partial N]$ vanishes on the level of the overall integrand, as anticipated in \eqref{bdy_cancellation}.

We saw that to achieve the boundary cancellation $S_{\text{bdy}}[\partial M]-S_{\text{bdy}}[\partial N]=0$, we only need to specify the conformal boundary of the subtraction manifold $N$, with the first sub-leading term in the Fefferman-Graham expansion descending from it, as in \eqref{eq:second_order_induced_metric}. In general, to completely fix the subtraction manifold would require some unclear to us global physical principle: it might be the case that which $N$ is chosen depends on the observable of interest. For asymptotically $\text{AdS}$ solutions the established background subtraction procedure used in the literature to obtain a finite on-shell action has $N \equiv \text{empty} \text{ AdS}_5$. For completeness, we show how to write empty $\text{AdS}_5$ in the coordinates discussed above.
In canonical global coordinates the unit empty $\text{AdS}_5$ metric can be written as
\begin{align}
  \begin{aligned}
  \dd[]{s}^2_{\text{AdS}_5} ={}& \frac{\dd[]{\rho}^2}{\rho^2} - \qty(\frac{1}{\rho} + \frac{\rho}{4 r_3^2})^2 \dd[]{t}^2 + \qty(\frac{1}{\rho} - \frac{\rho}{4 r_3^2})^2 \dd[]{s}^2_{S^3} \,, \\
  \dd[]{s}^2_{S^3} ={}& \frac{r_3^2}{4} \qty[\qty(\dd[]{\tilde{\psi}} + \cos^{}{\theta} \dd[]{\varphi})^2 + \dd[]{\theta}^2 + \sin^{2}{\theta} \dd[]{\varphi}^2] \,,
  \end{aligned}
\end{align}
where $r_3$ is the radius of the $S^3$. This coordinate system is related to our Fefferman-Graham coordinates $(t,\rho,\psi,z, \bar{z})$ as
\begin{align}
  \tilde{\psi} ={}& - \frac{2}{r_3} \qty(\psi + b(z, \bar{z})) \,, \quad \theta = 2 \cot^{-1}\qty(\sqrt{z \bar{z}}) \,, \quad \varphi = \tan^{-1}{\qty(\frac{\bar{z} - z}{\iu(\bar{z} + z)})} \,. 
\end{align}
Choosing $r_3$ and $b(z, \bar{z})$ such that  
\begin{align}
  r_3 ={}& \frac{4}{u} \,, \quad \partial_z b = a_z - \frac{\iu}{u z} \frac{1 - z \bar{z}}{1 + z \bar{z}} \,, 
\end{align}
we find the Fefferman-Graham form of empty $\text{AdS}_5$
\begin{align}
  \dd[]{s}_{\text{AdS}_5}^2 ={}& \frac{\dd[]{\rho}^2}{\rho^2} + \frac{1}{\rho^2} \qty(\tilde{h}^{(0)}_{ij} + \tilde{h}^{(2)}_{ij} \rho^2 + \tilde{h}^{(4)}_{ij} \rho^2) \dd[]{x^i} \dd[]{x^j} \,, 
\end{align}
with
\begin{align}
  \begin{aligned}
  \tilde{h}_{ij}^{(0)} \dd[]{x^i} \dd[]{x^j} ={}& - \dd[]{t}^2 + (\dd[]{\psi} + a)^2 + 4 \eu^{w} \dd[]{z} \dd[]{\bar{z}} \,, \\
  \tilde{h}_{ij}^{(2)} \dd[]{x^i} \dd[]{x^j} ={}& \frac{8 \Box w + u^2}{96} \dd[]{t}^2 - \frac{8 \Box w + 7 u^2}{96} \qty(\dd[]{\psi} + a)^2 + \frac{16 \Box w + 5 u^2}{24} e^{w} \dd[]{z} \dd[]{\bar{z}} \,, \\
  \tilde{h}_{ij}^{(4)} \dd[]{x^i} \dd[]{x^j} ={}& \frac{u^2}{128}\qty(\frac{8 \Box w + u^2}{96} \dd[]{t}^2 + \frac{8 \Box w + 7 u^2}{96} \qty(\dd[]{\psi} + a)^2 - \frac{16 \Box w + 5 u^2}{24} e^{w} \dd[]{z} \dd[]{\bar{z}}) \,, 
  \end{aligned}
\end{align}
where $u$ and $w$ are given in \eqref{eq:u_and_w}. The first two expressions above agree with the leading \eqref{eq:conformal_boundary} and sub-leading \eqref{eq:second_order_induced_metric} induced metrics on $M$. The sub-sub-leading terms $h_{ij}^{(4)}$ and $\tilde{h}_{ij}^{(4)}$ start differing through the appearance of the non-boundary functions $k_I(z, \bar{z})$. Note that when empty $\text{AdS}_5$ is used as a subtraction manifold $F = 0$ throughout, but the gauge field $A$ has to be taken the same as the pure gauge part of \eqref{eq:conformal_boundary}.

\section{Discussion}
\label{sect:Discussion}

In this paper we discussed a powerful approach to equivariant localization in odd dimensions, elaborating on the mathematical results of \cite{Goertsches:2015vga}, and alternative to the Berline-Vergne-Atiyah-Bott  formula. The method applies to foliations and, as such, has many applications to holography where a natural foliation in supersymmetric solutions is provided by a Killing vector, constructed as bilinear in the Killing spinors. As a main application, we evaluated the regularised on-shell action of a large class of supersymmetric solutions of five dimensional minimal  gauged supergravity. Among other things, we derived for the first time from purely topological arguments the on-shell action of the asymptotically AdS$_5\times S^5$ supersymmetric black holes with equal electric charges considered in \cite{Chong:2005hr,Cabo-Bizet:2018ehj}. It is worth noticing that our result  not only reproduce the entropy function for the most general currently known black hole solution \cite{Chong:2005hr,Cabo-Bizet:2018ehj}, but it predicts that \emph{any} toric asymptically AdS$_5$  solution  with the same topology will have the same entropy (function).

We believe that the possible applications of our approach are vast and here we just scratched the tip of the iceberg. In this section we discuss some of the aspects that still need to be understood and generalized in the context of applications to supergravity and holography.

Perhaps the most important generalization that needs to be  addressed is  the inclusion of magnetic fluxes, that would be allowed by removing the Calabi-Yau condition that we assumed. We have seen that the requirement that the contact one-form $\eta$ is regular imposes some constraints on the topology of the supergravity gauge field. In particular, we can efficiently deal with solutions where there are no two-cycles or in which the gauge field has no magnetic flux along non trivial two-cycles (e.g. spindles). This excludes for example the case of  topological solitons, namely solutions whose toric fan lies on a plane orthogonal to Reeb vector, as the examples  discussed in appendix \ref{app:known}.

Except the case of the $\CBS$, the  general toric geometries that we discussed have a clear interpretation only in the Euclidean setting and it would be very interesting to see 
 whether  their on-shell action can be reproduced by new saddles of the superconformal index, in the large $N$ limit.  More generally, it would be important to understand whether there exist 
other geometries, in addition to the $\CBS$, that admit extremal limits and have a Lorentzian interpretation in terms of black lenses or other physical black objects.  It is possible that 
in order to describe these types of geometries we will need to incorporate magnetic fluxes. It would be fascinating to predict the entropy of such black objects, 
whose existence in gauged supergravity has remained up to now elusive,  purely on topological grounds.

We also expect that our approach   should lead to a substantial improvement  in the classification of general toric solutions of minimal gauged supergravity attempted by \cite{Lucietti:2022fqj} by including orbifolds and ``non-extremal horizons'' in the context of Euclidean complex solutions. 
From a mathematical point of view, it would be interesting to develop our results for proving existence or giving obstructions to the existence of K\"ahler metrics obeying the complicated 
equation (\ref{nastyPDE}).

Here we only considered supersymmetric solutions of minimal gauged supergravity with a time-like supersymmetric Killing vector. Bosonic supersymmetric solution of this theory were classified in \cite{Gauntlett:2003fk} and there exists a second class of solutions with a null Killing vector. It would be interesting to see how our results generalize to the null class. More generally, we started with the {\it Lorentzian} classification of supersymmetric solutions and we performed an analytic continuation to the Euclidean, with all the connected ambiguities. A more radical approach would be to start directly from Euclidean solutions and carry out the supersymmetry analysis from scratch, allowing for complex backgrounds.

Moreover, we only considered solutions with conformally flat asymptotic boundaries where we can easily apply the subtraction method to regularize the divergent action. We leave the extension to more general asymptotically locally AdS solutions, which would require either a generalized background subtraction or holographic renormalization, for future work. In particular, in our approach the subtraction manifold, which provides one (or, in general, multiple) additional loci, is instrumental to ensure the cancellation of the $X$-dependence. 
It would be interesting to see how this cancellation works within holographic renormalization.

Finally, in this paper we focussed on applications to minimal gauged supergravity in five dimensions. We expect that we can extend the results to gauged supergravity with vector and hypermultiplets and, also in that case, the on-shell action of supersymmetric solutions can be written as the integral of an equivariant form plus a boundary term.
This would allow one, for example, to derive the  on-shell action of the asymptotically AdS$_5\times M_5$ supersymmetric black holes with general charges from topological arguments only. More generally, our approach is not limited to five dimensions and it would be interesting to apply it to seven-dimensional supergravity as well as
to ten-dimensional backgrounds, including also the internal manifold, which is also often   
odd-dimensional and toric. This would allow one to make contact with the equivariant volume approach of \cite{Martelli:2023oqk} and with the work of \cite{Chen:2024erz} that obtained the master volume (of the internal SE space) from the giant graviton expansion of the index.

\section*{Acknowledgments}

We  acknowledge partial support by the INFN.  EC, VD and DM are partially  supported by a grant Trapezio (2023) of the Fondazione Compagnia di San Paolo. AZ is partially supported by the MIUR-PRIN grant No. 2022NY2MXY. We thank   the authors of \cite{Cassani:2024kjn} for helpful discussions, in particular during the Eurostrings 2024 conference.
EC and VD would like to thank the Galileo Galilei Institute for hospitality during the development of this work.

\appendix

\section{Toric geometry}
\label{app:toric}

In this section we review the geometry of   symplectic toric orbifolds following the general formalism developed in \cite{Guillemin1994KaehlerSO,abreu2000kahler,Abreu:2001to,Lerman:1995aaa}.
Let $M_{2k}$ be a $2k$-dimensional orbifold equipped with a symplectic form $\Kform$.
A $\text{U}(1)^k$ action on $M_{2k}$ is Hamiltonian if it is possible to define moment maps $\mu^i:M_{2k}\to\bR$,
which are smooth $\text{U}(1)^k$ invariant functions that satisfy
\begin{equation}
	\iota_\xi\Kform=-\,\xi_i\,\dd\mu^i\:.
\end{equation}
If a $\text{U}(1)^k$ Hamiltonian action exists, $(M_{2k},\Kform)$ is then said to be a toric orbifold.
If $M_{2k}$ is compact, its image under the moment maps $\mu^i$ is a compact convex polytope $\cP$ of the form
\begin{equation}
\label{polytope_explicit}
	\cP=\{\,y_i\in\bR^k\:|\:l_a(y)\,\geq0,\,a=1,\ldots,d\,\}\:,\qquad l_a(y)\equiv y_i\,v_i-\lambda_a\:,
\end{equation}
where the vectors $v^a\in\bZ^k$, $a=1,\dots,d$,  are associated to each facet of $\cP$, and satisfy the following property:
each vertex of $\cP$ is the  intersection of exactly $k$ facets, corresponding to the vectors $v^{a_1},\ldots,v^{a_k}$,
and the order of the orbifold singularity of $M_{2k}$ at each vertex is $|\det(v^{a_1},\ldots,v^{a_k})\,|$. The isotropy group of each facet is $\text{U}(1)$. In a suitable  basis of $2\pi$-periodic angles $\tau_i$, the vector field $v^a=v^a_i \partial_{\tau_i}$ degenerates at the facet corresponding to $v^a$. The vertices are then the fixed point of the $\text{U}(1)^{k}$ action.
For orbifolds, the vectors $v^a$ are not necessarily primitive.  We can  define for each $v^a$ a primitive vector $\hat v^a$ and a positive integer $n_a$ such that $v^a= n_a \hat v^a$.
The structure group of every point in the inverse image of $\mathcal{F}_a$ is $\mathbb{Z}_{n_a}$ and the integer $n_a$ is called the  label of the facet $\mathcal{F}_a$. $M_{2k}$ is a smooth manifold 
if and only if all the labels $n_a$ are one and
for each vertex $|\det (v^{a_1},\ldots, v^{a_k})|=1$. When $M_{2k}$ is non compact, $\cP$ is a non compact polytope and not necessarily convex \cite{Lerman2015}, although it is in the case of Calabi-Yau cones \cite{Martelli:2005tp,Martelli:2006yb}. In this paper we assume convexity unless specified.

The localization theorems depends only on the topology, but it is sometimes useful to introduce a metric on the manifold. For K\"ahler toric orbifolds, the metric can be written as 
\begin{equation}
\label{toricmetricappendixA}
\dd s^2 =G_{ij} (y) \dd y_i \dd y_j + G^{ij} (y) \dd\tau^i \dd\tau^j 
\end{equation}
where $G_{ij} = \partial_{y_i}\partial_{y_j} G(y)$ and $G^{ij}$ is its inverse matrix. The function $G(y)$ is referred to as symplectic potential. The K\"ahler form  reads
\begin{equation}
\label{X1toricappendixA}
\Kform^1 = \dd y_i \wedge \dd \tau^i\, , 
\end{equation}
from which it is clear that $y_i$ are the moment maps associated to the torus action, that are defined up to additive constants. We also record the expression for the Ricci scalar of the four-dimensional metric 
\begin{equation}
R_\gamma = -\partial_{y_i}\partial_{y_j}  G^{ij}\, , 
\end{equation}

A canonical metric on $M_{2k}$ is given by  $G(y)=\frac12 \sum_{a=1}^d l_{a} \log l_{a}$ \cite{Guillemin1994KaehlerSO} 
so that
\begin{equation}\label{metric} G_{ij}=\frac{\partial^2 G}{\partial y_i \partial y_j} = \frac12 \sum_{a=1}^d \frac{v_i^{a}  v_j^{a}}{l_{a}} \, , \end{equation}
where the poles at the facets are chosen such that  the metric is smooth up to orbifold singularities.  A general K\"ahler metric on $M_{2k}$ is  obtained by replacing the symplectic potential $G(y)$ for the canonical metric with $G(y)+h(y)$, where $h(y)$ is smooth on the whole $\cP$.  In the neighbourhood of a vertex $p_A$ associated with the vectors $\{v^{a_1},\ldots ,v^{a_k}\}$
\begin{equation}\label{facetsG} G_{ij} =  \frac12 \frac{v_i^{a_1}  v_j^{a_1}}{l_{a_1}} + \frac12  \frac{v_i^{a_2}  v_j^{a_2}}{l_{a_2}} +\ldots + \frac12 \frac{v_i^{a_k}  v_j^{a_k}}{l_{a_k}}+\ldots \, ,\end{equation}
up to regular pieces. This can be inverted to give
\begin{equation}\label{facetsGinv} G^{ij} = \frac{2}{d_A^2}\left( (u_{A}^{a_1})_i (u_{A}^{a_1})_j l_{a_1}  + (u_{A}^{a_2})_i (u_{A}^{a_2})_j l_{a_2}+\ldots  + (u_{A}^{a_k})_i (u_{A}^{a_k})_j l_{a_k}  \right ) +\ldots \, ,\end{equation}
where the  vectors $u^{a_i}_A$   have integer entries and satisfy $u^{a_i}_A\cdot v^{a_j} = d_A \delta_{ij}$ as well as
\begin{equation} \label{uv}v^{a_1}_i (u^{a_1}_{A})_j+v^{a_2}_i (u^{a_2}_{A})_j+\ldots +v^{a_m}_i (u^{a_m}_{A})_j=d_A \delta_{ij}\, .\end{equation}
We introduced the notation $d_A=|\det(v^{a_1},\ldots,v^{a_k})\,|$.

The counter-image of each facet $\mathcal{F}_a$ is a $\mathbb{T}$-invariant divisor $D_a$ in $M_{2k}$ with an associated line bundle $\cL_a$.  One can prove that  \cite{Guillemin1994KaehlerSO}
\begin{equation}
\label{kform}
[\Kform]  = - 2 \pi  \sum_a \lambda_a \big[ c_1(\cL_a)\big] \, ,
\end{equation}
where the equation holds  in co-homology. Another useful relation is
\begin{equation}
	[\cR]=2\pi\sum_a\big[c_1(\cL_a)\big]\:,
\end{equation}
where $\cR$ is the Ricci two-form.
A representative of the first Chern class of $\cL_a$ is given by  \cite{Guillemin1994KaehlerSO,Abreu:2001to,Martelli:2023oqk}%
\begin{equation}\label{CC} c_1(\cL_a) = \dd \left ( \mu_a^i\dd \tau^i \right ) \, , \qquad\qquad 
\mu_a^i =\mu_a^i  (y) =  -\frac{1}{4\pi }\frac{G^{ij}v_j^a}{l_a}\, . 
\end{equation}
From \eqref{facetsGinv} we see that the restriction of the moment maps \eqref{CC} to the fixed point $p_A$ is given by
\begin{align}  \mu^{a}_i \Big |_{p_A} = \begin{cases} -\frac{1}{2\pi}  \frac{ u_i^{a}}{d_{A}}  \, \qquad &{\rm if} \,  a\in (a_1,\ldots , a_k)  \\
0 \, \qquad &{\rm if} \, a\notin (a_1,\ldots , a_k) \end{cases}
\end{align}
From $G^{ij}G_{jk} =\delta^i_k$ we also find the useful relation
\begin{align}\label{ide} \sum_{a=1}^d \mu_{a}^i v_k^{a} = -\frac{\delta_k^i}{2\pi} \, .\end{align}

In the main text we used these formulas mostly for the special case $k=2$, corresponding to the four-dimensional base of the foliation. There is the same number of fixed points and divisors and we can use the same letter, $a$, to label them. The $a$-th fixed point $p_a$ lies at the intersection of the divisors $D_{a-1}$ and $D_a$ and it is associated with the vectors $(v^{a-1},v^a)$. To compare with formula \eqref{mm} we just set
\begin{equation}\label{dictionary} A\equiv a \, ,\qquad (u_A^{a_1},u_A^{a_2})\equiv(u_a^1,u_a^2) \, ,\qquad (v^{a_1},v^{a_2})\equiv (v^{a-1},v^a)\, ,\qquad \tau^i   \equiv  \phi_i^B\, . \end{equation}

\subsection{The geometric interpretation of the weights in the localization formula}

In this subsection we provide an  alternative derivation of the relation \eqref{weights_VS_Chern} using the toric geometry of the four-dimensional base of the foliation.

The Riemmanian metric $\gR$ on $M$ with $\bT$ isometry can be split into a component along the orbits of $\xi$ and a
component on the base as follows:
\begin{equation}
	\gR=\gR(\xi,\xi)^{-1}\,\big((\gR)_{\mu\nu}\,\xi^\mu\dd x^\nu\big)^2+\gR^B\:.
\end{equation}
Since the metric $\gR$ is arbitrary, we can choose it so that $\gR^B$ is K\"ahler toric.\footnote{The Chern classes (intended as cohomology classes, rather than particular choices of representatives) do not depend on the choice of metric.}
Using the expression \eqref{flat_metric_5d} for $\gR$ near $L_a$ and carefully expanding both $\gR(\xi,\xi)$ and $(\gR)_{\mu\nu}\,\xi^\mu\dd x^\nu$,
one arrives at the following expression for $\gR^B$ near the point $p_a$ (which is the projection on the base of $L_a$) :
\begin{equation}
\label{base_metric_approx}
	\gR^B\big|_{\text{near }p_a}\approx\sum_{i=1,2}
		\bigg[\big(\dd\rLa^i\big)^2+(\rLa^i)^2\Big(\dd\phiLa^i-\frac{\xiLa^i}{\xiLa^0}\,\dd\phiLa^0\Big)^2\:\bigg]\:.
\end{equation}
For the dense subset \eqref{orbifold_assumption} of values of $\xi$ for which the base is an orbifold, we can compare the above expansion of $\gR$
with the expansion of a K\"ahler toric metric near $p_a$%
\begin{equation}
\label{4d_toric_approx}
	g_{\text{4d toric}}\big|_{\text{near }p_a}\approx\,2\,
		\Big[\dd\big(l_{a-1}(y)^{\frac12}\big)^2+\dd\big(l_a(y)^{\frac12}\big)^2+
		l_{a-1}(y)\,\big(d_a^{-1}\,\underline u^1_a\cdot\dd\underline\phi^B\big)^2+
		l_a(y)\,\big(d_a^{-1}\,\underline u^2_a\cdot\dd\underline\phi^B\big)^2\:\Big]\:,
\end{equation}
which follows from \eqref{facetsG} and \eqref{facetsGinv} using the identification \eqref{dictionary}. 
Comparison between \eqref{base_metric_approx} and \eqref{4d_toric_approx} leads to the identifications
\begin{align}
	\rLa^1=\sqrt{2\,l_{a-1}(y)}\:,\qquad\rLa^2=\sqrt{2\,l_a(y)}\:,\qquad
		d_a^{-1}\,\underline u^{i}_a\cdot\dd\underline\phi^B=\dd\phiLa^i-\frac{\xiLa^i}{\xiLa^0}\,\dd\phiLa^0\:.
\end{align}
If we contract the third expression with the vector $X$, we find \eqref{weights_VS_Chern}, thus deriving the relation between
restrictions of equivariant Chern classes and weights in a different manner.

We also see that
\begin{equation}
	\dd\rLa^1\wedge\dd\rLa^2=\frac{d_a}2\,\big(l_{a-1}(y)\,l_a(y)\big)^{-\frac12}\cdot\dd y_1\wedge\dd y_2=
		(\text{positive function})\cdot\dd y_1\wedge\dd y_2\:,
\end{equation}
which implies that the ordering $(\phiLa^0,\rLa^1,\phiLa^1,\rLa^2,\phiLa^2)$ corresponds to the same orientation for all $a=1,\ldots,d$
if the signs $\sa$ introduced in \eqref{xiLa_determinant} are all equal. With the appropriate convention for the orientation of $M$
we can take them to be $\sa=+1$ for all $a$.

\section{Complexified CCLP solution}
\label{app:known}

In this appendix we will recall the salient features of the local solution found in \cite{Chong:2005hr} (CCLP) focussing on the three parameter family preserving supersymmetry, largely 
following \cite{Cabo-Bizet:2018ehj}. In addition to colllecting the key ingredients of the solution, below we will describe in detail 
various coordinate systems  and their  roles in highlighting different global aspects of the solution. It is well known that the generic three parameter solution is necessarily complex \cite{Cabo-Bizet:2018ehj} and may be interpreted as a Euclidean ``black saddle'' contributing to the gravitational path integral. A two-parameter sub-family describes a supersymmetric and extremal black hole\footnote{A further one-parameter sub-family corresponds to the supersymemetric extremal black hole previously discovered in \cite{Gutowski:2004ez}.}, in which the solution becomes purely real and admits a Lorentzian interpretation.  A \emph{different} two-paramater sub-family gives rise to a solution with distinct global properties, which is referred to as ``topological soliton''\footnote{Alternative names for such solutions are ``bubbling geometries'' or ``microstate geometries''.}. In Euclidean setting 
this describes a space-time with a ``bolt'' with topology $\spindle \times S^1$ where $\spindle$ is a spindle,  including  $S^2\times S^1$ as a special case  \cite{Cassani:2015upa}. The latter admits also a Lorentzian interpretation with global structure $\mathbb{R}_t \times M_4$ where $M_4={\cal O}(-1)\to S^2$  is the total space of a line bundle on $S^2$.
For both classes of global solutions, we will derive the sets of (three) vectors of the fan used in the main body of the paper and we will describe their projection onto
 the base.

The metric and gauge field written in the original coordinates of \cite{Chong:2005hr} read
\begin{align}\label{eq:cclp_non_susy}
  \begin{aligned}
  \dd[]{s}^2 ={}& - \frac{\Delta_\theta \qty[(1 + r^2/\ell^2)\varrho^2 \dd[]{t} + 2q \nu] \dd[]{t}}{\Xi_a \Xi_b \varrho^2} + \frac{2q \nu \varpi}{\varrho^2} + \frac{\curly{f}}{\varrho^4} \qty(\frac{\Delta_\theta \dd[]{t}}{\Xi_a \Xi_b} - \varpi)^2 \\
  & + \frac{\varrho^2}{\Delta_r} \dd[]{r}^2 + \frac{\varrho^2}{\Delta_\theta} \dd[]{\theta}^2 + \frac{r^2 + a^2}{\Xi_a} s_{\theta}^2 \dd[]{\phi}^2 + \frac{r^2 + b^2}{\Xi_b} c_{\theta}^2 \dd[]{\psi}^2 \,, \\
  A ={}& \frac{\sqrt{3}}{\curly{x}} \frac{q}{\varrho^2} \qty(\frac{\Delta_\theta \dd[]{t}}{\Xi_a \Xi_b} - \varpi) + \alpha \dd[]{t} + \dd[]{\widetilde{\lambda}} \,, 
  \end{aligned}
\end{align}
where we have denoted $s_{\theta} = \sin^{}{\theta}$, $c_{\theta} = \cos^{}{\theta}$ and, using a notation slightly modified with respect to  \cite{Chong:2005hr}, we have defined the following
functions and one-forms:
\begin{align}
  \begin{aligned}
    \nu ={}& b s_{\theta}^2 \dd[]{\phi} + a c_{\theta}^2 \dd[]{\psi} \,, & \varpi ={}& \frac{a s_{\theta}^2}{\Xi_a} \dd[]{\phi} + \frac{b c_{\theta}^2}{\Xi_b} \dd[]{\psi} \,, \\
    \Delta_r ={}& \frac{(r^2 + a^2)(r^2 + b^2)(1 + r^2/\ell^2) + q^2 + 2ab q}{r^2} - 2m \,, & \curly{f} ={}& 2m \varrho^2 - q^2 + 2ab q \frac{\varrho^2}{\ell^2}\, , \\
    \Delta_\theta ={}& \qty(\ell^2 - a^2 c_{\theta}^2 - b^2 s_{\theta}^2)/\ell^2 \,,  & \varrho^2 ={}& r^2 + a^2 c_{\theta}^2 + b^2 s_{\theta}^2 \,, \\
    \Xi_a ={}& \qty(\ell^2 - a^2)/\ell^2 \,, & \Xi_b ={}& \qty(\ell^2 - b^2)/\ell^2 \,.
  \end{aligned}
\end{align}
The gauge shift constant $\alpha$ and the function $\widetilde{\lambda}(\theta,\phi,\psi)$ in \eqref{eq:cclp_non_susy} are fixed by demanding global regularity 
of the solution and its Killing spinors  \cite{Cabo-Bizet:2018ehj}.
The angular coordinates $\phi,\psi$ have $2\pi$-periodicity, $t$ is the time coordinate,  $\theta \in [0,\pi/2]$ and the radial coordinate $r\in [r_+,\infty]$, where $r_+$ is at the largest positive root of $\Delta_r$. In the generic four-parameters non-supersymmetric solution, $r=r_+$ is the location of  an ordinary  Killing horizon.

Requiring supersymmetry through the relation  \begin{align}\label{eq:susy_cond_cclp}
  q ={}& \frac{\ell m}{\ell + a + b} \,, 
\end{align}
reduces the four independent parameters $(a,b,q,m)$ to three, that may be taken to be $(a,b,m)$, or $(a,b, \widetilde{m})$, or $(a,b,r_+)$, where the parameters $m$, $\widetilde{m}$, and $r_+$ can be traded for one another as
 \begin{align}
 \label{mmtilderelation}
 \widetilde{m} = -1 +\frac{\ell^2 m}{(a+b) (a+\ell) (b+\ell) (a+b+\ell)} = -1 - \frac{(a - \iu \, r_+)(b - \iu \, r_+)(\ell - \iu \, r_+)}{(a + b)(\ell + a)(\ell + b)} \,. 
 \end{align}
Demanding that $r_+$ is real forces one to work with a \textit{complex} supersymmetric geometry despite the fact that one has not performed the usual Wick rotation of the time coordinate $t$.
In this context, one could more generally allow for  complex values of the parameters  $(a,b, r_+)$, however, it is useful to keep these real, so that there exists a two-parameter sub-family, 
obtained upon setting $\widetilde{m}=0$,  which is real, Lorentzian, and possesses an extremal Killing horizon at $r= \eval{r_+}_{\widetilde{m} = 0} \equiv r_* = \sqrt{\ell (a+b) + a b}$.

By further performing the analytic continuation of the time coordinate as
\begin{align}\label{eq:TEuclidean}
  t ={}& - \iu \tau \, , 
\end{align}
and compactifying $\tau$ on a circle one obtains what may be referred to  as
 a complex black saddle ($\CBS$), using the nomenclature introduced in \cite{Bobev:2020pjk}.
In this setting, there is an  $\text{U}(1)^3$ action generated by three commuting  Killing vectors and one can study  regularity of the solution by requiring that this  is smooth in the neighborhood of degenerating Killing vectors\footnote{Equivalently, one can perform this analysis in the non-supersymmetric real Euclidean solution and then impose supersymmetry on the results. This can is obtained 
taking $a$, $b$ to be pure imaginary, while keeping $m,r_+$ real. In particular, the quantities $\beta$, $\iu\Omega_1$ and $\iu \Omega_2$ are real  \cite{Cabo-Bizet:2018ehj}.}.

\subsection{Complex black saddle}

At this point the analysis bifurcates in two globally distinct geometries: below we discuss the generic case, while in the next subsection we address a
 special case leading to the so-called topological soliton.  One finds that generically  there are three 
  Killing vectors whose norm, computed with respect to the metric \eqref{eq:cclp_non_susy}, vanish on the loci  $r = r_+$, $\theta = 0$, $\theta = \pi/2$, respectively.  If no particular relation is imposed among the three parameters $(a,b,r_+)$, the degenerating Killing vectors are 
  \begin{align}
  \begin{aligned}
    V_{r = r_+} ={}& \frac{\beta}{2\pi\iu} \qty(\iu \partial_\tau + \Omega_1 \partial_\phi + \Omega_2 \partial_\psi) \,, \qquad V_{\theta = 0} = \partial_\phi \,, \qquad V_{\theta = \pi/2} = \partial_\psi \,, 
  \end{aligned}
\end{align}
where 
 \begin{align}
\Omega_1 = \frac{(\ell-\iu r_+) (\ell(a + b) + ab + \iu a r_+)}{\ell(a-\iu r_+) (\ell(a + b) + ab + \iu \ell r_+)}\, , \qquad \Omega_2 
= \frac{(\ell-\iu r_+) (\ell(a + b) + ab + \iu b r_+)}{\ell(b-\iu r_+) (\ell(a + b) + ab + \iu \ell r_+)}\, ,
\end{align}
and
 \begin{align}
\beta = \frac{2\pi\iu \ell(a - \iu r_+)(b - \iu r_+)(\ell(a + b) + a b + \iu \ell r_+)}{\qty(\ell(a + b) + a b - 2\iu(a + b + \ell)r_+-3 r_+^2)(\ell(a + b) + a b - r_+^2)}\, .
\end{align}
Below we will refer to the coordinates  $\psi_i = (\tau,\phi,\psi)$, $i = 0,1,2$, as to ``$\psi-$coordinates'' and to 
$\partial_{\psi_i} = \qty(\partial_\tau, \partial_\phi, \partial_\psi)$ as to ``$\psi-$basis''. In this basis we order the degenerating Killing vectors as
 \begin{align}
 \label{fanBHpsibasis}
 \qty(V^0)_{\psi\mathrm{-basis}} =  \left(0,1,0\right) \, ,\quad  \qty(V^1)_{\psi\mathrm{-basis}} =  \frac{\beta}{2\pi\iu}   \left(\iu,\Omega_1,\Omega_2 \right) \,  ,\quad  \qty(V^2)_{\psi\mathrm{-basis}} =  \left(0,0,1\right) \,, 
\end{align}
and the supersymmetric Killing vector bilinear reads
 \begin{align}
 \label{reebpsibasis}
(\xi )_{\psi-\mathrm{basis}} =
 \qty(\iu, \frac{1}{\ell},\frac{1}{\ell})\, .
\end{align}

Smoothness of the metric requires that the coordinates $\phi,\psi$ are $2\pi$-periodic, while the coordinate $\tau$ obeys twisted periodicity:
\begin{align}
  \begin{aligned}
    (\tau,\phi,\psi) \sim {}& \qty(\tau + \beta, \phi - \iu\Omega_1 \beta, \psi - \iu\Omega_2 \beta) \,, \quad \phi \sim \phi + 2\pi \,, \quad \psi \sim \psi + 2\pi \,. 
  \end{aligned}
\end{align}
In order to diagonalize this action, it is useful to introduce a new set of \textit{independently} $2\pi$-periodic coordinates $\phi_i$, $i=0,1,2$, defined as
\begin{align}
  \mqty(\phi_0 \\ \phi_1 \\ \phi_2) ={}& B \mqty(\tau \\ \phi \\ \psi) \,, \quad B = \mqty(\frac{2\pi}{\beta} & 0 & 0 \\ \iu\Omega_1 & 1 & 0 \\ \iu\Omega_2 & 0 & 1) \,. 
\end{align}
The vectors $V^a$ in the $\phi-$basis are obtained from those in (\ref{fanBHpsibasis}) by multiplying them with $B$ and read
 \begin{align}
  \qty(V^0)_{\phi\mathrm{-basis}} =  \left(0,1,0\right) \, ,\quad  \qty(V^1)_{\phi\mathrm{-basis}} =  \left(1,0,0 \right) \,  ,\quad  \qty(V^2)_{\phi\mathrm{-basis}} =  \left(0,0,1\right) \,, 
\end{align}
which are the vectors of the fan we used in \eqref{BH_vectors}, modulo an inconsequential reordering. Similarly, the supersymmetric Killing vector in the $\phi$-basis reads
 \begin{align}
(\xi )_{\phi-\mathrm{basis}} =  B (\xi )_{\psi-\mathrm{basis}} = \frac{1}{\beta}  \left(2 \pi \iu ,- \omegafuga ,- \omegafugb \right) \,, 
\end{align}
where we have defined the fugacities
\begin{align}
\omegafuga \equiv \beta \qty(\Omega_1 - \frac{1}{\ell}) \,, \qquad 
\omegafugb \equiv \beta \qty(\Omega_2 - \frac{1}{\ell} ) \,. 
\end{align}
These fugacities play an important role from multiple vantage points. To begin with, the on-shell action of the $\CBS$ can be computed by direct integration of the Lagrangian making use of the explicit metric. Then it can be renormalized using either background subtraction or holographic renormalization, see for example \cite{Chen:2005zj, Cabo-Bizet:2018ehj, Bobev:2022ocx}. Remarkably, the final result can be recast entirely in terms of $\omegafuga, \omegafugb$
\begin{align}
  \widehat{I} =\frac{\pi\ell^3}{108 G_5}\frac{\qty(\omegafuga + \omegafugb - 2\pi\iu)^3}{\omegafuga \omegafugb} \,.
\end{align}
This key expression was reproduced in \cref{sect:black hole localization} in an entirely metric agnostic fashion.

\subsection{Topological soliton}

The topological soliton may be viewed as a special case of the general three-parameter  $\CBS$ solution.
However, crucially, it has a \emph{different topology} from the generic solution and in particular in Lorentzian signature there is no vanishing time-like Killing vector, namely there isn't a Killing horizon. The solution is then globally $\mathbb{R}_t\times \Morb_4$, where $\Morb_4$ is  a non-compact orbifold with a spindle bolt (with topology discussed in detail in \cite{Crisafio:2024fyc}), that  for specific values of the parameters reduces to the smooth manifold $\mathcal{O}(-1)\to S^2$ \cite{Cassani:2015upa}. In the Euclidean setting, the solution becomes $S^1_\tau\times \Morb_4$, with radius of the Euclidean time circle an arbitrary modulus of the solution.

The qualitative difference from the  $\CBS$ solution is encoded in the distinct shape of the
 fan describing the degeneration loci of the three $\text{U}(1)$ directions in the metric, as we now describe.   Starting from the metric in (\ref{eq:cclp_non_susy}) in the original  coordinates\footnote{This computation can be done either in Lorentzian signature or in Euclidean, after setting $t=-\iu \tau$, but as will become clear momentarily, this is irrelevant.}
$(t,\phi,\psi)$ and computing the norm of a generic Killing vector, one can check that for the special value of the parameter
$r_+$ given by
\begin{align}
  r_+^2 ={}& r_0^2 \equiv - \qty(a + b + \frac{ab}{\ell})^2\, , 
\end{align}
the Killing vectors $V_{\theta = 0} = \partial_\phi$, $V_{\theta = \pi/2} = \partial_\psi$ are still degenerating as in the $\CBS$ solution, but 
there is a different  Killing vector vanishing at $r=r_0$, given by\begin{align}
V_{r=r_0} =   \partial_\phi + \frac{(b-\ell) (a b+2 a\ell+b \ell)}{(a-\ell) (a b+ a \ell +2b\ell)}\, \partial_\psi \, .
\end{align}
We denote this critical value of  the radius by $r_0$ as in \cite{Cassani:2015upa} to distinguish it from the non-extremal horizon radius $r_+$ 
and the extremal horizon radius $(r_*)^2= a b+a\ell+b \ell$. Using (\ref{mmtilderelation}) 
we obtain the corresponding special value of $\widetilde{m}$ that we denote by $\widetilde{m}_0$, which reads
\begin{align}
\label{mtTS}
\widetilde{m}_0 = -1 - \frac{(a b+a \ell+2 b \ell) (a b+2 a \ell+b \ell)}{\ell^3 (a+b)}\, . 
\end{align}

A standard regularity analysis\footnote{This was done in \cite{Cassani:2015upa}, but there it was demanded that there are no orbifold singularities, namely that  $n=v=1$. Here we have included these more general possibilities.} of the metric near to the loci of degeneration of the Killing vectors  shows
that taking the coordinates $\phi,\psi$ to be  $2\pi$-periodic, the metric is smooth in the orbifold sense by requiring the two conditions 
\begin{align}\label{TSquantization}
\frac{(b-\ell) (a b+2 a\ell+b \ell)}{(a-\ell) (a b+ a \ell +2b\ell)} =  n \in \mathbb{N} \, ,\qquad  \frac{\ell (a-\ell) (a b+ a \ell+2 b \ell)}{\left(3 \ell^2 +5 a \ell +5 b \ell+3 a b\right) (a b+a \ell+b \ell)} = v  \in \mathbb{N} \, ,
   \end{align}
where the former implies that the topology of the locus $r=r_0$ is that of spindle with one orbifold singularity $\spindle_{[n,1]}$, also known as a teardrop, while the latter ensures that the normal directions are a $\mathbb{C}/\mathbb{Z}_v$ singularity.  It is simple to see that the two equations (\ref{TSquantization}) may be solved for the parameters $a,b$ in terms of the two integers $n,v$, thus providing the ``quantization conditions''. The resulting solution  then depends on $n,v$, with the metric being smooth in the orbifold sense\footnote{Generically the  metric might become complex away from the degeneration loci, but it should be by now clear that this is  not a concern.}.  The coordinate $\tau$ can have an arbitrary periodicity $\tau \sim \tau +\beta_0$, since the vector field  $\partial_\tau$ is never vanishing, as stated at the beginning of this section.   In this case there is no need to introduce new coordinates  to diagonalize the torus action
and  the degenerating Killing vectors read
\begin{align}
  \qty(V^0)_{\psi\mathrm{-basis}} =  \left(0,1,0\right) \, ,\quad  \qty(V^1)_{\psi\mathrm{-basis}} =  v \left(0,1,n  \right) \,  ,\quad  \qty(V^2)_{\psi\mathrm{-basis}} =  \left(0,0,1\right) \,  . 
\end{align}
Notice that the we have included a highest  common factor $v$ in $V^1$, that equivalently may also be indicated with a ``label'' in the toric diagram  \cite{Crisafio:2024fyc}).
The supersymmetric Killing vector is exactly the same as that in (\ref{reebpsibasis}), since it does not depend on the parameters $a,b,m$.

Notice that the external vectors are exactly the same as those in (\ref{fanBHpsibasis}) while, crucially, the internal one
is different and coplanar with the other two, consistently with the fact that globally the geometry is the  direct  product of a four dimensional 
orbifold ${\cal O}(-1)\to \spindle_{[n,1]}$ with a never-shrinking circle $S^1_\tau$.  We will make a few more comments about the global structure of this solution at the end of subsection \ref{spindlebolts}.
 Notice that this fan does not  obey the ``CY condition'' discussed in section \ref{sect:reg gauge}, 
which unfortunately at present precludes the possibility of applying the localization formula discussed in the paper.

\subsection{Orthotoric  coordinates}

Let us now discuss the   complexified CCLP solution using the set of orthotoric coordinates\footnote{We have relabelled $\eta^{\text{there}} = x$ and $\xi^{\text{there}} = \rho$ to avoid clashing with our one-form $\eta$ and to signify that $\rho$ is a ``radial'' coordinate.}
 employed in \cite{Cassani:2015upa}. Specifically, these are local coordinates on the transverse four-dimensional K\"ahler foliation, using the canonical decomposition of the metric (\ref{eq:5d_metric}) stemming from the general analysis of supersymmetric solutions in the time-like class. Although the 
four dimensional  ``base'' is in general not globally defined, below we will discuss that with due care it is possible to regard it as an orbifold. This is precisely the counterpart of the difference between irregular and quasi-regular Sasaki-Einstein manifolds. In the notation of section \ref{sec:timelike_susy}, the four-dimensional K\"ahler metric $\gamma_{mn}$ has line element
\begin{align}
\label{ortometric}
\ell^{-2} \dd s^2_\gamma =  \frac{x - \rho}{\mathcal{F}(\rho)} \dd[]{\rho}^2 + \frac{\mathcal{F}(\rho)}{ x - \rho}\qty(\dd[]{\Phi} + x \dd[]{\Psi})^2 + \frac{x - \rho}{\mathcal{G}(x)} \dd[]{x}^2 + \frac{\mathcal{G}(x)}{ x - \rho}\qty(\dd[]{\Phi} + \rho \dd[]{\Psi})^2 \,
\end{align}
where
\begin{align}
  \mathcal{G}(x) ={}& g_0 (x^2 - 1) (x - g_1) \,, \qquad \mathcal{F}(\rho) = - \mathcal{G}(\rho) + m_0 (\rho - m_1)^3 \,,
\end{align}
and the parameters  $(g_0, g_1, m_2, m_1)$ are related to the three independent parameters $(a,b, \widetilde{m})$ as
\begin{align}
  g_0 ={}& - \frac{4}{\widetilde{m}} \,, \quad g_1 = \frac{2\ell^2 - a^2 - b^2}{a^2 - b^2} \,, \quad m_2 = -4 \qty(1 + \frac{1}{\widetilde{m}}) \,, \quad m_1 =  - \frac{2\ell + a + b}{a - b} \,.
\end{align}
The remaining data of the five-dimensional metric are the  one-form 
\begin{align}
\label{omegaorto}
  \omega ={}& \frac{\ell (\mathcal{F}''' + \mathcal{G}''')}{48(x - \rho)^2} \qty[\qty(\mathcal{F} + \frac{(x - \rho)}{2} \qty(\mathcal{F}' - \frac{1}{2} \mathcal{F}'''(\rho - m_1)^2)) \qty(\dd[]{\Phi} + x \dd[]{\Psi}) + \mathcal{G} \qty(\dd[]{\Phi} + \rho \dd[]{\Psi})] \notag \\
  & - \ell \frac{\mathcal{F}''' \mathcal{G}'''}{288}  \qty((x + \rho)\dd[]{\Phi} + x \rho \dd[]{\Psi}) -\frac{2\ell}{\tilde m} \dd[]{\Psi} \,,
\end{align}
and the function 
\begin{align}
  f ={}& \frac{24(x - \rho)}{\mathcal{F}'' + \mathcal{G}''} \,.
\end{align}
Here primes denote derivatives with respect to the variable on which the functions depend. Note that we have included a closed term in $\omega$ that locally may be reabsorbed by shifts of the coordinate $y$. However, the above form of $\omega$ is uniquely fixed by comparing the five-dimensional metric to the globally regular expression in the original coordinates, eq. 
 (\ref{eq:cclp_non_susy}). The gauge field may be obtained from the general form (\ref{eq:susy_relations}), where the connection $P$ of the Ricci-form $\mathcal{R} = \dd[]{P} $ is given by 
\begin{align}
P = - \frac{1}{2(x - \rho)} \qty(\mathcal{F}' \qty(\dd[]{\Phi} + x \dd[]{\Psi}) + \mathcal{G}' \qty(\dd[]{\Phi} + \rho \dd[]{\Psi})) \,,
\end{align}
while the K\"ahler form $ \Kform^1$ reads
\begin{align}
\label{X1orto}
 \Kform^1  =  \ell^2\dd \left[  (x + \rho)\dd[]{\Phi} + x \rho \dd[]{\Psi}\right] \,. 
\end{align}
The relation to the coordinates  $(t,r,\theta,\phi,\psi)$ of \cite{Chong:2005hr} is given by 
\begin{align}
\label{ortotocclp}
  \begin{aligned}
    \sin^{2}{\theta} ={}& \frac{1}{2}(1 - x) \,, \\
    r^2 ={}& \frac{1}{2}(a^2 - b^2) \widetilde{m} \rho + \qty[\ell(a + b) \qty(\widetilde{m} + 1) + ab] + \frac{1}{2} (a + b)^2 \widetilde{m} \,, \\
    t ={}& y \,, \\
    \phi ={}& \frac{y}{\ell} - \frac{4\qty(\ell^2 - a^2)}{\qty(a^2 - b^2) \widetilde{m}} \qty(\Phi - \Psi) \,, \\
    \psi ={}& \frac{y}{\ell} - \frac{4 \qty(\ell^2 - b^2)}{\qty(a^2 - b^2) \widetilde{m}} \qty(\Phi + \Psi) \,.
  \end{aligned}
\end{align}

In order to facilitate the chain of coordinate changes, we find it useful to define $\Psi_i \equiv  (y,\Phi,\Psi)$, $i=0,1,2$, to dub these ``$\Psi-$coordinates'', and 
to repackage the angular part of the change of coordinates as
\begin{align}
  \mqty(\tau \\ \phi \\ \psi) ={}& C \mqty(y \\ \Phi \\ \Psi) \,, \qquad C = \left(
\begin{array}{ccc}
 \iu & 0 & 0 \\
 \frac{1}{\ell} & -\gamma_a & \gamma_a \\
 \frac{1}{\ell} & -\gamma_b & -\gamma_b \\
\end{array}
\right) \,, \qquad
\begin{array}{c}
  \gamma_a \equiv \displaystyle{\frac{4(\ell^2 - a^2)}{(a^2 - b^2) \widetilde{m}}} \,, \\
  \gamma_b \equiv \displaystyle{\frac{4(\ell^2 - b^2)}{(a^2 - b^2) \widetilde{m}}} \,. 
\end{array}
\end{align}
The above change of coordinates applies both to the $\CBS$, when $\widetilde{m}$ is viewed as in independent parameter, as well as to the topological soliton, if one sets $\widetilde{m}=\widetilde{m}_0$ as in \eqref{mtTS}. From (\ref{ortotocclp}) one can also infer the ranges of the coordinates: $x\in [-1,1]$, while $\rho\in [\rho_+,\infty)$ for the $\CBS$ and $\rho \in [\rho_0 , \infty)$ for the topological soliton, with the lower bounds being 
\begin{align}
  \begin{aligned}
  \rho_+ ={}& - \frac{a^2 + b^2 + \ell(a + b) - \iu \, (2\ell + a + b) r_+}{(a - b)(\ell + a + b - \iu \, r_+)} \, ,& \text{for } \CBS\,, \\
\rho_0 ={}& -\frac{(a+b) \left(2 \ell (a+b)+a b+3 \ell^2\right)}{(a-b) \left(2 \ell (a+b)+a b+\ell^2\right)}\, , & \text{for topological soliton}\, .  \\
  \end{aligned}
\end{align}
Notice that the coordinate change (\ref{ortotocclp}) becomes singular in the extremal limit, $\tilde m=0$. To remedy this  one can 
 consider the scaling limit \cite{Cassani:2015upa}
\begin{align}
\label{scalinglimit}
\Phi = \varepsilon \tilde \phi \, , \qquad  \Psi = \varepsilon \tilde \psi \, ,\qquad \rho = - \frac{\tilde\rho}{\varepsilon}\, ,
\end{align}
with $\tilde m\to 0$, $\varepsilon \to 0$, keeping fixed 
\begin{align}
\frac{\tilde m}{\varepsilon} \equiv k \, . 
\end{align}
In the limit $\varepsilon \to 0$ the metric (\ref{ortometric}) remains finite and assumes the so-called ``Calabi type'' form
\begin{align}
\ell^{-2} \dd s^2_\gamma =   \frac{{\tilde \rho}} {{\tilde {\cal F}} (\tilde\rho)}  \dd {\tilde \rho}^2 + \frac{\tilde  {\cal{F}}( \tilde \rho) }{ \tilde \rho}\qty(\dd[]{\tilde \phi} + x \dd[]{\tilde \psi})^2 +
\tilde \rho\left( \frac{ \dd[]{x}^2 }{\tilde {\cal{G}}(x)} + \tilde {\cal{G}}(x)   \dd[]{\tilde \psi}^2 \right)\, , 
\end{align}
where the functions  $\tilde {\cal F} (\tilde\rho)$ and  $\tilde {\cal G}(x)$ are defined by 
\begin{align}
  {\cal F} (\rho) = \varepsilon^{-3} \tilde {\cal F} (\tilde\rho) + {\cal O}(\varepsilon^{-2})\, \qquad \quad   {\cal G} (x) =  \varepsilon^{-1}\tilde  {\cal G} (x) + {\cal O}(1)\,  ,
\end{align}
for $\varepsilon \to 0$.  A similar scaling limit may be performed when $b\to a$ and is discussed in  \cite{Cassani:2015upa}.

\subsection{Darboux  coordinates}

The final set of coordinates that we discuss is the Darboux coordinates that exist, at least locally, for any toric symplectic foliation. This will allow us to make contact with 
the general discussion of the transverse geometry in the main part of the paper and to provide a global four dimensional point of view on these solutions.  In order to cast the four-dimensional metric (\ref{ortometric}) in the form (\ref{toricmetric})
it is convenient to compare the K\"ahler form (\ref{X1orto})  to (\ref{X1toric}).
This  shows that  the angular variables must be related by a linear transformation
\begin{align}
\label{ortototoricangles}
\Phi & = a_1 \tau^1 + a_2 \tau^2 \, ,  \notag\\
\Psi & = b_1 \tau^1 + b_2 \tau^2 \, , 
\end{align}
so that the relation between  $x,\rho$  and the moment map  coordinates is
\begin{align}
\label{ortototoricangles}
y_1 & =\ell^2 \left(  a_1 (x+\rho) + b_1 x\rho \right)\, ,  \notag\\
y_2 & =\ell^2 \left(  a_2 (x+\rho) + b_2 x\rho \right) \, , \
\end{align}
where a priori $a_1, b_1, a_2, b_2$ are arbitrary real constants, with $c_D\equiv a_1 b_2 - a_2 b_1\neq 0$.
Indeed, 
comparing the metric (\ref{toricmetric}) with the orthotoric form (\ref{ortometric}), we find 
\begin{align}
\label{Gcomponentsorto}
\begin{aligned}
G^{11} &   =  \ell^2 \frac{(a_1+ b_1x )^2 \mathcal{F}(\rho)  + (a_1 + b_1 \rho )^2 \mathcal{G}(x)  }{x-\rho }\,, \\
G^{22} &   =    \ell^2 \frac{(a_2+ b_2x )^2 \mathcal{F}(\rho)  + (a_2 + b_1 \rho )^2 \mathcal{G} (x) }{x-\rho }\, , \\
G^{12} & = \ell^2 \frac{ (a_1+ b_1 x) (a_2+ b_2 x)\mathcal{F}(\rho) + (a_1+b_1 \rho ) (a_2+ b_2 \rho )\mathcal{G}(x)}{x-\rho}\, , 
\end{aligned}
\end{align}
without any restriction on the constants $a_i,b_i$. Note that 
\begin{align}
\mathrm{det} (G^{ij}) = \ell^4 c_D \mathcal{F}(\rho) \mathcal{G} (x) \, , 
\end{align}
showing that the metric on the K\"ahler base degenerates precisely at the zeroes of $\mathcal{G} (x)$ and of 
$\mathcal{F}(\rho)$. With a little more work it is also possible to find the symplectic potential $G(y)$, as shown in \cite{Lucietti:2022fqj}. However, we did not find this illuminating, so we will not record this here. 
For our purposes, it is more interesting to look at the angular part of the change of coordinates and establish a direct link between the Darboux coordinates and the $\phi$-coordinates and $\psi$-coordinates, where global properties of the $\CBS$ and  the topological soliton geometries are transparent, respecively. 
We dub $\tau^i \equiv  (y,\tau^1,\tau^2)$, $i = 0,1,2$, ``$\tau-$coordinates'' and $\partial_{\tau^i}$ the associated basis for the torus action. Then
\begin{align}
  \mqty(y \\ \Phi \\ \Psi) ={}& D \mqty(y \\ \tau^1 \\ \tau^2) \,, \qquad D = \mqty(1 & 0 & 0 \\ 0 & a_1 & a_2 \\ 0 & b_1 & b_2) \,. 
\end{align}
The relation between the $\phi-$coordinates and the  $\tau-$coordinates is then given by
\begin{align}\label{eq:matA}
  \mqty(\phi_0 \\ \phi_1 \\ \phi_2) ={}& A \mqty(y \\ \tau^1 \\ \tau^2) \,, \qquad A = BCD = \left(
\begin{array}{ccc}
 \frac{2 \pi \iu}{\beta } & 0 & 0 \\
 -\frac{\omegafuga}{\beta} & \gamma_a (b_1- a_1) & \gamma_a (b_2-a_2) \\
 -\frac{\omegafugb}{\beta} & -\gamma_b (a_1+b_1) & - \gamma_b (a_2+b_2) \\
\end{array}
\right)\, ,
\end{align}
and we note that 
\begin{align}
\mathrm{det}(A) = \frac{4\pi \iu}{\beta} c_D \gamma_a \gamma_b \, . 
\end{align}
Thus, the angular volumes of the two coordinate systems are related as
\begin{align}
  \dd[]{\phi_0} \wedge \dd[]{\phi_1} \wedge \dd[]{\phi_2} ={}& 2 c_D \gamma_a \gamma_b \dd[]{\qty(\frac{2\pi \iu}{\beta}y)} \wedge \dd[]{\tau^1} \wedge \dd[]{\tau^2} \,. 
\end{align}
The $\phi-$coordinates have canonical $2\pi$-periodicities. Demanding that $\qty(\widetilde{y} = \frac{2\pi\iu}{\beta} y, \tau^1, \tau^2)$ also have canonical $2\pi$-periodicities fixes
\begin{align}\label{eq:cD}
  c_D ={}& a_1 b_2 - a_2 b_1 = \frac{1}{2\gamma_a \gamma_b} \,. 
\end{align}
As a check,  the supersymmetric Killing vector correctly reads
\begin{align}
 \label{reebtaubasis}
(\xi )_{\tau-\mathrm{basis}} =  A^{-1} (\xi )_{\phi-\mathrm{basis}} =  \left(1,0,0\right) \, . 
\end{align}
For the topological soliton we only need the transformation
\begin{align}
  \mqty(\tau \\ \phi \\ \psi) ={}& A_0 \mqty(y \\ \tau^1 \\ \tau^2) \,, \qquad A_0 = CD =  \left(
\begin{array}{ccc}
 1 & 0 & 0 \\
 \frac{1}{\ell} & \gamma_a (b_1- a_1) & \gamma_a (b_2-a_2) \\
 \frac{1}{\ell} & -\gamma_b (a_1+b_1) & - \gamma_b (a_2+b_2) \\
\end{array}
\right)\, , 
\end{align}
where here $\widetilde{m}=\widetilde{m}_0$ and we note that  $\mathrm{det}(A_0) = 2 c_D \gamma_a \gamma_b$. 
 It is now straightforward to determine the vectors of the fan in the $\tau-$basis as $(V^a )_{\tau-\mathrm{basis}} = A^{-1} (V^a)_{\phi-\mathrm{basis}}$ for the $\CBS$: 
\begin{align}\label{eq:VtauCBS}
  \begin{aligned}
  (V^0)_{\tau\mathrm{-basis}} ={}& \qty(0, - \gamma_b (a_2 + b_2), \gamma_b(a_1 + b_1)) \,, \\
  (V^1)_{\tau\mathrm{-basis}} ={}&  \qty(\frac{\beta}{2\pi\iu}, -c_1 (a_2 + \rho_+ b_2), c_1 (a_1 + \rho_+ b_1)) \,, \\
  (V^2)_{\tau\mathrm{-basis}} ={}&  \qty(0, \gamma_a(a_2 - b_2), -\gamma_a(a_1 - b_1)) \,, 
  \end{aligned}
\end{align}
where
\begin{align}
  c_1 ={}& \frac{4(\ell^2 - a^2)(\ell^2 - b^2)}{(\ell(a + b) + ab - 2\iu(a + b + \ell)r_+ - 3 r_+^2)(\ell(a + b) + a b - r_+^2)} \,, 
\end{align}
and $(V^a )_{\tau-\mathrm{basis}} = A_0^{-1} (V^a)_{\psi-\mathrm{basis}}$ for the topological soliton:
\begin{align}\label{eq:Vtausoliton}
  \begin{aligned}
    (V^0)_{\tau\mathrm{-basis}} ={}& (0, - \gamma_b (a_2 + b_2), \gamma_b (a_1 + b_1)) \,, \\
    (V^1)_{\tau\mathrm{-basis}} ={}& \qty(0, - \frac{4\ell^3(\ell - b)}{(\ell(a + b) + ab)^2}\qty(a_2 + \rho_0 b_2), \frac{4\ell^3(\ell - b)}{(\ell(a + b) + ab)^2}(a_1 + \rho_0 b_1)) \,, \\
    (V^2)_{\tau\mathrm{-basis}} ={}& (0, \gamma_a(a_2 - b_2), -\gamma_a(a_1 - b_1)) \,. \\
  \end{aligned}
\end{align}
From these expressions we can then read off the corresponding two-dimensional fans. The interpretation of the third components of the fans in this basis is discussed in section \ref{sect:symptoric} and more details are presented in Appendix 
 \ref{comparisonwithlucietti} below. As a preliminary to  discussing  global aspects of the four dimensional transverse geometry below, we introduce the notation
\begin{align}\label{eq:def2dfan}
  (V^a )_{\tau-\mathrm{basis}} = (t^a ,v^a) \,,
\end{align}
where $v^a =(v^a_1,v^a_2)$ are two-dimensional vectors, consistently with the notation used  in section \ref{Five-dimensional toric geometries}.

\section{Four dimensional view}

\label{comparisonwithlucietti}

The purpose of this appendix is to discuss global aspects of the four-dimensional transverse geometry which, with appropriate choices of the supersymmetric Killing vector, 
 may be interpreted as a toric orbifold. 
We will also show how to reconstruct the three-dimensional fans from the point of view of the toric geometry of the base, using 
the explicit solution of \cite{Chong:2005hr}  as a concrete example. This will allow us to  make contact with the approach of 
 \cite{Lucietti:2022fqj}, who attempted a classification of supersymmetric solutions of five-dimensional gauged supergravity with a toric symmetry,  employing the formalism of toric symplectic geometry.
This is closely related to the setup that we consider in this paper, with one important difference: the solutions considered in  \cite{Lucietti:2022fqj} are always Lorentzian and therefore possible horizons are necessarily  \emph{extremal}.  We will show that the extremal CCLP black hole arising as a limit of the general $\CBS$ solution 
 fits in the Darboux coordinates as discussed in  \cite{Lucietti:2022fqj}.
On the other hand, 
toric geometries of the topological soliton type, i.e.  without horizons, 
 fall directly into the formalism of  \cite{Lucietti:2022fqj}.

\subsection{Spindle bolt geometries}
\label{spindlebolts}

We start from the 2d vectors of the $\CBS$ \eqref{eq:VtauCBS} and we calculate 
\begin{align}
\mathrm{det} (v^0,v^1) =   \frac{\omegafugb}{2\pi\iu}  \,, \qquad \mathrm{det} (v^1,v^2) = \frac{\omegafuga}{2\pi\iu}  \, , \qquad \mathrm{det} (v^0,v^2)  =  1  \, .
\end{align}
This means that if we formally take $ \frac{\omegafugb}{2\pi \iu}=n_2, \frac{\omegafuga}{2\pi \iu}=n_1  \in \mathbb{N}$, the two-dimensional fan 
 describes  a  ${\cal O}(-1)$ orbifold line bundle over a spindle $\spindle_{[n_2,n_1]}$.  In the actual solution this cannot be arranged as  $\omegafuga,\omegafugb\in \mathbb{R}+\iu \mathbb{R}$, thus this  can be interpreted only as an analytic continuation. We can now go back to the fan and check that with this assumption on $\omegafuga,\omegafugb$ the vectors can be taken in $\mathbb{Z}^2$ and they correctly describe the orbifold ${\cal O}(-1) \to \spindle_{[n_2,n_1]}$. 
  To do this we just need to show that  it is possible to choose   $a_1,a_2,b_1,b_2$ so that   the matrix $A$ in \eqref{eq:matA} belongs (up the factor of $\frac{\beta}{2\pi \iu}$ 
  that may be removed by dividing the first colum), to $\text{SL}(3,\mathbb{Z})$ consistently with the fact that it relates two sets of angular coordinates with canonical $2\pi$-periodicities.  
 In particular, we require that 
 \begin{align}
\left(
\begin{array}{ccc}
1 & 0 & 0 \\
 -n_1& \gamma_a (b_1- a_1) & \gamma_a (b_2-a_2) \\
 -n_2& -\gamma_b (a_1+b_1) & - \gamma_b (a_2+b_2) \\
\end{array}
\right)\in \text{SL}(3;\mathbb{Z})\, ,
\end{align}
whose general solution is given by 
\begin{align}
  \begin{aligned}
    a_1  ={}& -\frac{1}{2}\left(\frac{k_1}{\gamma_a}+  \frac{k_3}{\gamma_b}   \right) \, , &  a_2  ={}& -  \frac{1}{2}\left(\frac{k_2}{\gamma_a} + \frac{k_4}{\gamma_b}   \right) \, ,\\
    b_1  ={}& \frac{1}{2}\left(\frac{k_1}{\gamma_a} - \frac{k_3}{\gamma_b}   \right) \, , & b_2  ={}&\frac{1}{2}\left(\frac{k_2}{\gamma_a} - \frac{k_4}{\gamma_b}   \right) \, ,
  \end{aligned}
 \end{align}
 with $k_1, k_2,  k_3, k_4\in \mathbb{Z}$, such that   $k_1 k_3 - k_2 k_3=1$.  For simplicity and without loss of generality we may set $k_2=k_3=0$ and  $k_1=k_4=1$ so that 
\begin{align}
\label{simplesolution}
a_1  =  -b_1= -\frac{1}{2\gamma_a}  \, , \qquad a_2  = b_2= -   \frac{1}{2\gamma_b}    \, .
\end{align}
Plugging these back into \eqref{eq:VtauCBS} we obtain
\begin{align}
  v^0 ={}& (1,0) \,, \qquad v^1 = \qty(\frac{\omegafuga}{2\pi\iu}, \frac{\omegafugb}{2\pi\iu}) \longrightarrow (n_1, n_2) \,, \qquad v^2 = (0,1) \,, 
\end{align}
where the arrow represents the analytic continuation to integer $\frac{\vb*{\omega}_{1,2}}{2\pi\iu}$. Thus, we indeed describe the orbifold $\mathcal{O}(-1) \rightarrow \mathbb{\Sigma}_{[n_2, n_1]}$, as claimed.
 
In four dimensions  explicit supergravity solutions on ${\cal O}(-t) \to \spindle_{[n_2,n_1]}$ orbifolds have been constructed recently in  
 \cite{Crisafio:2024fyc}.  In the present context such  ``spindle bolt'' geometry must materialize in the four-dimensional K\"ahler base with metric (\ref{ortometric}). We will be brief as this is standard regularity analysis. In particular, let us look at the metric induced on the compact toric divisor $D_{1} = \{ \rho=\rho_+\}$: since   $v^1_1  \partial_{\tau^1}+v^1_2  \partial_{\tau^2} = 0$
then on $D_1$ we must have that $\tau_2 =\Gamma \tau_1$ with
\begin{align}
\Gamma = -\frac{v^1_2}{v_1^1} = \frac{a_1+\rho_+ b_1}{a_2+\rho_+ b_2} \,.
\end{align}
 The metric induced on $D_1$ near to the south ($x=-1$) and north ($x=1$) poles then reads
  \begin{align}
\ell^{-2} \dd s^2_{D_1} \approx   \frac{C_\mp}{x\mp 1} \dd[]{x}^2 + \frac{x\mp 1}{ C_\mp} 4 (a_1 + \rho_+ b_1)^2 \dd \tau_1^2\, , 
\end{align}
 where $C_\mp =\frac{\mp 1 -\rho_+} {{\cal G}'(\mp 1)}$. Changing coordinate to $R^2=(x\mp 1)/C_\mp$ and  
 taking $\Delta \tau^1=\frac{2\pi}{2c_1(a_1+\rho_+ b_1)} $ then shows  that at the poles there are conical singularities with periodicities of  $\frac{2\pi}{n_1}$ and 
 $\frac{2\pi}{n_2}$, respectively.

 As a consistency check, we can work out the topology of the three-dimensional locus  $\D_1$ corresponding to $\rho=\rho_+$, that is given by a $\text{U}(1)$-bundle over the spindle 
  $D_1= \spindle_{[n_2,n_1]}$. The connection of this bundle is given by the restriction of the global angular form 
$(\dd y+ \omega)|_{D_1}$. Introducing  fiber coordinate $\tilde y = 2\pi \iu y/\beta$ with canonical $2\pi$-periodicity, 
 the correctly normalized first Chern class is  computed to be  
\begin{align}
\label{firstchernCBS}
\frac{1}{2\pi} \int_{D_1}\dd (\dd \tilde y +\frac{2\pi \iu }{\beta} \omega) = \frac{(2\pi\iu)^2}{\omegafuga \omegafugb} \longrightarrow \frac{1}{n_1 n_2} \, ,
\end{align}
where the arrow represents the analytic continuation to integer $\frac{\vb*{\omega}_{1,2}}{2\pi\iu}$. Thus after the analytic continuation we have a $\mathbb{Z}^2$-valued 2d fan and the three-dimensional locus $\D_1$ (i.e. the non-extremal horizon) has the correct $S^3$ topology.
 
 A similar analysis can be performed for the topological soliton solution. 
  In this case 
\begin{align}
\mathrm{det} (v^0,v^1) =    \frac{nv}{\mathrm{det}(A_0)} \,, \quad \mathrm{det} (v^1,v^2) =   \frac{v}{\mathrm{det}(A_0)} \, , \quad \mathrm{det} (v^0,v^2)  =  \frac{1}{\mathrm{det}(A_0)} \, , 
\end{align}
therefore picking $c_D$ as before so that $\mathrm{det}(A_0)=1$  the topology of the base is now exactly (i.e. not just as an analytic continuation) that of a 
${\cal O}(-1)$ bundle over  of a spindle $\spindle_{[n,1]}$. 
Moreover, we now find that 
\begin{align}
\label{firstchernTS}
 \frac{1}{2\pi} \int_{D_1}\dd  \omega  =  0\, , 
\end{align}
implying that the $\text{U}(1)$ bundle over  $\spindle_{[n,1]}$ is trivial and therefore the three-dimensional locus at $\rho=\rho_0$ is now 
 $\D_1= \spindle_{[n,1]}\times S^1$. Again, it is  possible to choose the coefficients $a_1,a_2,b_1,b_2$ so that the matrix $A_0 \in \text{SL}(3,\mathbb{Z})$ and  plugging these values back into 
  \eqref{eq:Vtausoliton} we get the 
 $\mathbb{Z}^2$-valued 2d fan
 \begin{align}
v^0 = (1,0)  \, , \qquad v^1 = (v,vn)  \, , \qquad v^2 = (0,1)  \, ,
\end{align}
describing the orbifold ${\cal O}(-1) \to \spindle_{[n,1]}$, with the normal bundle to the spindle being $\mathbb{C}/\mathbb{Z}_v$.

Note that the topology of the base is qualitatively the same orbifold for the $\CBS$ and the topological soliton and the difference of the two solutions is completely 
encoded in the first Chern classes of the fibrations, namely (\ref{firstchernCBS}) vs (\ref{firstchernTS}). 
Another difference from the $\CBS$ case is that in the topological soliton geometry there is a two-cycle $[D_1]  \in H_2({\cal O}(-1) \to \spindle_{[n,1]})$, 
represented by the zero section of the line bundle, namely the spindle. Therefore, the flux of the gauge field through this cycle has to be correctly quantized. A short computation gives
\begin{align}
 \frac{1}{2\pi} \int_{D_1} {\cal R}  =  -\frac{\curly{x} \sqrt{3}} {\ell}  \frac{1}{2\pi} \int_{D_1} F =   3 \, \frac{2b \ell + a b +  a \ell}{\ell -b}\, , 
\end{align}
which, using the quantization conditions (\ref{TSquantization}),  indeed can be rewritten as 
\begin{align}
 \frac{1}{2\pi} \int_{D_1} {\cal R}  =\frac{1+n}{n} +\frac{1}{vn}\, ,
    \end{align}
 in agreement with the general formula for the integral of the Ricci form in the orbifold geometry; see \emph{e.g.} eq. (3.131) in  \cite{Crisafio:2024fyc}.

\subsection{Fans from moment maps}

One of the main advantages of symplectic toric geometry is that there exist moment maps for the torus action and the fan associated to a manifold
 can be extracted from the image of the moment maps, using the classic Delzant construction, or some generalizations (to orbifolds, cones, etc). 
Let us show how this works for the $\CBS$ and the topological soliton, starting from the moment maps (\ref{ortototoricangles}). 
Setting $x=1$ ($\theta =0$) in (\ref{ortototoricangles}) we have
\begin{align}
y_1 & =   \ell^2  ( \rho (a_1+b_1) +a_1)  \, , \notag\\
y_2 & =  \ell^2 ( \rho (a_2+b_2) +a_2) \, ,
\end{align}
which for $\rho\in [\rho_c,\infty)$ gives a half-line on the line 
\begin{align}
(a_2+ b_2) y_1 - (a_1+ b_1) y_2  & =    \ell^2  (a_1 b_2 - a_2 b_1)   \, .
\end{align}
Similarly, setting $x=-1$ ($\theta =\pi/2$) in (\ref{ortototoricangles}) we have
\begin{align}
y_1 & =   \ell^2  ( \rho (a_1-b_1) -a_1)  \, , \notag\\
y_2 & =  \ell^2 ( \rho (a_2-b_2) -a_2) \, ,
\end{align}
which  for $\rho\in [\rho_c,\infty)$ gives a half-line on the line 
\begin{align}
(a_2- b_2) y_1 - (a_1- b_1) y_2  & =    \ell^2  (a_1 b_2 - a_2 b_1)   \, .
\end{align}
Finally, setting $\rho=\rho_c$, gives a segment on the line
\begin{align}
(a_2+\rho_c b_2) y_1 - (a_1+\rho_c b_1) y_2  & =    \ell^2 \rho_c^2 (a_1 b_2 - a_2 b_1)   \, ,
\end{align}
connecting the two half-lines at the points  $(\ell^2  ( \rho_c (a_1+b_1) +a_1) ,  \ell^2 ( \rho_c (a_2+b_2) +a_2))$
and $(\ell^2  ( \rho_c (a_1-b_1) -a_1) ,  \ell^2 ( \rho_c (a_2-b_2) -a_2))$.
The three two-dimensional vectors corresponding to the polytope are then precisely 
proportional to  vectors $v^a$ appearing in \eqref{eq:VtauCBS} (for $\rho_c=\rho_+$) or in \eqref{eq:Vtausoliton} (for $\rho_c=\rho_0$), and with the correct normalizations they must eventually coincide with those.

As discussed in section \ref{sect:symptoric}, from the fan $v^a$ it is now possible to reconstruct the 3d fan $V^a$, by computing $t^a = v^a_i \omega_i |_{D_a}$. 
Using (\ref{omegaorto}) we find that for the $\CBS$
\begin{align}
t^0  = 0\, , \qquad t^1=\frac{\beta}{2\pi \im} \, , \qquad t^2 = 0 \, ,
\end{align}
consistently with \eqref{eq:VtauCBS}, while for the topological soliton we find
\begin{align}
t^0 = t^1 =  t^2 = 0 \, ,
\end{align}
consistently with \eqref{eq:Vtausoliton}.

Let us now impose  that the solution is extremal, by demanding that $\tilde m\to 0$.  The change of coordinates 
(\ref{ortotocclp}) from the original CCLP ones to the orthotoric coordinates is singular and  we then 
define
\begin{align}
a_i  \equiv \varepsilon \tilde a_i  \, , \qquad b_i  \equiv \varepsilon \tilde b_i  \, , 
\end{align}
and  consider the scaling limit as in 
(\ref{scalinglimit}). In particular, taking $a_i,b_i$ as in (\ref{simplesolution}) and setting  $k=-2/(a^2-b^2)$ for simplicity we find
\begin{align}
\tilde \rho = r^2 - r_*^2 \, ,
\end{align}
where $r_* ^2\equiv \ell (a+b) + a b $ is the extremal horizon, so that $r=r_*$ corresponds to $\tilde \rho=0$.
In the limit we find that the symplectic coordinates become 
\begin{align}
y_1 & =  \frac{\ell^2}{2(a^2-\ell^2)}  \sin^2 \theta  \,\tilde  \rho\, , \qquad y_2 =   \frac{\ell^2}{2(b^2-\ell^2)}\cos^2  \theta  \, \tilde \rho \, ,  
\end{align}
 so that the extremal horizon corresponds to the origin $(y_1,y_2)=(0,0)$, as shown in  \cite{Lucietti:2022fqj}.
 Recalling  that in the extremal limit $\Omega_1=\Omega_2=1$, we get 
\begin{align}
 \tau_1 = \phi_1 \, , \qquad \tau_2 = \phi_2    \, , 
\end{align}
meaning that the $2\pi$-periodic coordinates $\phi_1,\phi_2$ are exactly the angular coordinates of the symplectic toric description.  
 
The image of the above moment maps is clearly a quadrant of $\mathbb{R}^2$ with the origin being the extremal horizon. This may be thought of as obtained from 
the polytope of the non-extremal $\CBS$ solution after collapsing the segment corresponding to the compact divisor to a point. It is interesting that this is precisely analogous to the ordinary 
blow up/blow down procedure in symplectic toric geometry. This is schematically shown in \cref{fig:CBS_extremalBH}. 

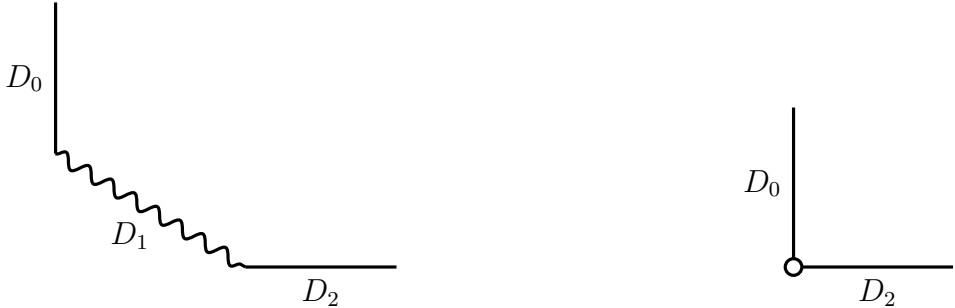
\begin{figure}[h]
\centering
\begin{subfigure}[b]{0.49\textwidth}
\centering
\begin{tikzpicture}
  \def\unit{1}
  \def\x{2.5\unit}
  \def\y{1.5*\unit}
  \def\leg{2*\unit}
  \def\axlength{4*\unit}
  \coordinate (O) at (0,0);
  \coordinate (B) at ($(O) + ({0},{\y})$);
  \coordinate (C) at ($(O) + ({\x},{0})$);
  \coordinate (A) at ($(B) + ({0},{\leg})$);
  \coordinate (D) at ($(C) + ({\leg},{0})$);
   

  \draw[very thick] (A) -- (B) node[midway, left] {$D_0$}; 
  \draw[very thick, decorate, decoration = snake] (B) -- (C) node[midway, below left, xshift=4] {$D_1$}; 
  \draw[very thick] (C) -- (D) node[midway, below] {$D_2$};

\end{tikzpicture}
\end{subfigure}
\hfill
\begin{subfigure}[b]{0.49\textwidth}
\centering
\begin{tikzpicture}
  \def\unit{1}
  \def\x{2.5\unit}
  \def\y{1.5*\unit}
  \def\leg{2*\unit}
  \def\axlength{4*\unit}
  \def\radius{0.1\unit}
  \coordinate (O) at (0,0);
  \coordinate (B) at ($(O) + ({0},{\radius})$);
  \coordinate (C) at ($(O) + ({\radius},{0})$);
  \coordinate (A) at ($(B) + ({0},{\leg})$);
  \coordinate (D) at ($(C) + ({\leg},{0})$);
   

  \draw[very thick] (A) -- (B) node[midway, left] {$D_0$}; 
  \draw[very thick] (O) circle (\radius);
  \draw[very thick] (C) -- (D) node[midway, below] {$D_2$};
\end{tikzpicture}
\end{subfigure}
\caption{The two-dimensional polytope of the non-extremal $\CBS$ solution (on the left) and of the extremal solution (on the right). The latter is obtained  by collapsing the segment corresponding to the compact divisor to a point.}
\label{fig:CBS_extremalBH}
\end{figure}

\section{Details of the Fefferman-Graham expansions}
\label{app:FG expansions}
In this appendix we redo the analysis of \cite{BenettiGenolini:2016tsn}, where the supersymmetry conditions are imposed asymptotically to the total 5d solution order by order. The difference with \cite{BenettiGenolini:2016tsn} is that we demand that the 5d solution is asymptotically $\text{AdS}_5$, as opposed to asymptotically locally $\text{AdS}_5$, thus the boundary is conformally flat. We also provide a supplementary \texttt{Mathematica} notebook containing the higher order expressions which were not explicitly given in \cite{BenettiGenolini:2016tsn}. For convenience we set $\ell = 1$ throughout.

\paragraph{A generic $\text{U}\qty(1)^3$ ansatz.} We begin with the 5d metric \eqref{eq:5d_metric}, assuming a $\text{U}\qty(1)^3$ isometry. This means that the base can be generically written as
\begin{align}\label{eq:base_metric}
  \dd[]{s}^2_\gamma ={}& U(r, z, \bar{z})^2 \qty[ \frac{\dd[]{r}^2}{r^2} + 4 r^2 W(r, z, \bar{z})^2 \dd[]{z} \dd[]{\bar{z}}] + \frac{r^4}{U(r, z, \bar{z})^2} (\dd[]{\hat{\psi}} + \phi)^2 \,, 
\end{align}
where $r$ denotes the radial coordinate, $z$ is a complex coordinate and the three $\text{U}\qty(1)$'s are generated by $(\partial_y, \partial_{\hat{\psi}}, \partial_\varphi)$, with $z = \abs{z} \eu^{- \iu \varphi}$ and $(U,W)$ are real functions. Through a redefinition of $\hat{\psi}$, the one form $\phi$ can always we brought to
\begin{align}
  \phi ={}& \phi_z(r, z, \bar{z}) \dd[]{z} + \overline{\phi_z(r, z, \bar{z})} \dd[]{\bar{z}} \,. 
\end{align}
Similarly through a redefinition of $y$ in \eqref{eq:5d_metric} the one form $\omega$ can always be brought to
\begin{align}
  \omega ={}& c(r, z, \bar{z}) (\dd[]{\hat{\psi}} + \phi) + C_z(r, z, \bar{z}) \dd[]{z} + \overline{C_z(r, z, \bar{z})} \dd[]{\bar{z}} \,. 
\end{align}
Choosing to order the coordinates as $(\hat{\psi},r, z, \bar{z})$ fixes the orientation on the base to
\begin{align}
  \star_\gamma 1 ={}& 2\iu r^3 U^2 W^2 \dd[]{z} \wedge \dd[]{\bar{z}} \wedge \dd[]{\hat{\psi}} \wedge \dd[]{r} \,.
\end{align}
Then, the conditions $\star_\gamma 1 = - \frac{1}{2} \X^1 \wedge \X^1$ and $\star_\gamma \X^1 = - \X^1$ determine
\begin{align}
  \X^1 ={}& \pm \qty(2\iu r^2 U^2 W^2 \dd[]{z} \wedge \dd[]{\bar{z}} + r \dd[]{r} \wedge (\dd[]{\hat{\psi}} + \phi)) \,. 
\end{align}
We choose the $ + $ sign and proceed. The conditions \eqref{eq:XI_conditions} uniquely fix
\begin{align}
  \X^2 + \iu \X^3 ={}& - 2 W \qty(U^2 \dd[]{r} \wedge \dd[]{\bar{z}} + \iu r^3 \qty(\dd[]{\bar{z}} \wedge \dd[]{\hat{\psi}} - \phi_z \dd[]{z} \wedge \dd[]{\bar{z}})) \,. 
\end{align}
Then, using the reality of $U$, the condition \eqref{eq:P_condition} uniquely fixes
\begin{align}
  P ={}& - \frac{1}{U^2 W} \partial_r \qty(r^3 W) (\dd[]{\hat{\psi}} + \phi) - \iu \qty(\dd[]{\bar{z}} \partial_{\bar{z}} - \dd[]{z} \partial_z) \log{W} \,,  
\end{align}
and further imposes
\begin{align}\label{eq:phi_conditions}
  \begin{aligned}
    \partial_r \phi_z ={}& \frac{2\iu}{r^3} U \partial_z U \,, \quad \partial_r U = \frac{\iu \qty(\partial_{\bar{z}}\phi_z - \partial_{z} \overline{\phi_{z}}) - 4U^2 W \qty(W + r \partial_r W)}{4r U W^2} \,. 
  \end{aligned}
\end{align}
Imposing these relations, and their derivatives with respect to $(r,z, \bar{z}$), the differential of $\phi$ is uniquely fixed to
\begin{align}
\label{dd_phi}
  \dd[]{\phi} ={}& \frac{2\iu}{r} \partial_r \qty(r^2 U^2 W^2) \dd[]{z} \wedge \dd[]{\bar{z}} + \iu \qty(\dd[]{\bar{z}} \partial_{\bar{z}} - \dd[]{z} \partial_z) U^2 \wedge \frac{\dd[]{r}}{r^3} \,,
\end{align}
and the Ricci scalar on the base reads
\begin{align}
  R_\gamma ={}& - \frac{2}{r^2 U^2 W^2} \qty[\partial_z \partial_{\bar{z}} \log{W} + \partial_r \qty(rW \partial_r \qty(r^3 W)) + W \partial_r \qty(r^3 W)] \,. 
\end{align}
By imposing that expression \eqref{dd_phi} satisfies $\dd(\dd\phi)=0$, one obtains the following integrability equation:
\begin{align}\label{eq:integrability}
  \partial_z \partial_{\bar{z}} U^2 + r^3 \partial_r \qty[\frac{1}{r} \partial_r \qty(r^2 U^2 W^2)] ={}& 0 \,.  
\end{align}
Finally, supersymmetry demands that
\begin{align}\label{eq:omega_conditions}
  \begin{aligned}
  \dd[]{\omega} + \star_\gamma \dd[]{\omega} ={}& \frac{R_\gamma}{24} \qty(\mathcal{R} - \frac{1}{4} R_\gamma X^1) \,, \\
  \qty(\dd[]{\omega})_{mn} (\X^1)^{mn} ={}& - \frac{1}{12} \qty(\frac{1}{2} \nabla_\gamma^2 R_\gamma + \frac{2}{3} (R_\gamma)_{mn} (R_\gamma)^{mn} - \frac{1}{3} R_\gamma^2) \,. 
  \end{aligned}
\end{align}
On the total 5d space we order the coordinates as $(y, r, \hat{\psi}, z, \bar{z})$, such that the base and the total space have opposite orientation as in \eqref{eq:opposite_orientation}. 
For completeness, we reiterate that the total 5d metric and gauge field are given by
\begin{align}
  \begin{aligned}
    \dd[]{s}^2_g ={}& - f^2 (\dd[]{y} + \omega)^2 + f^{-1} \dd[]{s}^2_\gamma \,, \\
    A ={}& - \frac{\sqrt{3}}{\curly{x}} \qty[f (\dd[]{y} + \omega) + \frac{1}{3} P - f_0 \dd[]{y} + \dd[]{\lambda(\hat{\psi},z, \bar{z})}] \,,
  \end{aligned}
\end{align}
with $f = - 24/R_\gamma$.

\paragraph{Large-$r$ asymptotic expansion.} Next we expand the functions appearing above at large-$r$ as
\begin{align}\label{eq:large_r_expansion}
  \begin{aligned}
    U ={}& U_0 + U_2 \frac{1}{r^2} + U_4 \frac{1}{r^4} + U_6 \frac{1}{r^6} + \dots \,, \\
    W ={}& W_0 + W_2 \frac{1}{r^2} + W_4 \frac{1}{r^4} + W_6 \frac{1}{r^6} + \dots \,, \\
    \phi_z ={}& \phi_{z,0} + \phi_{z,2} \frac{1}{r^2} + \phi_{z,4} \frac{1}{r^4} + \phi_{z,6} \frac{1}{r^6} + \dots \,, \\
    c ={}& c_{-2} r^2 + c_0 + c_2 \frac{1}{r^2} + c_4 \frac{1}{r^4} + c_6 \frac{1}{r^6} + \dots \,, \\
    C_z ={}& C_{z,-2} r^2 + C_{z,0} + C_{z,2} \frac{1}{r^2} + C_{z,4} \frac{1}{r^4} + C_{z,6} \frac{1}{r^6} + \dots \,, \\
  \end{aligned}
\end{align}
where the coefficients $\qty(U_i, W_i, c_i)$ are real functions of $(z, \bar{z})$, and the coefficients $\qty(\phi_{z,i}, C_{z,i})$ are complex functions of $(z, \bar{z})$. First, the integrability condition \eqref{eq:integrability} fixes $\qty(U_4,U_6)$ in terms $(U_0, U_2, W_0, W_2, W_4, W_6)$ and further demands
\begin{align}
  \partial_z \partial_{\bar{z}} U_0 ={}& - \frac{\partial_{\bar{z}} U_0 \partial_z U_0}{U_0} \,. 
\end{align}
Then conditions \eqref{eq:phi_conditions} fix $\qty(\phi_{z,2}, \phi_{z,4}, \phi_{z,6})$ in terms of $(U_0,U_2)$ and further demand
\begin{align}
  \partial_z \overline{\phi_{z,0}} - \partial_{\bar{z}} \phi_{z,0} ={}& 4\iu U_0^2 W_0^2 \,. 
\end{align}
Finally, the conditions \eqref{eq:omega_conditions} fix $\qty(c_{-2}, c_0, c_2, c_4, c_{6}, C_{z, - 2}, C_{z,2}, C_{z,4},C_{6,z})$ in terms of

\noindent $(C_{z,0}, U_0, U_2, W_0, W_2, W_4, W_6)$. So far we have set the logarithmic terms in the large-$r$ expansion to zero but we have not fully demanded conformal flatness of the boundary. Doing so amounts to setting
\begin{align}
  U_0 ={}& \frac{1}{2} \sqrt{u} \,, \qquad W_0 = \eu^{w/2} = \frac{2}{u} \frac{1}{1 + z \bar{z}} \,,
\end{align}
where $u$ is a constant and $w$ is a convenient function appearing in the main text. The second condition above follows since conformal flatness demands $R_{\text{2d}} = \frac{u^2}{2}$, where $R_{\text{2d}}$ is the Ricci scalar of the 2d piece of the boundary metric: $\dd[]{s}^2_2 = 4 \eu^w \dd[]{z} \dd[]{\bar{z}}$. With these demands in place at leading order the 5d metric is
\begin{align}
  \dd[]{s}^2_g ={}& \frac{\dd[]{r}^2}{r^2} + r^2 \dd[]{s}_{\text{bdy}}^2 \,, 
\end{align}
and the conformal boundary is given by
\begin{align}
  \dd[]{s}^2_{\text{bdy}} ={}& \qty(1 - \frac{\iu}{\sqrt{u} \, \eu^{w/2}} \qty(\dd[]{C_0})_{z \bar{z}}) (\dd[]{\hat{\psi}} + \phi_0)^2 - 2 (\dd[]{y} + C_0)(\dd[]{\hat{\psi}} + \phi_0) + 4 \eu^w \dd[]{z} \dd[]{\bar{z}} \,,
\end{align}
where $C_0 = C_{z,0} \dd[]{z} + \overline{C_{z,0}} \dd[]{\bar{z}}$. For simplicity, and because it is sufficient for our purposes, we set $C_{z,0} = 0$ and redefine the remaining unspecified functions as
\begin{align}
  \phi_{z,0} ={}& a_z \,, \quad U_2 = \eu^{w/2} k_1 \,, \quad W_2 = \eu^{w/2} k_2 \,, \quad W_4 = \eu^{w/2} k_3 \,, \quad W_6 = \eu^{w/2} k_4 \,. 
\end{align}
The final expressions for all the coefficients appearing in \eqref{eq:large_r_expansion} in terms of the constant $u$ and the functions $(k_1, k_2, k_3, k_4)$ are stored in the variable 
\noindent \texttt{expansionFunctions} in the supplementary notebook.

\paragraph{Bringing the asymptotic expansion to Fefferman-Graham form.} Finally, we bring the so presented large-$r$ asymptotic expansion to Fefferman-Graham form. To do so, first we perform the coordinate transformation
\begin{align}\label{eq:toFG1}
  y ={}& t \,, \quad \hat{\psi} = \psi + t \,. 
\end{align}
Then we perform the further asymptotic coordinate transformation
\begin{align}\label{eq:toFG2}
  \begin{aligned}
  r \rightarrow {}& \frac{1}{\rho} \qty[1 + m_{r,2} \rho^2 + m_{r,4} \rho^4 + \dots ] \,, \\
  \psi \rightarrow {}& \psi + m_{\psi,4} \rho^4 + \dots \,, \\ 
  z \rightarrow {}& z + m_{z,4} \rho^4 + \dots \,,
  \end{aligned}
\end{align}
were the coefficients appearing above are all functions of (the new) $(z, \bar{z})$ and the boundary is now located at $\rho = 0$. Demanding Fefferman-Graham form means that the $\rho \rho$ component of the metric is $1/\rho^2$ and there are no cross terms between $\rho$ and any other coordinate (to the order of interest) --- this fixes the coefficients $m$. Their expressions in terms of the constant $u$ and the functions $(k_1, k_2, k_3, k_4)$ are stored in the variable \texttt{mFunctions} in the supplementary \texttt{Mathematica} notebook.

\bibliographystyle{ytphys}
\bibliography{equivodd}

\end{document}